\documentclass{aa}
\usepackage{sidecap}
\usepackage{graphicx}
\usepackage{txfonts}
\usepackage{multirow}
\usepackage[colorlinks=true, linkcolor=blue, urlcolor=blue, citecolor=blue]{hyperref}

\begin{document}

\title{Dissecting the 3D chemo-dynamical structures of NGC~1381: A galaxy hosting an ancient slow bar with an accretion-related bulge and thick disc}
\titlerunning{3D chemo-dynamical structures of NGC~1381}

\author
{Yunpeng Jin\inst{1}\thanks{E-mail: jyp199333@163.com}
\and Ling Zhu\inst{2}\thanks{E-mail: lzhu@shao.ac.cn}
\and Shude Mao\inst{1}
\and Marie Martig\inst{3}
\and Francesca Pinna\inst{4,5}
\and Glenn van de Ven\inst{6}
\and Yuchen Ding\inst{3}}

\institute
{Department of Astronomy, Westlake University, Hangzhou, Zhejiang 310030, China
\and Shanghai Astronomical Observatory, Chinese Academy of Sciences, 80 Nandan Road, Shanghai 200030, China
\and Astrophysics Research Institute, Liverpool John Moores University, 146 Brownlow Hill, Liverpool L3 5RF, UK
\and Instituto de Astrof\'isica de Canarias, calle V\'ia L\'actea s/n, E-38205 La Laguna, Tenerife, Spain
\and Departamento de Astrof\'isica, Universidad de La Laguna, Avenida Astrof\'isico Francisco S\'anchez s/n, E-38206 La Laguna, Spain
\and Department of Astrophysics, University of Vienna, T\"urkenschanzstra\ss{}e 17, 1180 Wien, Austria}

\date{}

\abstract
{We applied the barred population-orbit superposition method developed and validated in our previous 2025 papers to construct 3D chemo-dynamical models for the barred S0 galaxy NGC~1381 (FCC 170) in the Fornax cluster. Based on the properties of stellar orbits in the models, we decomposed NGC~1381 into six components with distinct kinematics and morphologies: (1) a dynamically warm nuclear disc; (2) a rigidly rotating, boxy, peanut, or X-shaped bar; (3) a dynamically hot, spheroidal bulge; (4) a dynamically cold thin disc; (5) a vertically extended thick disc with slightly slower rotation than the thin disc; and (6) a dynamically hot, spatially diffuse stellar halo. The luminosity fractions of the dynamical components indicate that the luminosity of NGC~1381 is dominated by the bar ($f_{\rm bar}\sim30\%$) and thin disc ($f_{\rm thin}\sim28\%$), with notable contributions from the thick disc ($f_{\rm thick}\sim16\%$) and bulge ($f_{\rm bulge}\sim17\%$) and only minor contributions from the nuclear disc ($f_{\rm nucl}\sim5\%$) and halo ($f_{\rm halo}\sim5\%$). The nuclear disc, bar, and thin disc are metal-rich ($[Z/\rm H]\gtrsim0$), $\alpha$-poor ($\rm[Mg/Fe]\lesssim0.2$), and old ($\sim13\rm\,Gyr$), consistent with in situ formation in the early Universe. The bulge, thick disc, and stellar halo are metal-poor ($[Z/\rm H]\lesssim0$), $\alpha$-rich ($\rm[Mg/Fe]\gtrsim0.2$), and younger than or comparable in age to the in situ components, suggesting that they are related to ex situ formation through the accretion of minor mergers. Further analysis of stellar population gradients for these ex situ-related components reveals that the thick disc and stellar halo, with their flat metallicity and [Mg/Fe] gradients, are likely dominated by a similar ex situ population. In contrast, the bulge exhibits a negative metallicity gradient ($\nabla[Z/\rm H]_{bulge}<0$), which might originate from either predominantly ex situ stars or a mixture of in situ and ex situ stars; the bulge's ex situ fraction cannot be quantified with the current data. Our modelling also reveals the presence of a slow bar ($\mathcal{R}=2.40_{-0.27}^{+0.54}$), with a bar pattern speed of $\rm\Omega_p=34_{-7}^{+4}\,km\,s^{-1}\,kpc^{-1}$, a bar length of $R_{\rm bar}=2.24_{-0.22}^{+0.43}\rm\,kpc$, and a corotation radius of $R_{\rm CR}=5.38_{-0.28}^{+1.59}\rm\,kpc$, which is consistent with its ancient formation time.}


\maketitle
\nolinenumbers

\section{Introduction}
The classifications of galaxy structures originated from their morphologies and colours. For instance, the central regions of galaxies were initially classified as either red, spheroidal `classical bulges' or blue, flat `pseudobulges' linked to secular evolution (e.g. \citealp{Kormendy2004,Athanassoula2005}). However, later detailed morphological analyses have revealed the complexity of bulge regions (e.g. \citealp{Buta2015}). Recent observations have uncovered composite bulges in some galaxies, where distinct structures such as a classical bulge, a bar, and a nuclear disc can coexist (e.g. \citealp{Erwin2021,Tahmasebzadeh2024}).

In standard models of galaxy formation, bars, nuclear discs, and classical bulges have distinct origins: bars can form through internal instabilities in rotation-dominated discs (e.g. \citealp{Hohl1971,Ostriker1973,Efstathiou1982}) or external tidal perturbations (e.g. \citealp{Noguchi1996,Lokas2018,Yoon2019}); nuclear discs assemble from gas inflow towards galaxies' centres channelled by bars (e.g. \citealp{Athanassoula2003,Kormendy2004,Gadotti2011}); and classical bulges likely originate from dissipationless collapse or mergers (e.g. \citealp{Bournaud2007,Fisher2008,Gadotti2009,Hopkins2010}). 
However, how these formation mechanisms collectively influence the assembly history of galaxies, including the roles of in situ star formation and external accretion, remains poorly understood.

A key to understanding different structures lies in decomposing them using line-of-sight information. The traditional photometric decomposition techniques have been widely used over the past few decades. These techniques rely on strong assumptions such as bulges described by \citet{Sersic1968} profiles, discs modelled by exponential profiles, and bars approximated as prolate Ferrers bars (see \citealp{Binney2008}). However, these assumptions often deviate from reality. Previous studies have shown that the surface brightness profiles of discs can be down-bending or up-bending in their inner or outer regions (e.g. \citealp{MendezAbreu2017,Breda2020,Ding2023}), while real bars can be triaxial, exhibiting boxy, peanut, or X-shaped (hereafter BP/X-shaped) structures (e.g. \citealp{Shaw1987,Luetticke2000,Erwin2017,Li2017}). A recent study by \citet{Gadotti2026} demonstrated that whether photometrically decomposed bulges are identified as classical bulges or nuclear discs depends on both the optimisation algorithms used and the imaging data quality. All these findings indicate that traditional photometric decomposition techniques struggle to reliably distinguish between structures of different physical origins.

Compared to traditional photometric decomposition, dynamical decomposition can better trace the physical origins of different galaxy structures, as demonstrated in cosmological simulations (e.g. \citealp{Du2019,Du2020,Pulsoni2020,OrtegaMartinez2022,Zana2022,Cristiani2024}). In recent years, multiple integral field spectroscopy (IFS) surveys (e.g. \citealp{Bacon2001,Bacon2017,Cappellari2011,Sanchez2012,Bryant2015,Bundy2015}) have delivered spatially resolved spectra for thousands of nearby galaxies. From these spectra, one can derive stellar kinematic and population maps, enabling both the dynamical decomposition of real galaxies and the subsequent analysis of their chemical properties.

The dynamical decomposition of real galaxies can be performed using Schwarzschild's orbit-superposition method \citep{Schwarzschild1979,Schwarzschild1982,Schwarzschild1993}, in which 3D galaxy models are constructed via fitting of the luminosity distributions and stellar kinematic maps. This method has multiple implementations (e.g. \citealp{vdB2008,Long2018,Vasiliev2020,Neureiter2021,Quenneville2022,Dattathri2024}). The triaxial implementation developed by \citet{vdB2008} (see also the publicly available version DYNAMITE\footnote{\url{https://dynamics.univie.ac.at/dynamite_docs/}}; \citealp{Jethwa2020}) was initially designed without accounting for bars. It has been validated in simulations (e.g. \citealp{Zhu2018a,Jin2019}) and has been used to analyse dynamical structures of galaxies in IFS surveys (e.g. \citealp{Zhu2018a,Zhu2018b,Zhu2018c,Jin2020,Santucci2022,Santucci2023,Thater2025}).
Subsequently, this implementation was updated to support the modelling of stellar populations (e.g. \citealp{Poci2019,Poci2021,Zhu2020,Zhu2022a,Zhu2022b,Ding2023,Jin2024}) and to include bar structures for non-edge-on cases (inclination angle $\theta\lesssim80^\circ$; \citealp{Tahmasebzadeh2021,Tahmasebzadeh2022,Tahmasebzadeh2024}). The combination of dynamical modelling with stellar populations enables a powerful way to uncover the assembly history of galaxies \citep{vdV2025}. In recent work \citep{Jin2025a,Jin2025b}, we further enhanced this method by simultaneously incorporating bars and stellar populations in the modelling and extending it to handle edge-on ($\theta\gtrsim80^\circ$) galaxies. This improved method was validated using simulated galaxies from the Auriga simulations \citep{Grand2017,Grand2024} and can estimate the bar pattern speeds with a relative uncertainty of $\lesssim15\%$ for both side-on and end-on bars \citep{Jin2025a}. This method can also reliably characterise the luminosity distributions, ages, and metallicities of various dynamical structures, including bars, bulges, discs, and stellar halos \citep{Jin2025b}.

\begin{figure*}
    \sidecaption
    \includegraphics[width=12cm]{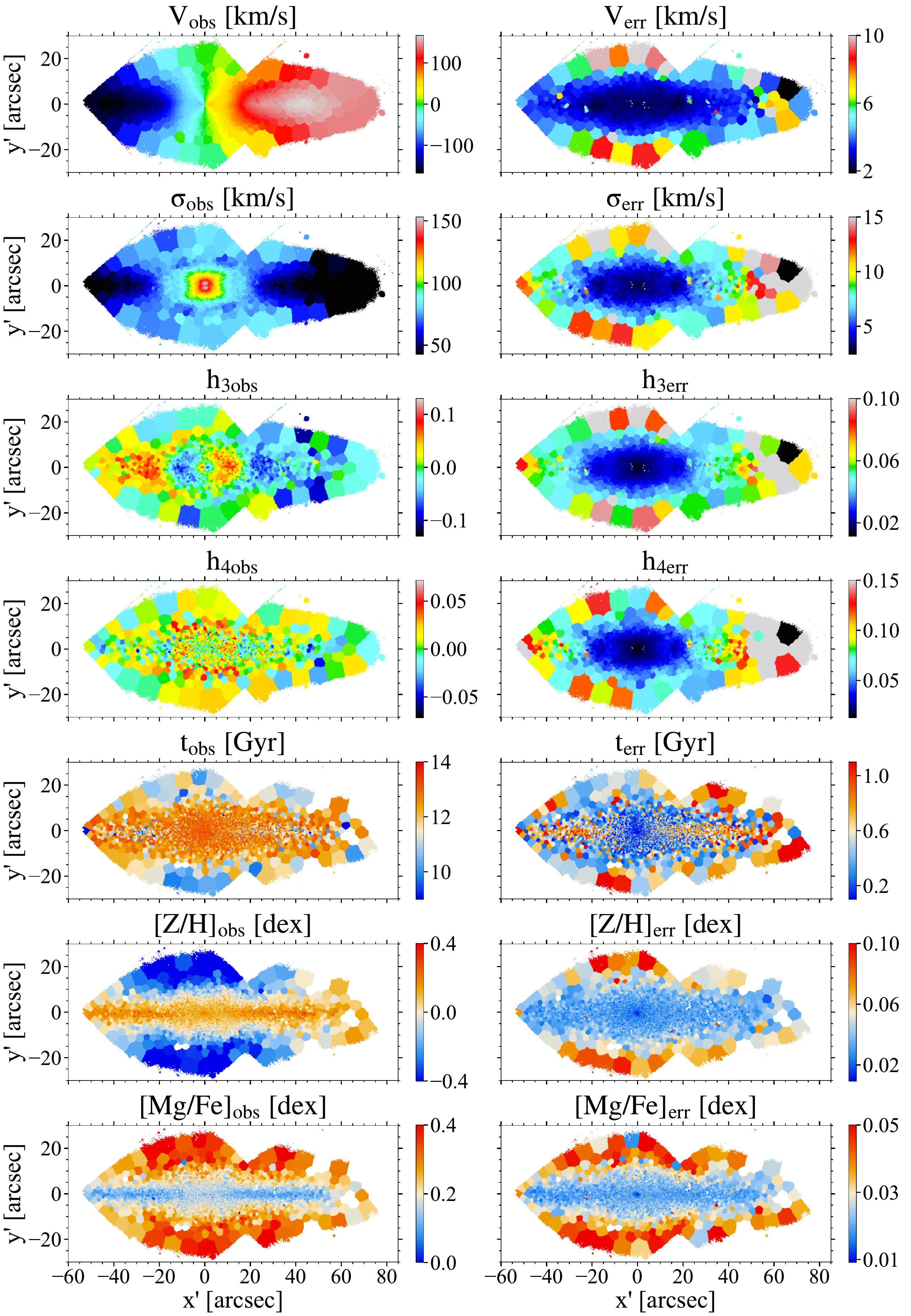}
    \caption{Stellar kinematic and population maps (\textit{left panels}) and their corresponding error maps (\textit{right panels}) for NGC~1381. \textit{From top to bottom}: Mean velocity, $V$; velocity dispersion, $\sigma$; third-order Gauss-Hermite coefficient, $h_3$; fourth-order Gauss-Hermite coefficient, $h_4$; stellar age, $t$; stellar metallicity, $[Z/\rm H]$; and [Mg/Fe] abundance.}
    \label{kinematic-chemo-data}
\end{figure*}

NGC~1381 (FCC 170) is an edge-on disc galaxy located in the central dense region of the Fornax cluster, with an estimated stellar mass of $2.25\times10^{10}\rm\,M_\odot$ \citep{Iodice2019a} and a distance of $19.57\rm\,Mpc$ (\citealp{Spriggs2021}; see also \citealp{Blakeslee2009} for a slightly different value). This galaxy was classified as an S0 galaxy \citep{Ferguson1989}. Subsequently, both optical and near-infrared images (e.g. \citealp{Luetticke2000,Bureau2006,Venhola2018}) revealed an obvious boxy bulge in its central region, which indicates the possible existence of a bar. The stellar kinematic maps confirm the presence of a bar by presenting an anti-correlated, correlated, and anti-correlated again $h_3$--$V$ relation along the galaxy's major axis (Fig.~\ref{kinematic-chemo-data}; see also \citealp{Pinna2019a,Poci2021}). The positive $h_3$--$V$ correlation is recognised as a robust bar diagnostic (e.g. \citealp{Bureau2005,Molaeinezhad2016,LiZhaoyu2018}), while the negative $h_3$--$V$ correlation at the innermost regions is the signature of a bar-induced nuclear disc (e.g. \citealp{Chung2004,Bureau2005,Gadotti2020,FraserMcKelvie2025}). This galaxy likely fell into the Fornax cluster approximately 8--12\,Gyr ago (e.g. \citealp{Iodice2019b,Ding2023}). During the galaxy's infall, interactions with the intracluster medium stripped the hot ionised gas from its stellar halo, cutting off the gas supply for star formation and resulting in strangulation-dominated quenching (e.g. \citealp{MorokumaMatsui2022,Martig2026}) and, consequently, old stellar populations in the BP/X-shaped regions nowadays (e.g. \citealp{Pinna2019a,MartinNavarro2021}). This implies that NGC~1381 probably formed its bar in the early Universe before falling into the cluster, with its internal structures preserved as fossil records until the present day.

The structures of NGC~1381 have been studied in previous works, but the different components in the galaxy's inner regions were not accurately distinguished. For example, based directly on the projected 2D morphology, kinematics, and stellar populations, \citet{Pinna2019a} divided the galaxy into distinct regions, each dominated by a different structure: a nuclear disc, a boxy structure, a thin disc, and inner and outer thick discs. Dynamical decompositions based on orbit-superposition modelling were later performed by \citet{Poci2021} and \citet{Ding2023}; however, their models did not explicitly include a non-axisymmetric bar and did not concentrate much on the nuclear disc. In this work, we modelled NGC~1381 using the barred population-orbit superposition method \citep{Jin2025a,Jin2025b}, which can effectively disentangle the bar and the dynamically hot bulge. This enabled a full dynamical decomposition of the galaxy into a nuclear disc, a bar, a bulge, a thin disc, a thick disc, and a stellar halo. We derived the 3D spatial distributions, kinematics, and chemical properties of these components, including the bar pattern speed and bar length.

The paper is organised as follows. In Sect.~\ref{sec2}, we introduce the photometric and spectroscopic data we used. In Sect.~\ref{sec3}, we describe how we constructed barred population-orbit superposition models for NGC~1381, including the stellar-orbit-based dynamical decomposition. In Sect.~\ref{sec4}, we characterise the bar, including its pattern speed, length, and corotation radius, as well as the luminosity distributions and chemical properties of all decomposed components. We discuss our results in Sect.~\ref{sec5} and summarise our conclusions in Sect.~\ref{sec6}.

Throughout the paper, the coordinate system ($x',y'$) represents the observing plane, with the galaxy's major axis aligned with the $x'$-axis. The intrinsic bar's rotating frame is denoted as ($x,y,z$), where the $x$-axis and $y$-axis coincide with the bar's major and minor axes, respectively. The inertial frame ($x_0,y_0,z$) coincides with $(x,y,z)$ at the initial time but deviates from it at later times due to the figure rotation about the $z$-axis.

\section{Photometric and spectroscopic data}
\label{sec2}
\begin{figure}
    \centering
    \includegraphics[width=9cm]{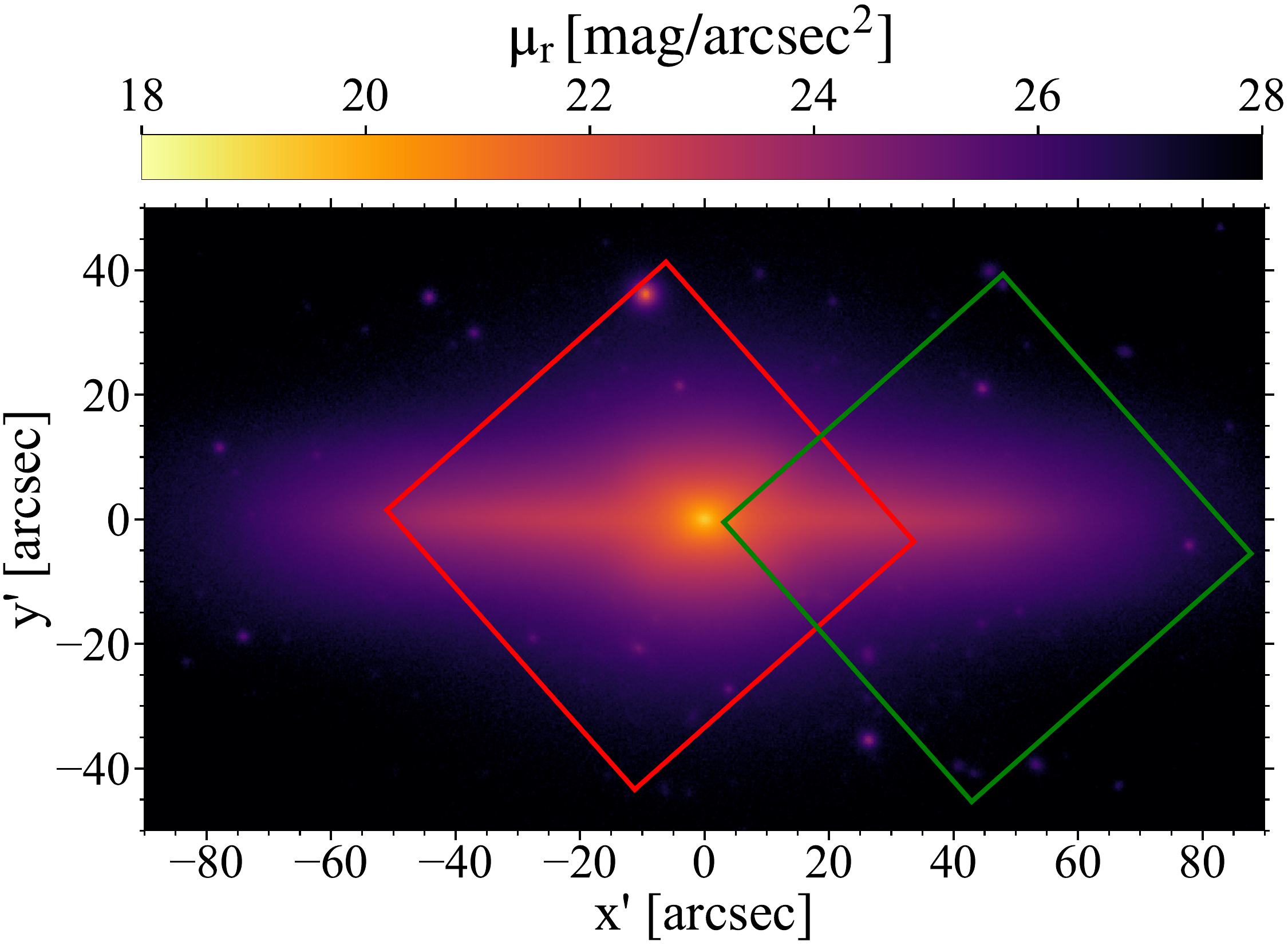}
    \caption{Imaging and MUSE coverage of NGC~1381 (FCC 170). The image shows the deep $r$-band imaging from FDS, with its major axis rotated to the $x'$-axis and its surface brightness indicated by the colour bar. The red and green squares represent the central and halo MUSE pointings from F3D, respectively.}
    \label{image-and-MUSE-pointing}
\end{figure}

\subsection{The Fornax deep survey and the Fornax3D project}
\label{sec2.1}
The Fornax Deep Survey (FDS; \citealp{Iodice2016}) is a deep multiband optical imaging survey covering the entire Fornax cluster out to the virial radius ($\rm\sim0.7\,Mpc$; \citealp{Drinkwater2001}). The observations were conducted using the 2.6-meter Very Large Telescope (VLT) Survey Telescope located at Cerro Paranal, Chile. We used the $r$-band deep imaging from FDS, which reaches a surface brightness level of $\rm\sim27\,mag\,arcsec^{-2}$ \citep{Venhola2018}.

The Fornax3D project (F3D; \citealp{Sarzi2018}) is an IFS survey targeting the 33 brightest galaxies within or near the virial radius of the Fornax cluster \citep{Ferguson1989}. The observations were conducted using the wide-field mode of the Multi-Unit Spectroscopic Explorer (MUSE) instrument on the VLT \citep{Bacon2010}. MUSE provides a field of view of $\rm1\times1\,arcmin^2$ with a spatial sampling of $\rm0.2\times0.2\,arcsec^2$. The wavelength coverage is 4650--9300$\,\AA$ with a spectral resolution of 2.5$\,\AA$ (full width at half maximum) at 7000$\,\AA$. NGC~1381 is one of the 33 target galaxies and was observed with two MUSE pointings. Fig.~\ref{image-and-MUSE-pointing} displays its $r$-band image and the regions covered by MUSE.

\subsection{Stellar kinematics and populations}
\label{sec2.2}

The stellar kinematic maps we adopted were derived following the same approach as \citet{Iodice2019b} but with an improved signal-to-noise ratio ($S/N$) of 100. These maps were generated by applying the full spectral fitting software pPXF \citep{Cappellari2004,Cappellari2017} to spectra binned with the Voronoi 2D method \citep{Cappellari2003}. The maps include the mean velocity ($V$), velocity dispersion ($\sigma$), higher-order velocity moments parametrised by the Gauss-Hermite coefficients ($h_3$ and $h_4$; \citealp{Gerhard1993,vdMarel1993,Rix1997}), and their corresponding errors. Since our dynamical modelling is triaxial, we further point-symmetrised the kinematic maps and their error maps to reduce noise and presented them in Fig.~\ref{kinematic-chemo-data}.

We employed three versions of stellar population maps: Version A originates from \citet{Martig2026}, using Voronoi-binned spectra with $S/N=60$; version B is based on \citet{Pinna2019a} but with $S/N$ improved to 100 ($S/N=40$ in the original work); version C is from \citet{MartinNavarro2021}, also with $S/N=100$. All versions were derived by applying the pPXF fitting to Voronoi-binned spectra, with stellar population synthesis models \citep{Vazdekis2010,Vazdekis2015} based on the MILES stellar libraries \citep{SanchezBlazquez2006} and BaSTI isochrones \citep{Pietrinferni2004,Pietrinferni2006}. Versions A and B are mass-weighted and assume a constant \citet{Kroupa2001} initial mass function (IMF), using spectra covering 4750--5500$\,\AA$ and a fixed regularisation parameter of 0.05, with the spatial coverage of the maps similar to that of the kinematic maps. In contrast, version C is light-weighted, allows for the variation in the IMF, and uses spectra covering 4800--6400$\,\AA$. Version C has a smaller spatial coverage because pixels with $S/N<5$ were not included during the Voronoi-binning process. Compared to versions A and B, version C has an offset of $\sim-0.3\rm\,dex$ in metallicity and $\sim0.05\rm\,dex$ in [Mg/Fe] abundance due to different settings when fitting the spectra, but the spatial trends in the maps are similar.

Figure~\ref{kinematic-chemo-data} shows the maps of stellar age, metallicity, and [Mg/Fe] abundance for version A, with their error maps derived from the Monte Carlo simulations of perturbed spectra. The stellar population maps for versions B and C are presented in the left panels of Figs.~\ref{chemo-maps-best-fitting-Francesca} and~\ref{chemo-maps-best-fitting-Nacho}, respectively. Versions B and C did not include error maps, and we thus adopted fixed uncertainties of 1\,Gyr for age, $\sim0.05\rm\,dex$ for metallicity, and $\sim0.03\rm\,dex$ for [Mg/Fe] abundance. These uncertainties generally agree with those of version A, as well as with estimates from previous studies of NGC~1381 and other galaxies with similar MUSE data (e.g. \citealp{Pinna2019a,Pinna2019b,Martig2021,Sattler2023,Sattler2025}). We treated data version A as the default version in our analysis. It will be shown later that the fixed uncertainties in versions B and C do not affect our main findings.

\section{Constructing barred population-orbit superposition models}
\label{sec3}
We constructed the barred population-orbit superposition models in three major steps. First, we built barred orbit-superposition models by simultaneously fitting the stellar luminosity distributions and kinematic data (Sect.~\ref{sec3.1}). Then, we categorised the stellar orbits in our models into different dynamical structures based on their orbital properties (Sect.~\ref{sec3.2}). Finally, we assigned ages and metallicities to the orbits by fitting the stellar age and metallicity maps (Sect.~\ref{sec3.3}). Our methodology for the first step follows \citet{Jin2025a}, but has been modified to include a supermassive black hole. For the second and third steps, it follows \citet{Jin2025b}, with updated orbit classification criteria during the decomposition. The model uncertainties we adopted throughout the paper are introduced in Sect.~\ref{sec3.4}.

\subsection{Constructing barred orbit-superposition models}
\label{sec3.1}
\begin{figure*}
    \centering
    \includegraphics[width=18cm]{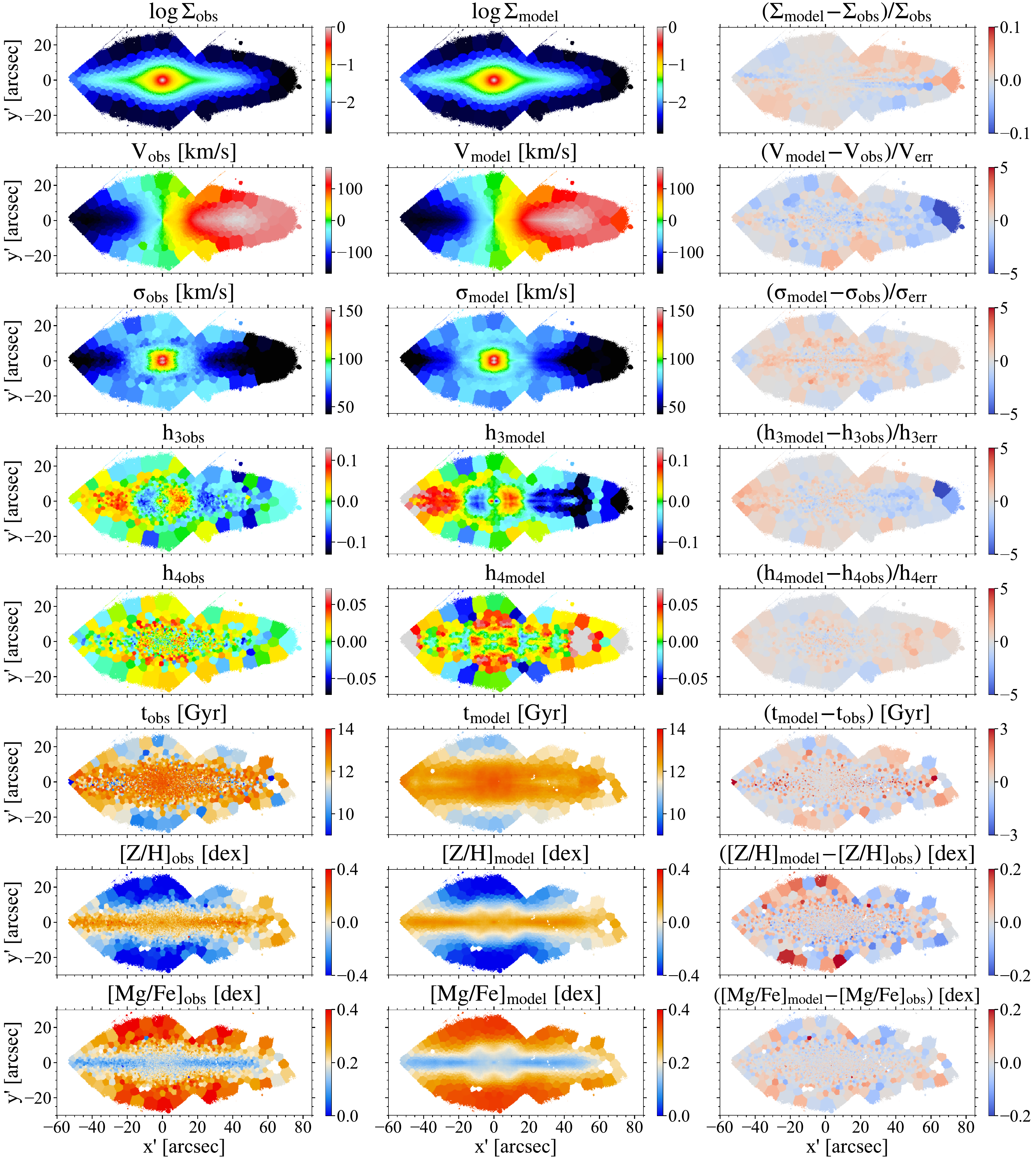}
    \caption{Surface brightness, stellar kinematic data, stellar population data, and best-fitting model for NGC~1381. \textit{From left to right}: Observations, model fittings, and residuals. \textit{From top to bottom}: Logarithmic normalised surface brightness, $\log\Sigma$; mean velocity, $V$; velocity dispersion, $\sigma$; third-order Gauss-Hermite coefficient, $h_3$; fourth-order Gauss-Hermite coefficient, $h_4$; stellar age, $t$; stellar metallicity, $[Z/\rm H]$; and [Mg/Fe] abundance. The residual maps represent the relative deviations for surface brightness, the standardised residuals for stellar kinematics, and absolute differences for stellar populations.}
    \label{kinematic-chemo-maps-best-fitting}
\end{figure*}

To construct a barred orbit-superposition model, the gravitational potential is first built from four components: (1) a triaxial stellar bar with constant figure rotation, $\rm\Omega_p$; (2) an axisymmetric stellar disc aligned with the bar's major axis; (3) a spherical dark matter halo; and (4) a central supermassive black hole. 

The stellar potential calculation begins with fitting the galaxy's surface brightness ($r$-band deep imaging from FDS) using the multi-Gaussian expansion (MGE) formalism \citep{Emsellem1994,Cappellari2002}, where each 2D Gaussian is assigned to either the triaxial bar component or the axisymmetric disc component. The MGE fitting results for NGC~1381, along with their corresponding contours, are presented in Table~\ref{table-mge-parameters} and Fig.~\ref{mge-contours}. These 2D Gaussians are then separately deprojected into 3D luminosity density using the viewing angles: the inclination $\theta$ (where $\theta=90^\circ$ corresponds to an edge-on view) and the bar azimuthal angle $\varphi$ (where $\varphi=0^\circ$ corresponds to an end-on bar view and $\varphi=90^\circ$ to a side-on view). By multiplying the 3D luminosity density by a constant mass-to-light ratio, $M_\star/L$, and solving Poisson's equation, the potentials of both the stellar bar and the stellar disc are derived. We note that the bar and disc mentioned here are used to construct the stellar potential and are not necessarily the same as the dynamically decomposed bar and disc discussed later. 

The dark matter potential follows a three-parameter generalised Navarro-Frenk-White profile (\citealp{Navarro1996,Zhao1996}; see also \citealp{Barnabe2012,Cappellari2013}). The central black hole's contribution is modelled using a Plummer potential \citep{vdB2008} with a fixed softening length $r_s=\rm0.001\,arcsec$, while the black hole mass $M_{\rm BH}$ is treated as a free parameter. Overall, the gravitational potential is determined by eight free parameters in the modelling: the inclination angle, $\theta$; the bar azimuthal angle, $\varphi$; the stellar mass-to-light ratio, $M_\star/L$; the bar pattern speed, $\rm\Omega_p$; the dark matter concentration, $c$; the virial mass, $M_{200}$; the inner density slope of dark matter, $\gamma$; and the black hole mass, $M_{\rm BH}$. We note that adopting a constant $M_\star/L$ in the modelling does not influence our main results, although \citet{MartinNavarro2021} reported the presence of an $M_\star/L$ gradient in this galaxy (see Appendix~\ref{sec-parameter-space}).

In the given gravitational potential determined by a set of free parameters, initial conditions for $\sim10^6$ orbits are sampled from the phase space and then integrated. After the integration, the weights of these orbits are solved through the non-negative least squares method \citep{Lawson1974}, by matching the MGE-fitted 2D surface brightness, the MGE-deprojected 3D luminosity density, and the observed kinematic data. The goodness of the model ($\chi^2$) is evaluated by computing the sum of squared residuals between model-fitted and observed kinematic maps, including the velocity ($V$), velocity dispersion ($\sigma$), third-order and fourth-order Gauss-Hermite coefficients ($h_3$ and $h_4$), which is written as
\begin{equation}
\begin{split}
    \chi^2=\sum_{i=1}^{N_{\rm obs}}\left[ \left(\frac{V_{\rm obs}^i-V_{\rm model}^i}{V_{\rm err}^i}\right)^2+\left(\frac{\sigma_{\rm obs}^i-\sigma_{\rm model}^i}{\sigma_{\rm err}^i}\right)^2 \right.\\
    \left.+\left(\frac{h_{\rm 3obs}^i-h_{\rm 3model}^i}{h_{\rm 3err}^i}\right)^2+\left(\frac{h_{\rm 4obs}^i-h_{\rm 4model}^i}{h_{\rm 4err}^i}\right)^2 \right].
\end{split}
\end{equation}

The best-fitting model with the minimum $\chi^2$ is found by iteratively searching the free parameter space ($\theta,\varphi,M_\star/L,\Omega_{\rm p},c,M_{200},\gamma$, $M_{\rm BH}$). Starting with initial guesses, models with lower $\chi^2$ are selected and new models are generated by exploring nearby regions in the parameter space. This process repeats until the minimum $\chi^2$ ($\chi^2_{\rm min}$) is reached, ensuring all surrounding models are calculated. The model with $\chi^2_{\rm min}$ is identified as the best-fitting model, which is taken as the default model in our analysis.

We present the surface brightness and stellar kinematic maps of the best-fitting model for NGC~1381 in Fig.~\ref{kinematic-chemo-maps-best-fitting}. The features of the stellar kinematics are well recovered, including the anti-correlated, correlated, and anti-correlated again $h_3$--$V$ relation along the galaxy's major axis, which correspond to the signatures of the nuclear disc, bar, and main disc, respectively. The parameter space we explored and the model-predicted enclosed mass profiles are displayed in Figs.~\ref{parameter-grids} and~\ref{mass-profiles}, respectively.

\subsection{Decomposing galaxies based on stellar orbits}
\label{sec3.2}
\begin{figure}
    \centering
    \includegraphics[width=9cm]{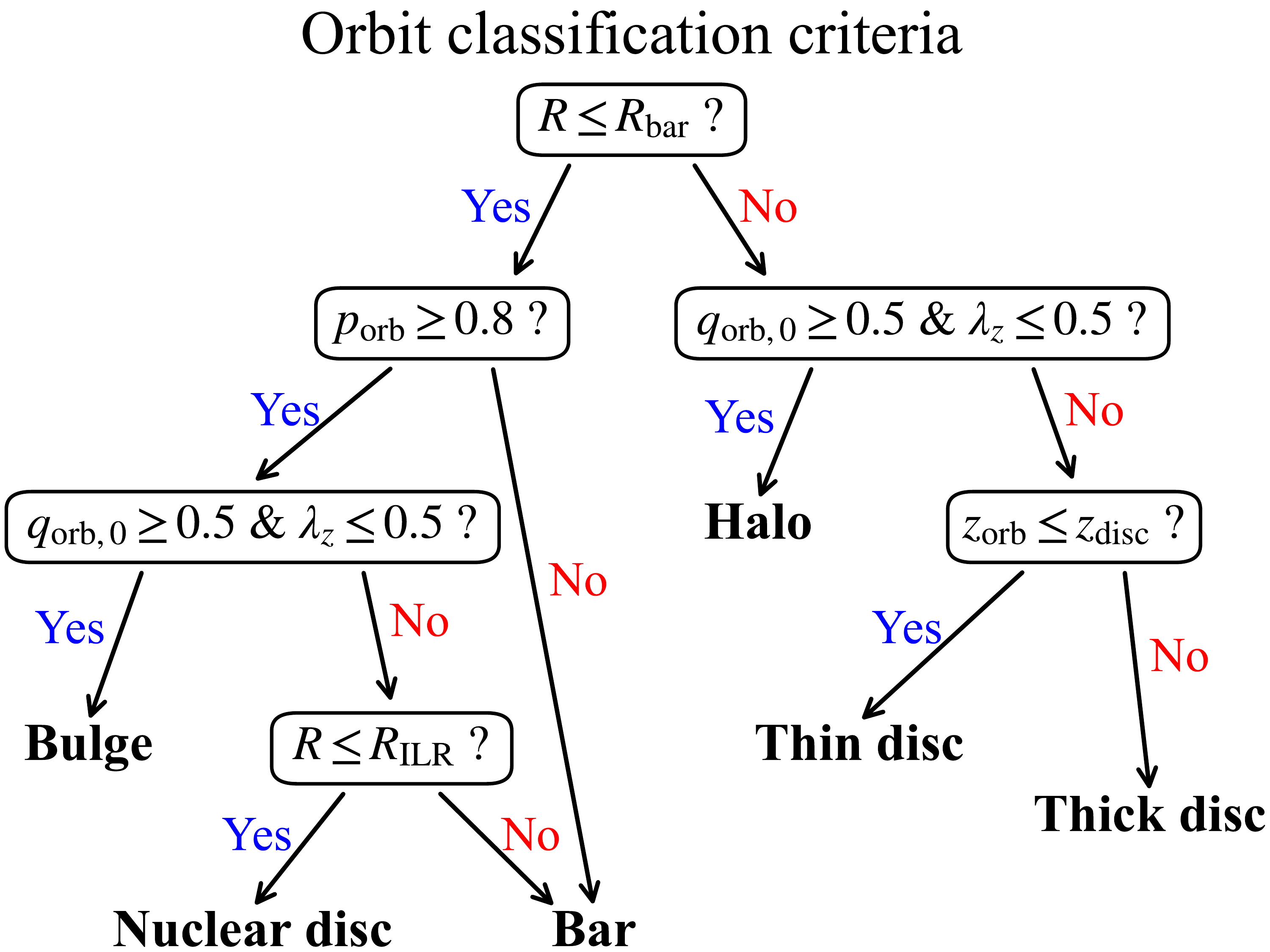}
    \caption{Orbit classification criteria used in our dynamical decomposition. Each box represents a criterion. Five properties are adopted to classify the orbits: (1) circularity, $\lambda_z$; (2) time-averaged radius, $R$; (3) axis ratio, $p_{\rm orb}$, in the $x$-$y$ plane (the bar's rotating frame); (4) axis ratio, $q_{\rm orb,0}$, in the $x_0$-$z$ plane (the inertial frame); and (5) vertical height, $z_{\rm orb}$. The dynamical bar length, $R_{\rm bar}$, and the radius of the inner Lindblad resonance, $R_{\rm ILR}$, are determined from our model results, while the separation threshold of vertical height, $z_{\rm disc}=7\rm\,arcsec$, is estimated from Fig.~4 of \citet{Pinna2019a}. Six components were derived from the decomposition: a nuclear disc, a bar, a bulge, a thin disc, a thick disc, and a stellar halo.}
    \label{orbit-classification-criteria}
\end{figure}
\begin{figure*}
    \centering
    \includegraphics[width=18cm]{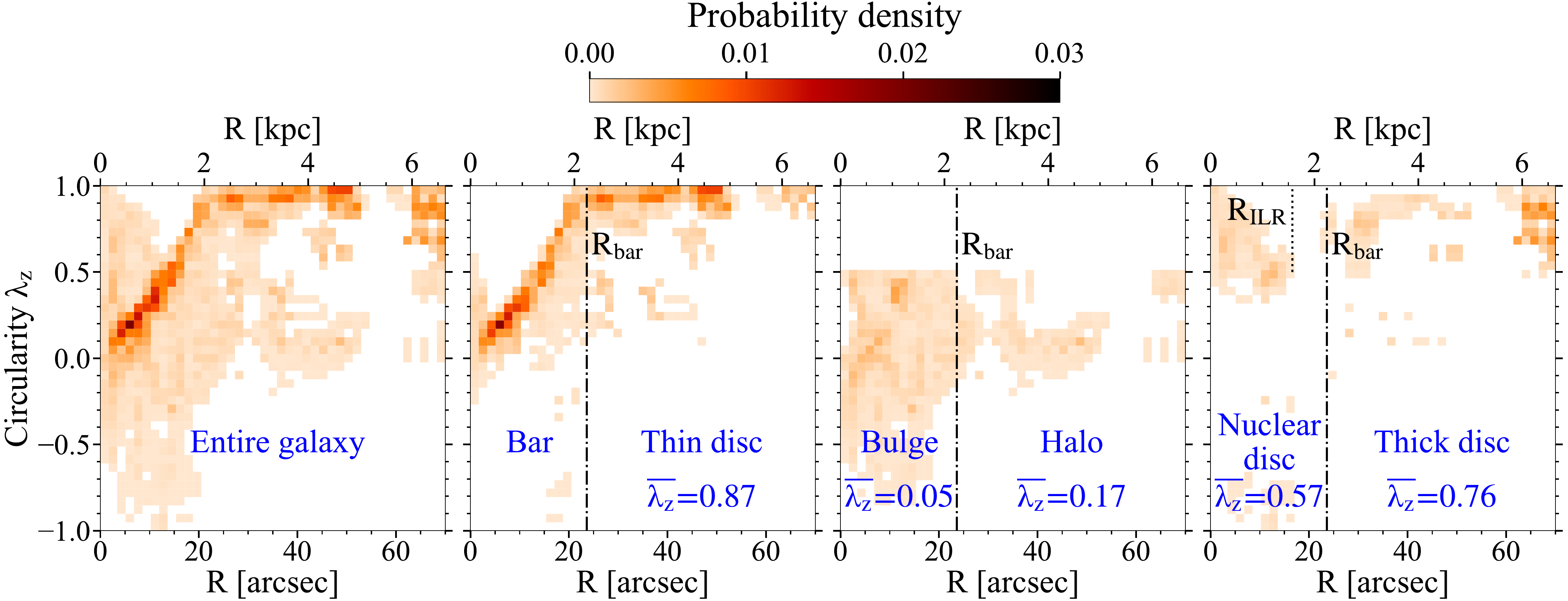}
    \caption{Probability density distributions of stellar orbits in the $\lambda_z$--$R$ phase space for the best-fitting model of NGC~1381. The leftmost panel shows the stellar orbit distribution for the entire galaxy. The three remaining panels each display two dynamical components, separated by a vertical dashed line indicating the bar length ($R_{\rm bar}=2.24\rm\,kpc$): (1) the bar and thin disc; (2) the bulge and halo; and (3) the nuclear disc and thick disc. The dotted line in the rightmost panel represents the radius of the inner Lindblad resonance ($R_{\rm ILR}=1.57\rm\,kpc$). The probability densities of all orbits within $R\le\rm70\,arcsec$ ($6.64\rm\,kpc$) are normalised to unity, with their values indicated by the colour bar. The mean circularities ($\overline{\lambda_z}$) of the thin disc, bulge, halo, nuclear disc, and thick disc are indicated.}
    \label{orbit-distributions}
\end{figure*}

Based on the kinematic and morphological properties of stellar orbits, we decomposed the galaxy into six components: a nuclear disc, a bar, a bulge, a thin disc, a thick disc, and a stellar halo. These properties include: (1) orbital circularity, $\lambda_z$; (2) time-averaged orbital radius, $R$; (3) axis ratio, $p_{\rm orb}$ in the $x$-$y$ plane (bar's rotating frame); (4) axis ratio, $q_{\rm orb,0}$ in the $x_0$-$z$ plane (inertial frame); and (5) vertical height, $z_{\rm orb}$.

The circularity, $\lambda_z$, quantifying the orbit's angular momentum $L_z$ in the inertial frame ($x_0,y_0,z$), is defined as
\begin{equation}
    \lambda_z=\overline{L_z}/(R\times\overline{V_{\rm rms}}),
\label{circularity-with-cross-terms}    
\end{equation}
where $\overline{L_z}=\overline{x_0v_{y0}-y_0v_{x0}}$, $R=\overline{\sqrt{x_0^2+y_0^2+z^2}}$, and $\overline{V_{\rm rms}}=\sqrt{\overline{v_{x0}^2+v_{y0}^2+v_z^2+2v_{x0}v_{y0}+2v_{x0}v_z+2v_{y0}v_z}}$. Dynamically cold orbits with $\lambda_z\sim1$ (rotation-dominated) mainly contribute to the disc components, while dynamically hot orbits with $\lambda_z\sim0$ (dispersion-dominated) or $\lambda_z<0$ (retrograde) usually belong to the bulge or stellar halo. A discussion of the influence of the cross terms in $\overline{V_{\rm rms}}$ is provided in Appendix~\ref{sec-circularity-definitions}.

In a typical barred galaxy, the circularity increases with radius in the bar-dominated inner region, whereas it remains dynamically cold at all radii in the disc-dominated outer regions (see the left panel of Fig.~\ref{orbit-distributions}). Therefore, the dynamical bar length, $R_{\rm bar}$, is defined as the radius where dynamically cold orbits ($\lambda_z\ge0.8$) begin to dominate. For NGC~1381, we calculated the cold orbit fraction, $f_{\rm cold}$, within a $\rm0.5\,kpc$ moving average and identified $R_{\rm bar}$ as the smallest radius at which $f_{\rm cold}\ge0.5$. In our models, we treat $R_{\rm bar}$ as a boundary to distinguish the galaxy's inner regions (nuclear disc, bar, and bulge) from the outer regions (thin disc, thick disc, and stellar halo), where $R_{\rm bar}$ equals $23.6\rm\,arcsec$ ($2.24\rm\,kpc$) for the best-fitting model. The bar length ($R_{\rm bar}$) provides a natural separation of the bar and the nuclear disc embedded within it from the outer thin disc, whereas the radial division between the bulge, thick disc, and halo remains somewhat uncertain. For simplicity, in this work we adopt $R_{\rm bar}$ as a divider for the bulge, thick disc, and halo, and the halo here can be understood as the less bound part of the bulge. We note that the time-averaged radius of the orbits ($R$) scales with the orbital binding energy ($E_{\rm bind}$). The rationale for using this boundary is further discussed in Appendix~\ref{sec-Ebind}.

The radius of the inner Lindblad resonance, $R_{\rm ILR}$, is defined as the radius at which a star's epicyclic frequency ($\kappa$) and azimuthal frequency ($\Omega(R)$) satisfy $\Omega(R)-\kappa/2 = \Omega_{\rm p}$. This resonance radius can be calculated directly from the gravitational potential and is consistently smaller than the bar length ($R_{\rm ILR}$<$R_{\rm bar}$). The orbits inside $R_{\rm ILR}$, which can be dynamically cold and/or perpendicular to the bar (e.g. \citealp{Contopoulos1977,Athanassoula1992a,Athanassoula1992b,Binney2008}), make up the nuclear disc in our modelling. For the best-fitting model, we derived $R_{\rm ILR}=16.6\rm\,arcsec$ ($1.57\rm\,kpc$).

Based on the time-averaged amplitudes of orbital trajectories, three orbital quantities are calculated: (1) the axis ratio, $p_{\rm orb}$, in the bar's rotating frame; (2) the axis ratio, $q_{\rm orb,0}$, in the inertial frame; and (3) the vertical height, $z_{\rm orb}$. These quantities are expressed as
\begin{equation}
    p_{\rm orb}=\frac{\overline{|y-\overline{y}|}}{\overline{|x-\overline{x}|}},\quad
    q_{\rm orb,0}=\frac{\overline{|z-\overline{z}|}}{\overline{|x_0-\overline{x_0}|}},\quad
    z_{\rm orb}=\overline{|z|}.
\end{equation}
The axis ratio, $p_{\rm orb}$, is adopted to distinguish bar orbits, which are elongated with the bar, from nuclear disc and bulge orbits. The axis ratio, $q_{\rm orb,0}$, together with $\lambda_z$, is used to separate the disc components from the bulge and stellar halo. The vertical height, $z_{\rm orb}$, distinguishes between the thin and thick discs, using a separation threshold of $z_{\rm disc}=7\rm\,arcsec$ ($0.66\rm\,kpc$) estimated from \citet{Pinna2019a}.

Combining these properties, we present the detailed orbit classification criteria in Fig.~\ref{orbit-classification-criteria}, and thus the dynamical decomposition of the galaxy can be achieved. Fig.~\ref{orbit-distributions} displays the decomposed stellar orbit distributions in the $\lambda_z$--$R$ phase space for the best-fitting model of NGC~1381 (see Appendix~\ref{sec-barred-vs-unbarred} for a comparison with the stellar orbit distributions of the unbarred model from \citealp{Ding2023}). Other models within the $1\sigma$ confidence level were decomposed using the same criteria (see Sect.~\ref{sec3.4} for the definition of uncertainties). This galaxy is divided into six components: a rigidly rotating bar ($\lambda_z$ strongly correlated with $R$); a dynamically hot bulge and halo ($\overline{\lambda_z}=0.05_{-0.07}^{+0.03}$ and $0.17_{-0.08}^{+0.13}$, respectively); a dynamically cold thin disc ($\overline{\lambda_z}=0.87_{-0.03}^{+0.02}$); a thick disc with slightly slower rotation than the thin disc ($\overline{\lambda_z}=0.76_{-0.03}^{+0.04}$); and a dynamically warm nuclear disc ($\overline{\lambda_z}=0.57_{-0.11}^{+0.12}$). The spatial distributions of luminosity-weighted mean velocity and velocity dispersion for each component are shown in Fig.~\ref{velocity-sigma-distributions}.

\subsection{Tagging the stellar orbits with stellar populations}
\label{sec3.3}
Following \citet{Jin2025b}, we separated the orbits belonging and not belonging to the bar into two distinct $\lambda_z$--$R$ phase spaces. We then used the Voronoi 2D binning method to divide orbits in each $\lambda_z$--$R$ phase space into different orbit bundles, assuming that each orbit bundle $k$ has a simple stellar population with age ($t_k$), metallicity ($[Z/{\rm H}]_k$), and [Mg/Fe] abundance ($[{\rm Mg/Fe}]_k$) to be determined.

By projecting the orbits onto the observation plane, the model-recovered, luminosity-weighted stellar population in the $i$-th aperture can be expressed as
\begin{equation}
    p_{\rm model}^i=\frac{\sum_{k=1}^{N_b} p_k f_k^i}{\sum_{k=1}^{N_b} f_k^i},
\end{equation}
where the quantity $p$ represents age, metallicity, or [Mg/Fe] abundance. Here, $N_b$ denotes the total number of orbit bundles, while $f_k^i$ represents the luminosity contribution from the $k$-th orbit bundle within the $i$-th aperture.

Similar to the kinematics, the chi-square difference between model-fitted maps and observed data is given by
\begin{equation}
    \chi_p^2=\sum_{i=1}^{N_{\rm obs}}\left(\frac{p_{\rm obs}^i-p_{\rm model}^i}{p_{\rm err}^i}\right)^2.
\end{equation}
We solved the stellar population parameter of each orbit bundle, $p_k$, by minimising $\chi_p^2$ for age, metallicity, and [Mg/Fe] abundance independently via the bounded-variable least squares method (implemented in Python's SciPy\footnote{\url{https://scipy.org/}}), with the boundaries of stellar population parameters set as
\begin{equation}
\small
\begin{cases}
    \min(t_{\rm obs}^i)\le t_k \le \min[\max(t_{\rm obs}^i)+\Delta(t_{\rm obs}^i), \rm14\,Gyr], \\[0.5em]
    \min([Z/\mathrm{H}]_{\rm obs}^i)\le [Z/\mathrm{H}]_k \le \max([Z/\mathrm{H}]_{\rm obs}^i)+\Delta([Z/\mathrm{H}]_{\rm obs}^i), \\[0.5em]
    \min([\mathrm{Mg/Fe}]_{\rm obs}^i)-\Delta([\mathrm{Mg/Fe}]_{\rm obs}^i)\le [\mathrm{Mg/Fe}]_k \le \max([\mathrm{Mg/Fe}]_{\rm obs}^i),
\end{cases}
\end{equation}
where $\Delta$ represents the standard deviation of the observed data. This $\Delta$ term is added because the galaxy's central region is a mixture of multiple stellar populations, allowing for the possible presence of intrinsically older, more metal-rich, and $\alpha$-poor stellar populations than the observed data. In contrast, no such $\Delta$ is needed for the younger, more metal-poor, and $\alpha$-rich stellar populations in the galaxy's outer regions, which are dominated by the thick disc. Moreover, the stellar populations there have larger uncertainties compared to the inner regions, so setting the minimum or maximum values of the observed maps already provides sufficient tolerance. The recovered stellar population maps of the best-fitting model are shown in Figs.~\ref{kinematic-chemo-maps-best-fitting} (version A),~\ref{chemo-maps-best-fitting-Francesca} (version B), and~\ref{chemo-maps-best-fitting-Nacho} (version C).

\subsection{Model uncertainties}
\label{sec3.4}
In our models, we estimated the uncertainties by perturbing the observed data maps. For purely kinematic properties, the uncertainties originate from the stellar kinematic maps; for properties related to stellar populations, the uncertainties are estimated by considering the contributions from both stellar kinematics and stellar populations.

The kinematic confidence level of the modelling is used to estimate the uncertainties of the free parameters ($\theta,\varphi,M_\star/L,\Omega_{\rm p},c,M_{200},\gamma,M_{\rm BH}$), as well as of the properties presented in Sects.~\ref{sec3.2},~\ref{sec4.1}, and~\ref{sec4.2}. This confidence level was calculated by perturbing the kinematic maps $(V_{\rm obs}^i,\sigma_{\rm obs}^i,h_{\rm 3obs}^i,h_{\rm 4obs}^i)$ 1000 times, adding Gaussian noise based on their error maps $(V_{\rm err}^i,\sigma_{\rm err}^i,h_{\rm 3err}^i,h_{\rm 4err}^i)$ estimated from the pPXF fitting. The perturbed kinematic maps were then point-symmetrised in the same way as the original maps. Fixing the best-fitting gravitational potential, we recomputed the orbit weights for each perturbed kinematic map and obtained 1000 new $\chi^2$ values, whose distribution is approximately Gaussian. The kinematic confidence levels were defined based on this distribution: the $1\sigma$, $2\sigma$, and $3\sigma$ confidence levels correspond to $\chi^2$ variations within the $\pm1\sigma$ ($68\%$), $\pm2\sigma$ ($95\%$), and $\pm3\sigma$ ($>99\%$) regions of this distribution, respectively. We took the $1\sigma$ confidence level ($\chi^2-\chi^2_{\rm min}\le\Delta\chi^2_{\rm CL}$) as the uncertainties. Approximately 1400 models (1400 combinations of free parameters) were located within the $1\sigma$ confidence level and were adopted to estimate the uncertainties.

The $1\sigma$ uncertainties of the model-predicted stellar ages incorporate contributions from two sources: the observed stellar kinematic maps and the observed stellar age maps. To quantify the uncertainties arising from kinematics ($\Delta t_{\rm kin}$), we first fixed the observed stellar age maps and calculated the stellar ages of each component for all models within the kinematic $1\sigma$ confidence level. We then took the standard deviation of these ages across different models as $\Delta t_{\rm kin}$. To estimate the uncertainties from the stellar age maps ($\Delta t_{\rm age}$), we fixed the orbit libraries and weights of the best-fitting model. We then perturbed the observed stellar age maps 1000 times according to their error maps (in the same way as we perturbed the kinematic maps) and recalculated the stellar ages for each component. The standard deviations of the resulting age distributions were taken as $\Delta t_{\rm age}$. Assuming that the uncertainties from kinematics and age maps are independent, the total $1\sigma$ uncertainties for stellar ages ($\Delta t_{\rm all}$) were derived using error propagation: $\Delta t_{\rm all}^2=\Delta t_{\rm kin}^2+\Delta t_{\rm age}^2$. The uncertainties of stellar metallicities ($[Z/\rm H]$) and [Mg/Fe] abundances were estimated similarly. These uncertainties are used in Sect.~\ref{sec4.3} (Figs.~\ref{mean-stellar-populations} and~\ref{metallicity-alpha-gradients}).

\section{Dynamical structures and their chemical properties}
\label{sec4}
\begin{figure}
    \centering
    \includegraphics[width=9cm]{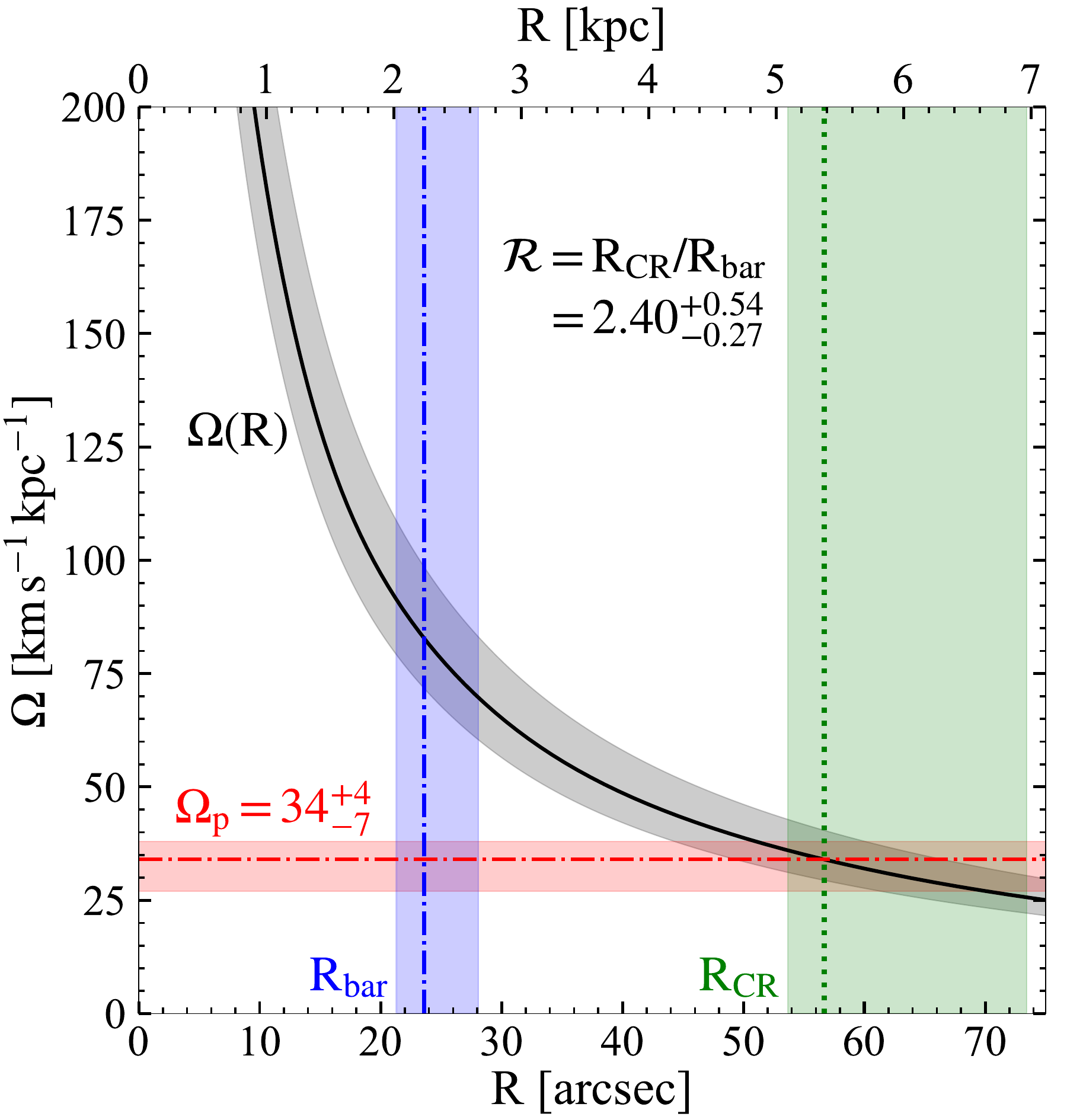}
    \caption{Analysis of corotation resonance for NGC~1381. The black solid line, red dashed line, blue dashed line, and green dotted line represent the angular velocity profile ($\Omega(R)$), bar pattern speed ($\rm\Omega_p$), bar length ($R_{\rm bar}$), and corotation radius ($R_{\rm CR}$) of the best-fitting model, respectively. The corresponding shaded regions indicate uncertainties calculated from models within the $1\sigma$ confidence level. The dimensionless bar rotation rate, $\mathcal{R}=R_{\rm CR}/R_{\rm bar}$, is labelled.}
    \label{corotation-radius}
\end{figure}
\begin{figure*}
    \centering
    \includegraphics[width=18cm]{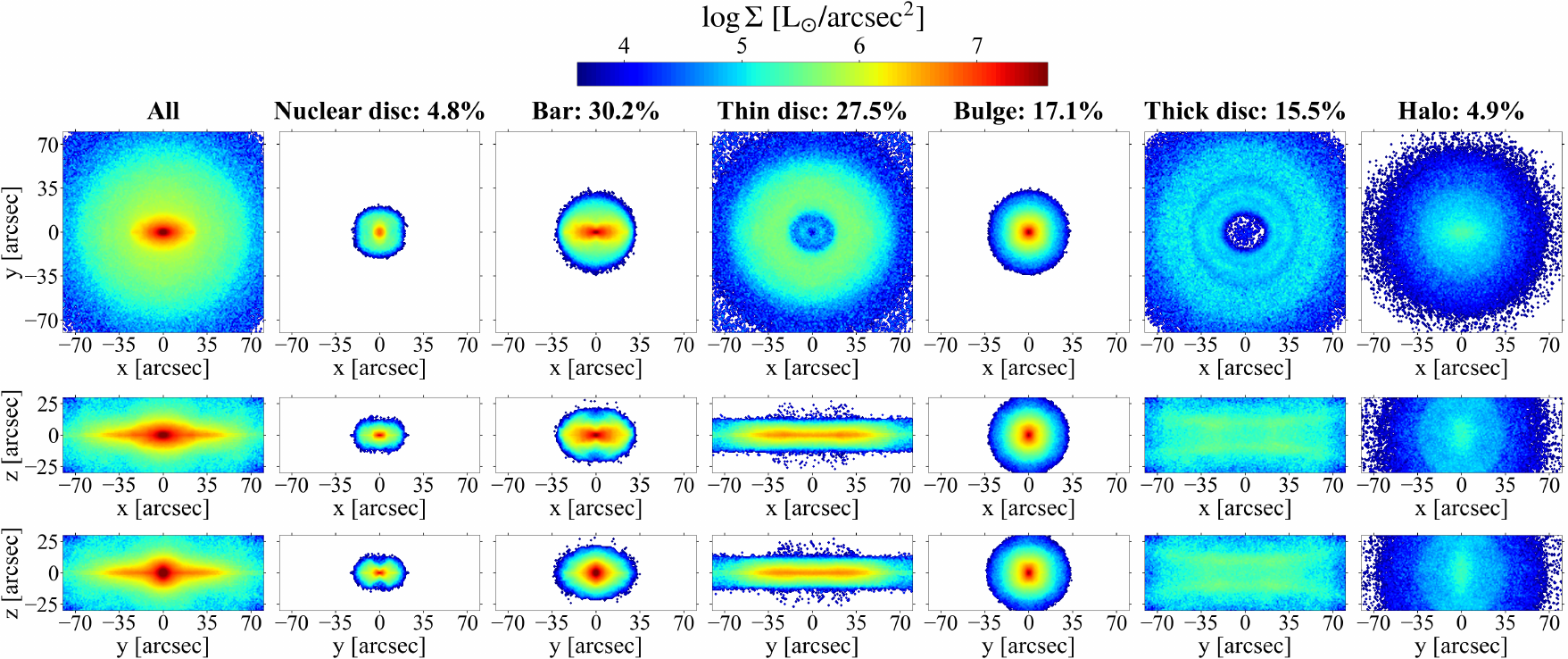}
    \caption{Luminosity distributions of dynamical components in the best-fitting model of NGC~1381. \textit{From top to bottom}: Logarithmic surface brightness on the $x$-$y$ (face-on), $x$-$z$ (edge-on and bar side-on), and $y$-$z$ (edge-on and bar end-on) planes, respectively. \textit{From left to right}: Entire galaxy, nuclear disc, bar, thin disc, bulge, thick disc, and stellar halo. The luminosity fraction of each component is labelled.}
    \label{luminosity-distributions}
\end{figure*}
\begin{figure*}
    \centering
    \includegraphics[width=18cm]{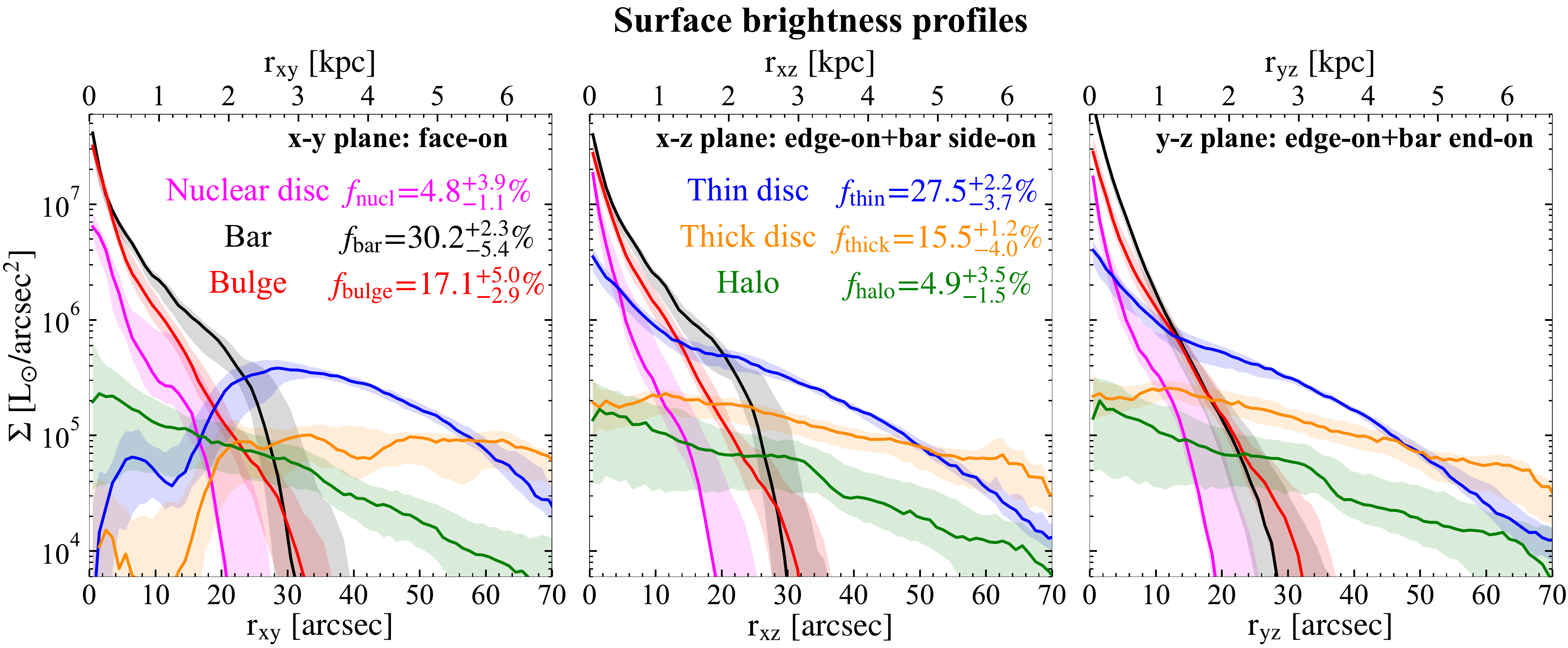}
    \caption{Surface brightness profiles for the dynamical components in the models of NGC~1381. \textit{From left to right}: Surface brightness profiles on the $x$-$y$ ($r_{xy}=\sqrt{x^2+y^2}$), $x$-$z$ ($r_{xz}=\sqrt{x^2+z^2}$), and $y$-$z$ ($r_{yz}=\sqrt{y^2+z^2}$) planes. The solid curves represent the best-fitting profiles of different components: nuclear disc (magenta), bar (black), bulge (red), thin disc (blue), thick disc (orange), and stellar halo (green). The corresponding shaded regions indicate uncertainties calculated from models within the $1\sigma$ confidence level. The luminosity fraction of each component, together with its uncertainty, is shown.}
    \label{surface-brightness-profiles}
\end{figure*}

In this section, we analyse the properties of NGC~1381's dynamical components derived from our barred population-orbit superposition models, including the bar pattern speed, bar length, and corotation radius (Sect.~\ref{sec4.1}); the 3D luminosity distributions of the different structures (Sect.~\ref{sec4.2}); and the stellar populations of each component (Sect.~\ref{sec4.3}).

\subsection{Bar pattern speed, bar length, and corotation radius}
\label{sec4.1}

From the gravitational potential $\Phi$ derived from the modelling, the angular velocity profile $\Omega(R)$ of the galaxy on the disc plane ($z=0$) is
\begin{equation}
    \Omega(R)=\frac{v_{\rm circ}(R)}{R}=\sqrt{\left. \frac{1}{R}\frac{\partial\Phi}{\partial R}\right |_{z=0}}.
\end{equation}
For each model within the $1\sigma$ confidence level, the bar pattern speed ($\rm\Omega_p$), being a free parameter of the potential, was already determined. We further calculated the dynamical bar length ($R_{\rm bar}$; see Sect.~\ref{sec3.2}) and $\Omega(R)$. The corotation radius, $R_{\rm CR}$, defined as the radius where the angular velocity matches the bar pattern speed ($\Omega(R)=\rm\Omega_p$), was then derived accordingly. Finally, we computed the dimensionless bar rotation rate, $\mathcal{R}=R_{\rm CR}/R_{\rm bar}$, which is adopted to distinguish fast bars ($1.0\le\mathcal{R}\le1.4$), slow bars ($\mathcal{R}>1.4$), and ultrafast bars ($\mathcal{R}<1.0$; \citealp{Debattista2000,Buta2009,Aguerri2015}). All these quantities are illustrated in Fig.~\ref{corotation-radius}. We derived $\rm\Omega_p=34_{-7}^{+4}\,km\,s^{-1}\,kpc^{-1}$, $R_{\rm bar}=23.6_{-2.3}^{+4.5}\rm\,arcsec$ ($2.24_{-0.22}^{+0.43}\rm\,kpc$), and $R_{\rm CR}=56.7_{-3.0}^{+16.7}\rm\,arcsec$ ($5.38_{-0.28}^{+1.59}\rm\,kpc$). These yield a slow bar with $\mathcal{R}=2.40_{-0.27}^{+0.54}$.

\subsection{Luminosity distribution of dynamical components}
\label{sec4.2}

Using the orbital weights and trajectories from the best-fitting model, we reconstructed the 3D luminosity distributions for the six dynamical components. Fig.~\ref{luminosity-distributions} displays these distributions projected onto the $x$-$y$, $x$-$z$, and $y$-$z$ planes. The 3D morphologies of different components are well distinguished: the nuclear disc occupies the galaxy's inner regions ($\lesssim20\rm\,arcsec$); the bar is BP/X-shaped; the thin disc dominates the galaxy's outer regions ($\gtrsim25\rm\,arcsec$); the bulge is spheroidal; the thick disc is vertically extended; and the stellar halo is radially diffuse. We note that the `holes' in the thin and thick disc components arise from our dynamical classification criteria: orbits with $R\le R_{\rm bar}$ (i) follow the bar's rigid rotation, (ii) belong to the dispersion-dominated bulge, or (iii) are rotation-dominated orbits within the inner Lindblad resonance that belong to the nuclear disc (see Fig.~\ref{orbit-distributions}). Such `holes' also emerged in our previous simulation tests, where our model results match the truth well \citep{Jin2025b}. The substructures, including the BP/X-shaped structure in the nuclear disc and the ring-like structure in the thick disc, are artefacts of model fluctuations, and they change within the $1\sigma$ confidence level across different models.

Similarly, we derived the luminosity distributions of each component for each model within the $1\sigma$ confidence level and plotted their surface brightness profiles on the $x$-$y$, $x$-$z$, and $y$-$z$ planes in Fig.~\ref{surface-brightness-profiles}. The profiles of the galaxy's inner components (nuclear disc, bar, and bulge) exhibit similarly steep radial declines. In contrast, the thin and thick discs display more complex trends: The inner regions of both discs ($r\lesssim20\rm\,arcsec$; $2\rm\,kpc$) exhibit `holes' in the $x$-$y$ plane, while they display approximately constant gradients in the $x$-$z$ and $y$-$z$ planes, with the thin disc being less vertically extended compared to the thick disc. The stellar halo profiles are approximately exponential across all projections. These complex surface brightness profiles derived from our dynamical models indicate that the underlying structures (with different physical origins) might deviate significantly from the assumptions adopted in traditional photometric decomposition methods.

We calculated the luminosity fractions of all components in the entire galaxy, which are displayed in Fig.~\ref{surface-brightness-profiles}: the nuclear disc with $f_{\rm nucl}=4.8_{-1.1}^{+3.9}\%$; the bar with $f_{\rm bar}=30.2_{-5.4}^{+2.3}\%$; the bulge with $f_{\rm bulge}=17.1_{-2.9}^{+5.0}\%$; the thin disc with $f_{\rm thin}=27.5_{-3.7}^{+2.2}\%$; the thick disc with $f_{\rm thick}=15.5_{-4.0}^{+1.2}\%$; and the stellar halo with $f_{\rm halo}=4.9_{-1.5}^{+3.5}\%$. These results demonstrate that the galaxy is dominated by the bar and thin disc, with some contributions from the thick disc and bulge, while the nuclear disc and halo account for minor fractions.

\subsection{Stellar populations of dynamical components}
\label{sec4.3}
\begin{figure*}
    \centering
    \includegraphics[width=18cm]{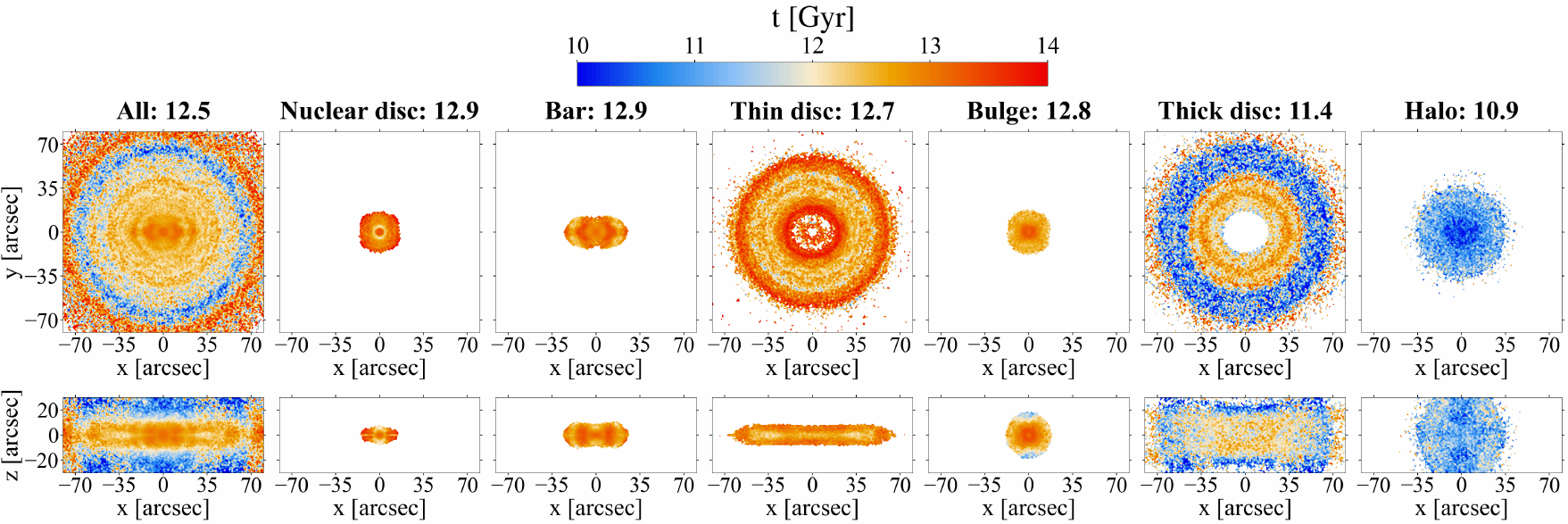}
    \caption{Spatial distributions of luminosity-weighted stellar ages for dynamical components in the best-fitting model of NGC~1381, projected onto the $x$-$y$ (\textit{top}) and $x$-$z$ (\textit{bottom}) planes. \textit{From left to right}: Entire galaxy, nuclear disc, bar, thin disc, bulge, thick disc, and stellar halo. For each component, only pixels above specified brightness thresholds are plotted. The mean stellar ages of the entire galaxy and its components are labelled.}
    \label{age-distributions}
\end{figure*}
\begin{figure*}
    \centering
    \includegraphics[width=18cm]{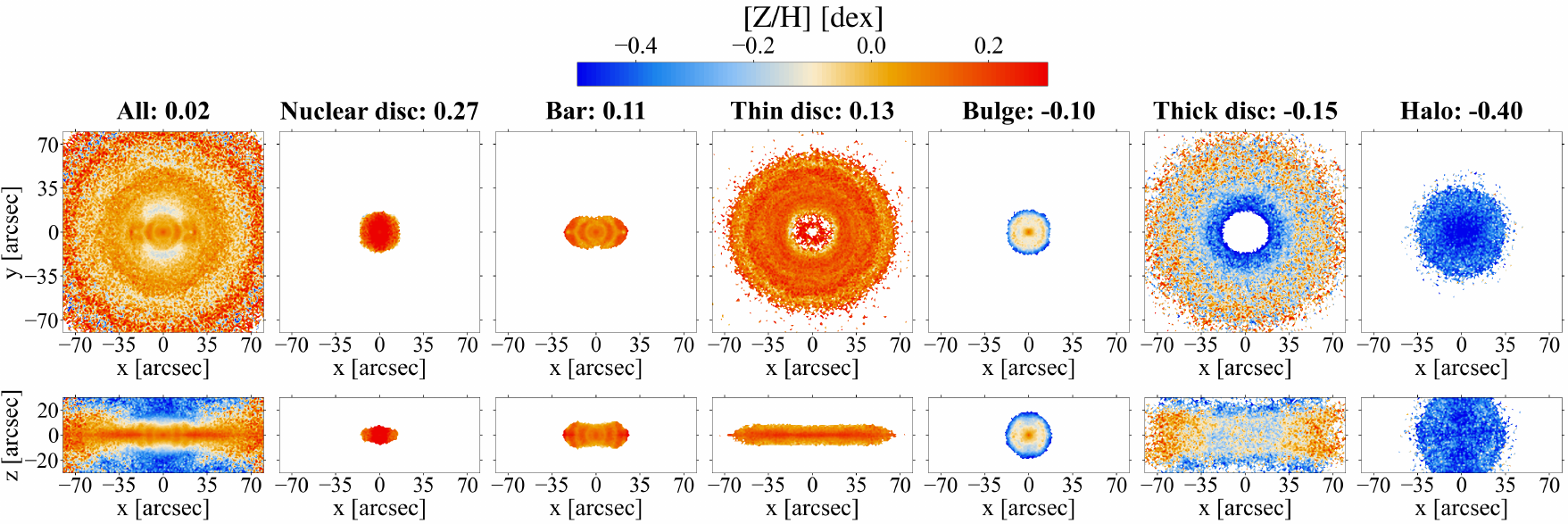}
    \caption{Same as Fig.~\ref{age-distributions} but for luminosity-weighted stellar metallicities ($[Z/\rm H]$).}
    \label{metallicity-distributions}
\end{figure*}
\begin{figure*}[!h]
    \centering
    \includegraphics[width=18cm]{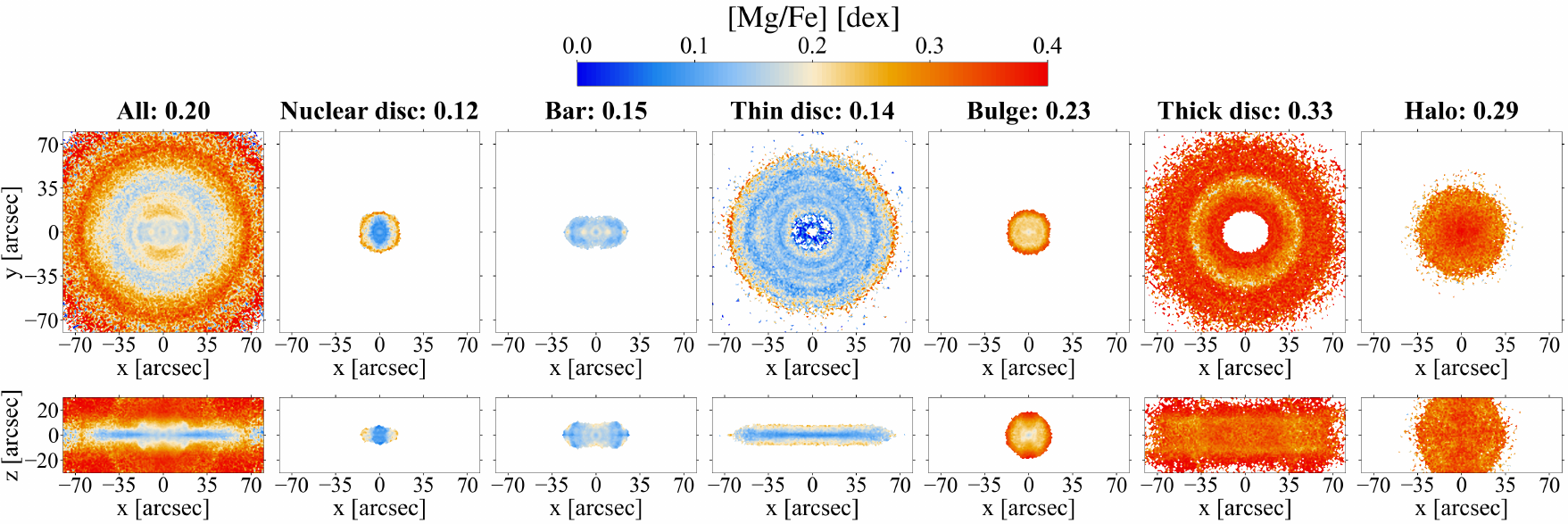}
    \caption{Same as Fig.~\ref{age-distributions} but for luminosity-weighted [Mg/Fe] abundances.}
    \label{alpha-distributions}
\end{figure*}
\begin{figure*}
    \centering
    \includegraphics[width=18cm]{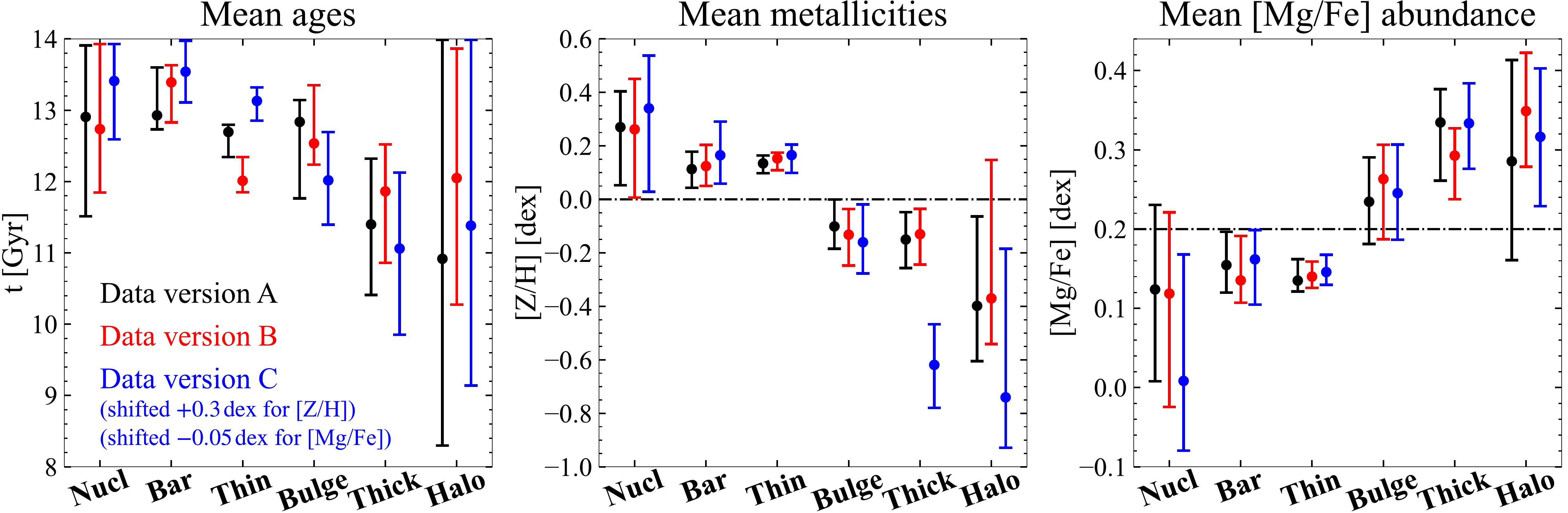}
    \caption{Luminosity-weighted mean stellar ages (\textit{left}), stellar metallicities (\textit{middle}), and [Mg/Fe] abundances (\textit{right}) of different components derived from the three data versions: A (black), B (red), and C (blue). To facilitate a direct comparison, the results for version C have been shifted by $0.3\rm\,dex$ in metallicity and $-0.05\rm\,dex$ in [Mg/Fe] abundance to account for systematic offsets relative to versions A and B. In each panel, the $x$-axis labels (from left to right) represent the nuclear disc, bar, bulge, thin disc, thick disc, and stellar halo. The dots represent the values of the best-fitting model, and the error bars indicate the $1\sigma$ confidence level. The horizontal dashed lines in the middle and right panels represent the reference values ($[Z/\rm H]=0.0$ and $\rm[Mg/Fe]=0.2$ for versions A and B; and $[Z/\rm H]=-0.3$ and $\rm[Mg/Fe]=0.25$ for version C).}
    \label{mean-stellar-populations}
\end{figure*}
\begin{figure*}
    \sidecaption
    \includegraphics[width=12cm]{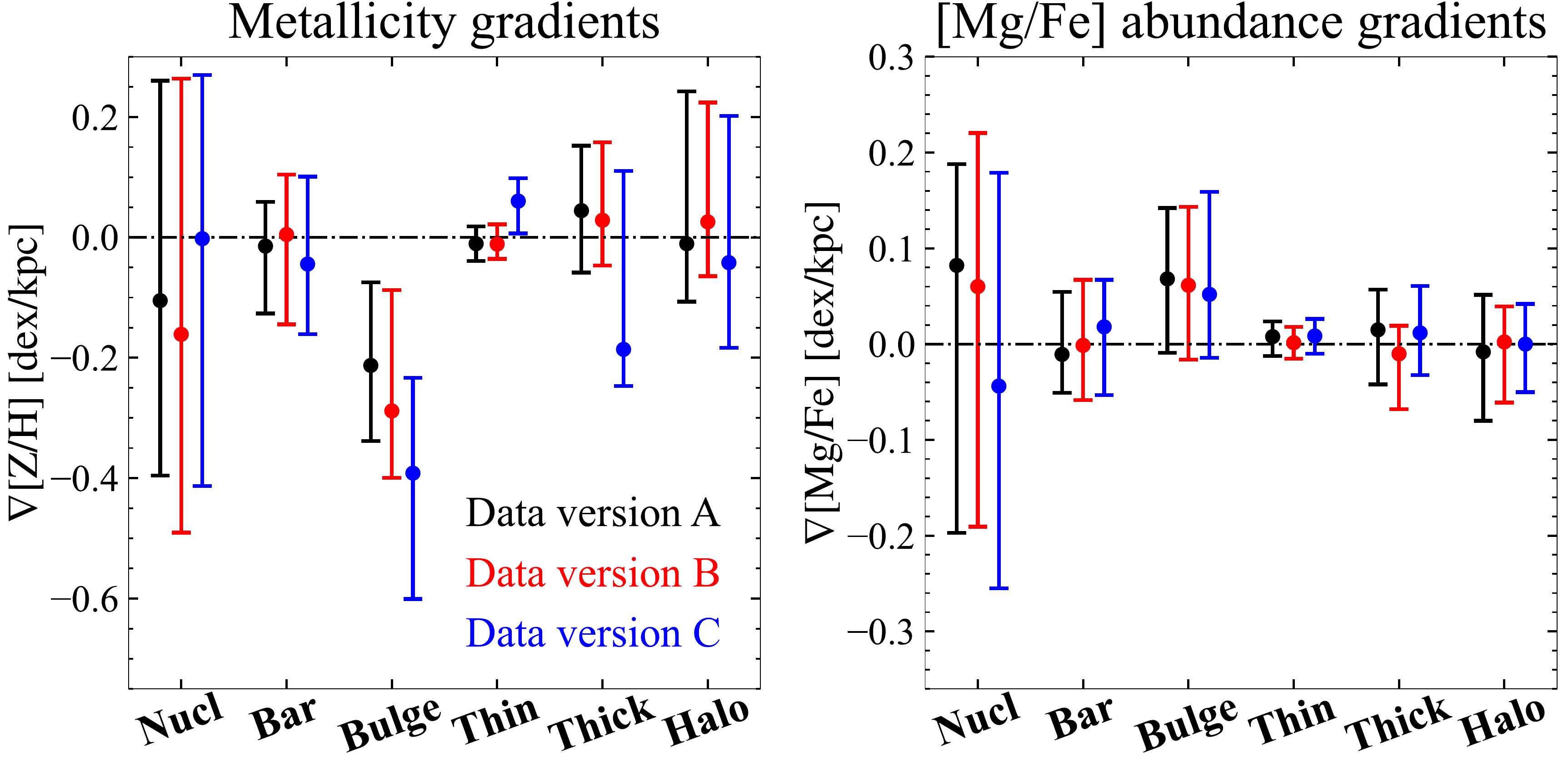}
    \caption{Stellar metallicity gradients (\textit{left}) and [Mg/Fe] abundance gradients (\textit{right}) of different components derived from the three data versions: A (black), B (red), and C (blue). The $x$-axis labels (from left to right) represent the nuclear disc, bar, bulge, thin disc, thick disc, and stellar halo. The dots represent the values of the best-fitting model, and the error bars indicate the $1\sigma$ confidence level.}
    \label{metallicity-alpha-gradients}
\end{figure*}

As mentioned in Sect.~\ref{sec3.3}, we derived the stellar age for each orbit in our models. We therefore calculated the 3D luminosity-weighted stellar age distributions for each decomposed component in the best-fitting model. Fig.~\ref{age-distributions} presents these distributions through their projections onto the $x$-$y$ and $x$-$z$ planes. We also show the mean stellar age for each component in the figure. All components are old, with the nuclear disc, bar, thin disc, and bulge sharing similar ages, while the thick disc and stellar halo are slightly younger.

Following our analysis of stellar ages, we similarly derived the luminosity-weighted stellar metallicity ($[Z/\rm H]$) and [Mg/Fe] abundance distributions in the best-fitting model and presented them in Figs.~\ref{metallicity-distributions} and~\ref{alpha-distributions}, respectively. The corresponding metallicity and [Mg/Fe] abundance profiles are shown in Fig.~\ref{metallicity-alpha-profiles}. Based on the mean $[Z/\rm H]$ and [Mg/Fe] values, the six dynamical components can be separated into two groups with distinct stellar populations: (1) the metal-rich ($[Z/\rm H]>0$) and $\alpha$-poor ($\rm[Mg/Fe]<0.2$) nuclear disc, bar, and thin disc; and (2) the metal-poor ($[Z/\rm H]<0$) and $\alpha$-rich ($\rm[Mg/Fe]>0.2$) bulge, thick disc, and stellar halo. These results indicate distinct physical origins for the two groups.

For each model within the $1\sigma$ confidence level, we calculated the luminosity-weighted mean stellar ages, metallicities, and [Mg/Fe] abundances of all dynamical components and presented them in Fig.~\ref{mean-stellar-populations}. For each data version (A, B, and C), obvious differences are revealed in the mean metallicities and [Mg/Fe] abundances: The bulge, thick disc, and stellar halo are consistently metal-poor and $\alpha$-rich compared to the nuclear disc, bar, and thin disc within the $1\sigma$ confidence level. Although version C has systematic offsets in stellar metallicity and [Mg/Fe] abundance compared to versions A and B, the relative differences in stellar populations between different components are consistent across all versions. This indicates that our main findings are not affected. All dynamical components are generally old (8--14\,Gyr), with the thick disc and stellar halo being slightly younger than the nuclear disc, bar, and thin disc. The bulge is either younger than (version C) or comparable in age to (versions A and B) the nuclear disc, bar, and thin disc.

To better investigate the origins of different structures, we further computed their stellar metallicity gradients ($\nabla[Z/\rm H]$) and [Mg/Fe] abundance gradients ($\nabla\rm[Mg/Fe]$). These gradients were determined by applying a luminosity-weighted linear regression to the radial profiles, and are shown in Fig.~\ref{metallicity-alpha-gradients}. We note that the age gradients are not shown due to their large uncertainties. For metallicity gradients, only the bulge exhibits a significant negative gradient ($\nabla[Z/\rm H]_{bulge}<0$ for all data versions), while all other components show no significant trends (uncertainties encompass zero). For [Mg/Fe] abundance gradients, none of the components display any significant positive or negative gradients.

\section{Discussion}
\label{sec5}
\subsection{An ancient slow bar with $\mathcal{R}>2$}
\label{sec5.1}
We derived a bar rotation rate of $\mathcal{R}=2.40_{-0.27}^{+0.54}$ for NGC~1381, corresponding to a slow bar. This rate depends directly on the bar length, $R_{\rm bar}$, the bar pattern speed, $\rm\Omega_p$, and the angular velocity profile, $\Omega(R)$. In this subsection, we discuss the possible reasons for such a slow bar in NGC~1381.

Statistical studies have reported that most bars in nearby galaxies are fast bars ($1.0\le\mathcal{R}\le1.4$), with average bar pattern speeds of approximately 20--$50\rm\,km\,s^{-1}\,kpc^{-1}$, average bar lengths of roughly 3--$6\rm\,kpc$, and average corotation radii of around 4--$10\rm\,kpc$ (e.g. \citealp{Rautiainen2008,Aguerri2015,Guo2019,Cuomo2020,GarmaOehmichen2022,RuizGarcia2024}). The discrepancies in mean values among these studies arise from differences in sample selection and methodology, with both bar lengths and corotation radii strongly correlated with stellar masses. However, because measuring bar pattern speeds requires high-quality data, most galaxies in these samples ($M_\star\gtrsim3\times10^{10}\rm\,M_\odot$) are more massive than NGC~1381 ($M_\star=1.65_{-0.38}^{+0.13}\times10^{10}\rm\,M_\odot$ from our models and $2.25\times10^{10}\rm\,M_\odot$ reported by \citealp{Iodice2019a}). Compared to these statistical averages, our modelling indicates that NGC~1381, with a smaller stellar mass, hosts a relatively short bar ($R_{\rm bar}=2.24_{-0.22}^{+0.43}\rm\,kpc$), but its pattern speed ($\rm\Omega_p=34_{-7}^{+4}\,km\,s^{-1}\,kpc^{-1}$) and corotation radius ($R_{\rm CR}=5.38_{-0.28}^{+1.59}\rm\,kpc$) are comparable to the typical values.

According to the current paradigm of bar dynamics, the formation of nuclear discs is driven by gas inflow that is channelled by bars (e.g. \citealp{Athanassoula2003,Kormendy2004,Gadotti2011}). Therefore, the stellar age of the nuclear disc, $t_{\rm nucl}$, is a good tracer of the bar formation time, $t_{\rm bar,form}$ (e.g. \citealp{deSaFreitas2023,deSaFreitas2025}). The nuclear disc of NGC~1381 is very old ($t_{\rm nucl}\sim13\rm\,Gyr$), indicating its formation in the early Universe. After the galaxy entered the cluster approximately 8--12\,Gyr ago (e.g. \citealp{Iodice2019b,Ding2023}), the star formation of the galaxy might have ceased due to strangulation-dominated environmental quenching (e.g. \citealp{MorokumaMatsui2022,Martig2026}), which has led to an ancient ($t_{\rm bar}\sim13\rm\,Gyr$ and $t_{\rm bar,form}\sim13\rm\,Gyr$) bar in the present day.

Theoretical studies have indicated that bars formed in rotation-dominated discs typically exhibit high pattern speeds during their early evolution, as demonstrated in simulations (e.g. \citealp{Semczuk2024,Habibi2024}). As shown in Fig.~\ref{mass-profiles}, the dark matter in NGC~1381 becomes dominant at radii comparable to or smaller than the bar length. Therefore, dynamical friction between the bar and dark matter could transfer the bar's angular momentum to the dark matter halo (e.g. \citealp{Tremaine1984b,Weinberg1985,Debattista2000}), significantly reducing the pattern speed, $\rm\Omega_p$. This scenario is also supported by studies of individual galaxies that contain slow bars (e.g. UGC~628, \citealp{Chemin2009}; NGC~4277, \citealp{Buttitta2022}; NGC~4371, \citealp{Tahmasebzadeh2024}; NGC~6951, \citealp{Lee2025}). The effect of dynamical friction might have been enhanced after the galaxy's infall into the cluster, when the bar could no longer gain angular momentum from gas inflows. Furthermore, $\Omega(R)$ in NGC~1381's outer regions is mainly affected by dark matter, leading to an $R_{\rm CR}$ comparable to that of nearby massive galaxies despite the galaxy's lower stellar mass.

In practice, we cannot fully exclude the influence of model uncertainties on the rotation rate, $\mathcal{R}$, but our conclusion that the galaxy hosts a slow bar remains robust. The angular velocity profile, $\Omega(R)$, was well constrained in our modelling, as the observed kinematic maps were fully utilised to constrain the gravitational potential. The bar pattern speed, $\rm\Omega_p$, can be recovered with a relative uncertainty of $\lesssim15\%$, as demonstrated in our previous simulation tests \citep{Jin2025a}. The dynamical bar length, $R_{\rm bar}$, is degenerate with the bar azimuthal angle, $\varphi$, since we observe only the projected bar (projected length $\sim R_{\rm bar}\sin{|\varphi|}$). Consequently, a smaller $|\varphi|$ leads to a longer inferred $R_{\rm bar}$. In this work, we derived $|\varphi|={30_{-14}^{+4}}^\circ$, corresponding to an end-on bar. The BP/X-shaped structure in the galaxy, as shown in Fig.~\ref{image-and-MUSE-pointing}, is difficult to detect from an extremely end-on view ($|\varphi|\lesssim15^\circ$), which indicates that the model-predicted $|\varphi|$ is unlikely to be overestimated and $R_{\rm bar}$ is unlikely to be underestimated. The robustness of $\Omega(R)$, $\rm\Omega_p$, and $R_{\rm bar}$ ensures the accuracy of the derived $\mathcal{R}$ in our analysis.

\subsection{The assembly history of NGC~1381}
\label{sec5.2}
We analysed the physical origins of different structures to investigate the assembly history of NGC~1381. The mean metallicities and [Mg/Fe] abundances shown in Fig.~\ref{mean-stellar-populations} demonstrate two distinct stellar populations in NGC~1381: (1) the metal-rich, $\alpha$-poor, and old nuclear disc, bar, and thin disc; and (2) the metal-poor and $\alpha$-rich bulge, thick disc, and stellar halo, which are younger than or comparable in age to the first group.

The stellar populations of the nuclear disc, bar, and thin disc, which are similarly metal-rich and $\alpha$-poor, support their in situ origins. The bar likely formed through the dynamical reshaping of the thin disc, while the nuclear disc might be the result of in situ star formation from gas inflow driven by the bar or other clumpy instabilities (e.g. \citealp{Neumayer2020}). For NGC~1381, the bar-driven scenario is a reasonable explanation (e.g. \citealp{Debattista2004,Athanassoula2013b}) and would produce a dynamically cold nuclear disc at its early stage. We note that the nuclear disc at present is dynamically warm ($\overline{\lambda_z}=0.57_{-0.11}^{+0.12}$), which might be caused by long-term dynamical heating from the bar, or simply by mixing with the dynamically hot bulge, given the relatively loose kinematic constraint we adopted ($\lambda_z>0.5$; see Fig.~\ref{orbit-classification-criteria}).

In contrast, the metal-poor and $\alpha$-rich stellar populations in the bulge, thick disc, and stellar halo may have either formed in situ at an early epoch through a rapid process---similar to the thick disc formation scenario proposed for the Milky Way (e.g. \citealp{Lehnert2014,Helmi2018,Yu2021})---or have been accreted later via minor mergers. However, unlike in the Milky Way, the thin disc and bar in NGC~1381 are older than or comparable in age to the bulge, thick disc, and halo. The presence of the bar and the strangulation-dominated (rather than rapid) quenching of NGC~1381 (e.g. \citealp{MorokumaMatsui2022,Martig2026}) indicate that the thin disc has experienced a sustained period of secular evolution and star formation. Therefore, early in situ components should be substantially older than the thin disc and bar; the observed age relationships do not support an early in situ origin for the bulge, thick disc, or halo.

We therefore suggest that a non-negligible fraction of the stars in the bulge, thick disc, and halo of NGC~1381 were accreted through minor mergers that occurred after the formation of the thin disc and bar. Their low metallicity and enhanced $\alpha$-element abundances are consistent with an origin in lower-mass systems like dwarf galaxies. Moreover, the fact that these accretion events did not disrupt the pre-existing thin disc and bar further supports their interpretation as minor mergers.

The similarly old ages of all six components suggest that NGC~1381 may have fallen into the Fornax cluster early, consistent with previous works (e.g. \citealp{Iodice2019b,Ding2023}). The cluster environment has therefore influenced the galaxy's subsequent evolution. In addition to strangulation-dominated quenching, other processes such as dynamical harassment or tidal interactions may have dynamically heated the galaxy's structures, especially the loosely bound outer regions. However, the clear difference in stellar populations between the thin and thick discs (see Fig.~\ref{mean-stellar-populations}) argues against the scenario in which the thick disc was built through the dynamical heating of the thin disc. The similar radial extents of the thin and thick discs (see Figs.~\ref{luminosity-distributions} and~\ref{surface-brightness-profiles}), together with the lack of obvious metallicity and [Mg/Fe] gradients in the thick disc (see Fig.~\ref{metallicity-alpha-gradients}), also argue against the outer regions of the thin disc having been transformed into thick disc stars via dynamical heating. Instead, the stellar population in the thick disc is highly similar to that in the halo, suggesting that their origins differ from that of the thin disc.

In Fig.~\ref{metallicity-alpha-gradients}, only the bulge shows a negative metallicity gradient ($\nabla[Z/\rm H]_{bulge}<0$ for all data versions). The metallicity gradients of other components, as well as all [Mg/Fe] abundance gradients, are not significant. These results indicate that, apart from the bulge, stars in each component share similar stellar populations, suggesting similar physical origins (in situ for the bar, thin disc, and nuclear disc; ex situ for the thick disc and halo). In contrast, the bulge's negative gradient can arise either from the main galaxy or from the accreted dwarf galaxies; its precise origin remains ambiguous given the limitations of the current data. Due to this negative gradient and the uncertainties in stellar age, we cannot fully rule out the possibility of a primordial bulge predating the bar formation, nor can we quantify the ex situ fraction of the bulge.

Our findings are consistent with \citet{Pinna2019a}, who defined the thick disc of NGC~1381 directly from the projected 2D image and attributed its origin to the accretion of ex situ stars. Compared to their results, our study performs a more physically motivated 3D decomposition, successfully distinguishing structures with different origins, particularly the bar, bulge, and nuclear disc. We further reveal that ex situ stars are present not only in the outer regions (thick disc and stellar halo) but also in the inner regions (bulge) of the galaxy.

Based on the analysis in this subsection and Sect.~\ref{sec5.1}, we propose the following assembly history for NGC~1381: The galaxy likely formed its bar from a rotation-dominated disc approximately 13\,Gyr ago. Then, the nuclear disc formed from gas inflow and/or radial migration of disc stars driven by the bar. The bar was fast initially but gradually slowed down due to dynamical friction with dark matter. The galaxy experienced a series of minor mergers, which built up its thick disc and stellar halo. These mergers also led to the formation (or growth) of the dynamically hot bulge. After the galaxy fell into the Fornax cluster 8--12\,Gyr ago, the gas supply in the outer regions of the galaxy was cut off (strangulation), resulting in the galaxy being finally quenched but with its internal structures well preserved. This environmental quenching might have further enhanced the bar's slowdown by halting the supply of angular momentum, leading to a slow bar with $\mathcal{R}>2$ at present.

\section{Summary}
\label{sec6}
We constructed 3D chemo-dynamical models for the barred S0 galaxy NGC~1381 in the Fornax cluster, using the barred population-orbit superposition method developed and validated in \citet{Jin2025a,Jin2025b}. We decomposed this galaxy into six components based on stellar orbits, namely a nuclear disc, a bar, a bulge, a thin disc, a thick disc, and a stellar halo. We analysed the stellar kinematics, luminosity distributions, and stellar populations of each component and inferred their physical origins. The main conclusions are as follows.

\begin{enumerate}
\item We decomposed NGC~1381 into six components with distinct kinematics and morphologies: (1) a dynamically warm nuclear disc ($f_{\rm nucl}=4.8_{-1.1}^{+3.9}\%$); (2) a rigidly rotating, BP/X-shaped bar ($f_{\rm bar}=30.2_{-5.4}^{+2.3}\%$); (3) a dynamically hot, spheroidal bulge ($f_{\rm bulge}=17.1_{-2.9}^{+5.0}\%$); (4) a dynamically cold thin disc ($f_{\rm thin}=27.5_{-3.7}^{+2.2}\%$); (5) a vertically extended thick disc with slightly slower rotation than the thin disc ($f_{\rm thick}=15.5_{-4.0}^{+1.2}\%$); and (6) a dynamically hot, spatially diffuse stellar halo ($f_{\rm halo}=4.9_{-1.5}^{+3.5}\%$). 

\item Our modelling reveals the presence of a slow bar ($\mathcal{R}=2.40_{-0.27}^{+0.54}$) in NGC~1381, with a bar pattern speed of $\rm\Omega_p=34_{-7}^{+4}\,km\,s^{-1}\,kpc^{-1}$, a bar length of $R_{\rm bar}=2.24_{-0.22}^{+0.43}\rm\,kpc$, and a corotation radius of $R_{\rm CR}=5.38_{-0.28}^{+1.59}\rm\,kpc$.

\item The stellar metallicities and [Mg/Fe] abundances exhibit clear differences between components: The nuclear disc, bar, and thin disc are metal-rich ($[Z/\rm H]\gtrsim0$ for versions A and B; $[Z/\rm H]\gtrsim-0.3$ for version C) and $\alpha$-poor ($\rm[Mg/Fe]\lesssim0.2$ for versions A and B; $\rm[Mg/Fe]\lesssim0.25$ for version C), while the bulge, thick disc, and stellar halo are metal-poor ($[Z/\rm H]\lesssim0$ for versions A and B; $[Z/\rm H]\lesssim-0.3$ for version C) and $\alpha$-rich ($\rm[Mg/Fe]\gtrsim0.2$ for versions A and B; $\rm[Mg/Fe]\gtrsim0.25$ for version C). The bulge shows a negative metallicity gradient ($\nabla[Z/\rm H]_{bulge}<0$), while the metallicity gradients of other components and the [Mg/Fe] abundance gradients of all components are not significant. All components in NGC~1381 are generally old (8--14\,Gyr). Compared to the nuclear disc, bar, and thin disc, the bulge, thick disc, and stellar halo are either younger or of comparable age.

\end{enumerate}

Based on these results, we infer that the nuclear disc, bar, and thin disc of NGC~1381 formed in situ in the early Universe. The bar was likely fast initially but gradually slowed due to dynamical friction with the dark matter halo and the lack of gas supply after the galaxy's infall into the cluster. In contrast, the thick disc and stellar halo likely formed via minor mergers. These mergers also contributed to the formation or growth of the dynamically hot bulge.

As a fossil record from the early Universe, NGC~1381 provides valuable insights into early galaxy evolution. This particular galaxy serves as a non-trivial local descendant of the barred galaxies at $z\gtrsim2$ discovered by JWST observations (e.g. \citealp{LeConte2024,Guo2025}). A valuable next step will be to statistically study barred galaxies in both cluster and non-cluster environments and to compare them with future IFS observations of high-redshift galaxies. Such a comparative analysis will advance our understanding of the evolution of barred galaxies from high redshifts to the present day.

\begin{acknowledgements} 
We thank Dr. Richard J. Long for useful discussions. This work is supported by the National Natural Science Foundation of China under Grant No. 12403017. This work is partly supported by the National Natural Science Foundation of China (Grant No. 11821303 to SM) and the CAS Project for Young Scientists in Basic Research, Grant No. YSBR-062 (LZ). FP acknowledges support from the Horizon Europe research and innovation programme under the Maria Sk\l{}odowska-Curie grant `TraNSLate' No. 101108180.

\end{acknowledgements}

\begin{appendix}
\onecolumn
\section{Stellar population maps for data versions B and C}
\begin{figure*}[!h]
    \centering
    \includegraphics[width=18cm]{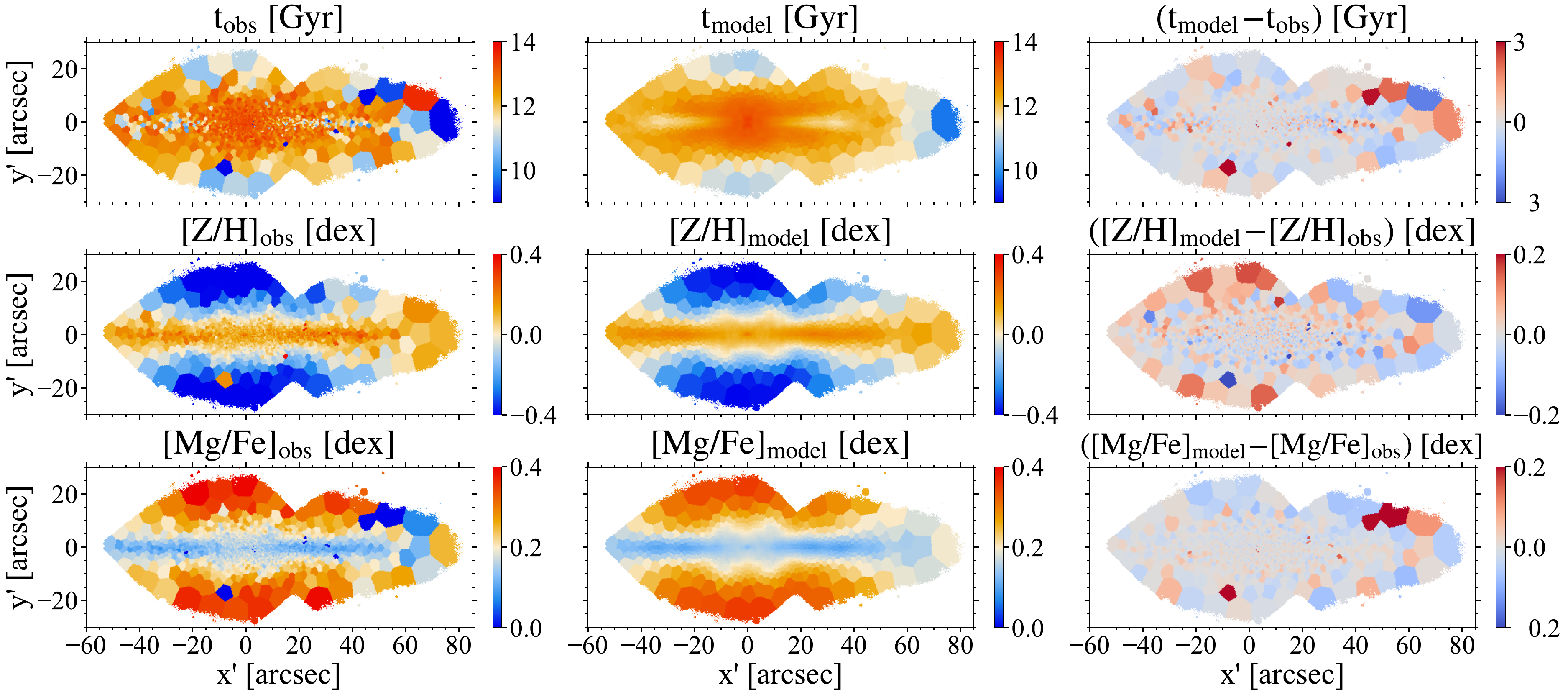}
    \caption{Same as the bottom three panels of Fig.~\ref{kinematic-chemo-maps-best-fitting} but for data version B.}
    \label{chemo-maps-best-fitting-Francesca}
\end{figure*}
\begin{figure*}[!h]
    \centering
    \includegraphics[width=18cm]{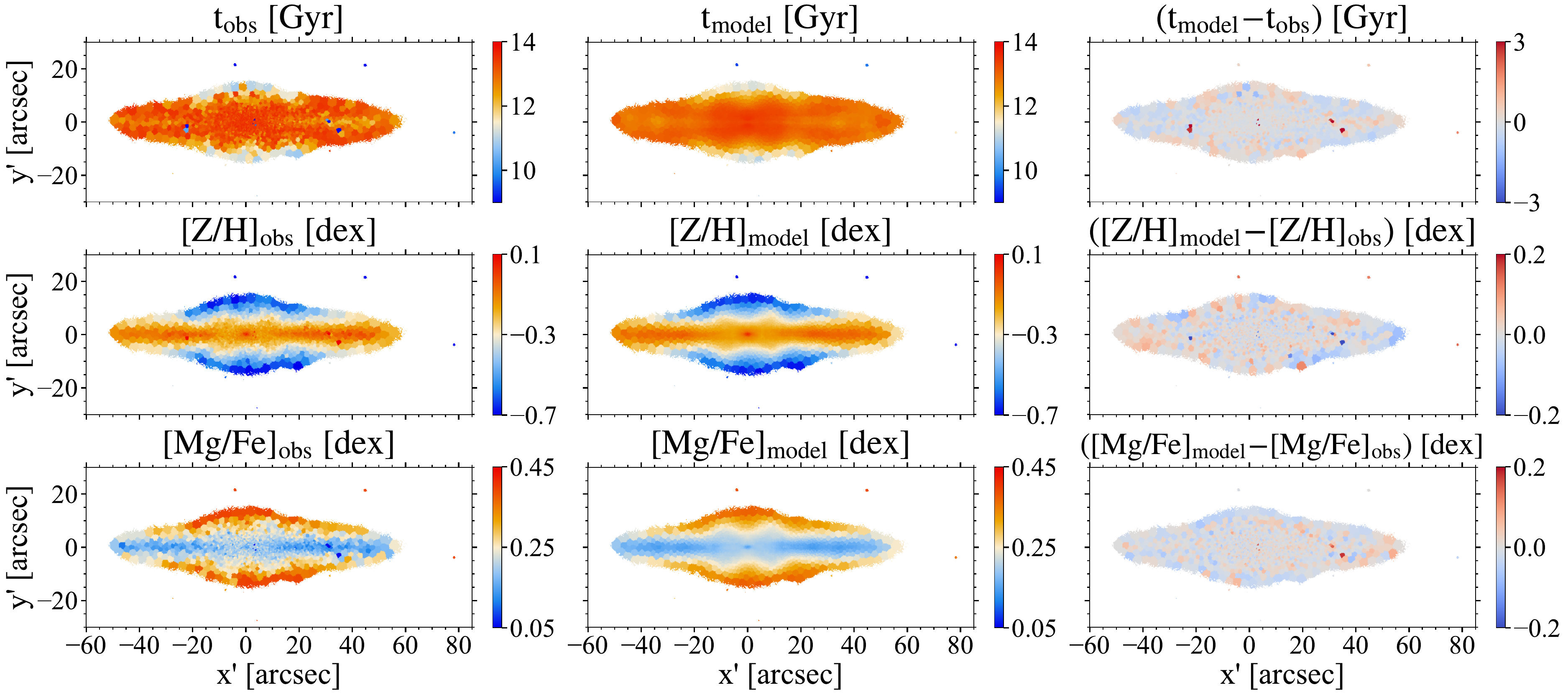}
    \caption{Same as the bottom three panels of Fig.~\ref{kinematic-chemo-maps-best-fitting} but for data version C.}
    \label{chemo-maps-best-fitting-Nacho}
\end{figure*}

The stellar population data and model fittings based on data version B (\citealp{Pinna2019a}) are shown in Fig.~\ref{chemo-maps-best-fitting-Francesca}, while those from version C (\citealp{MartinNavarro2021}) are presented in Fig.~\ref{chemo-maps-best-fitting-Nacho}. Compared to versions A and B, version C has systematic offsets in the stellar metallicity and [Mg/Fe] abundance, which originate from their different settings when fitting the spectra. Nevertheless, the overall trends in the maps are consistent, ensuring the robustness of our main conclusions.

\section{MGE fitting results}
\begin{figure}[!h]
    \centering
    \includegraphics[width=9cm]{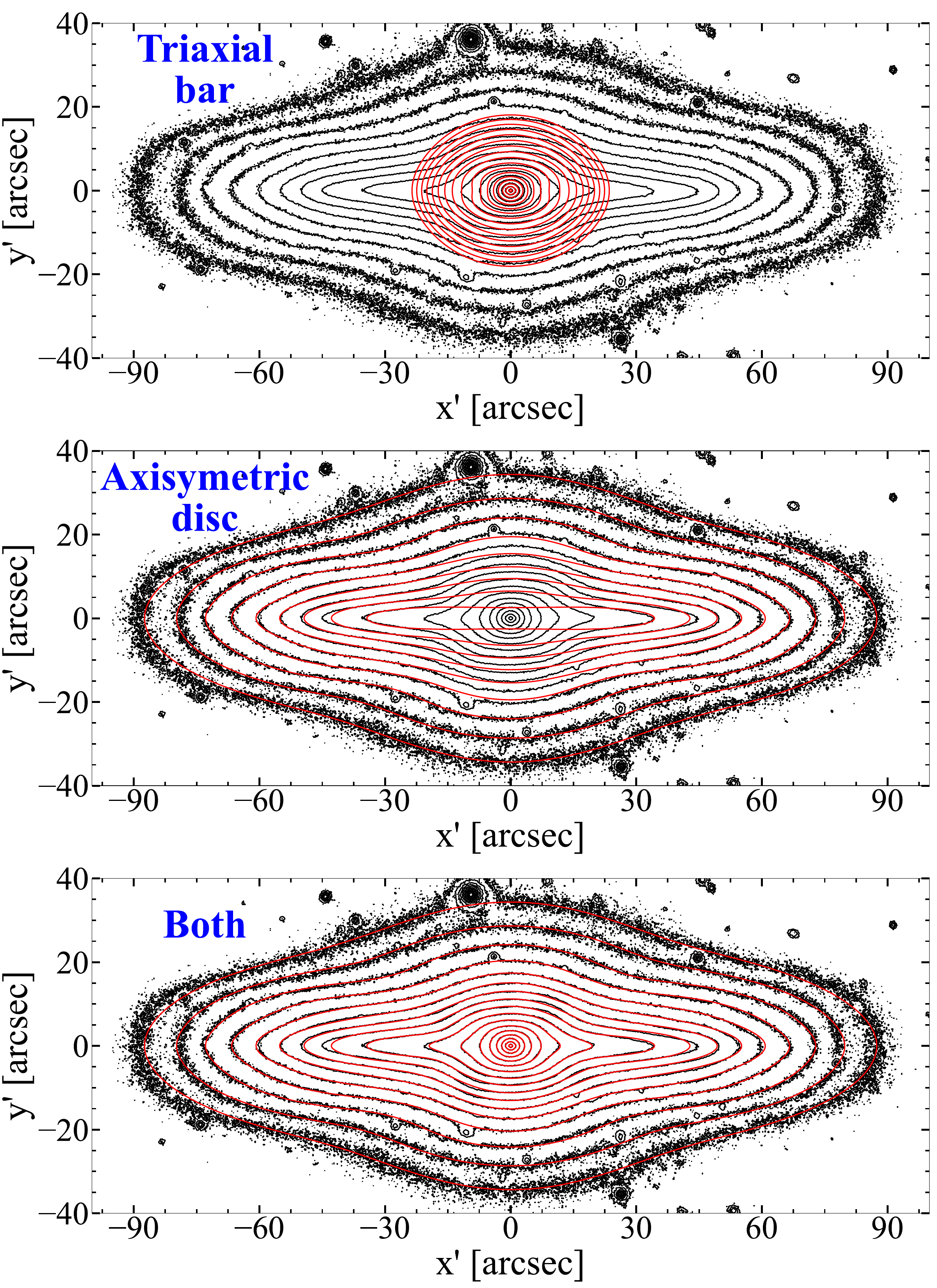}
    \caption{Observed surface brightness contours (black) and MGE-fitted contours (red) for NGC~1381. The red contours represent the triaxial bar component, the axisymmetric disc component, and the sum of the two components in the MGE fitting results, respectively (see Table~\ref{table-mge-parameters}). The contour interval is 0.5 magnitude.}
    \label{mge-contours}
\end{figure}

The $r$-band surface brightness of NGC~1381 was fitted by the MGE formalism \citep{Cappellari2002}, which can be expressed as
\begin{equation}
    \Sigma(x',y')=\sum_{j=1}^{N} \frac{L_j}{2\pi\sigma_j'^2 q_j'}\exp \left[-\frac{1}{2\sigma_j'^2} \left(x'^2+\frac{y'^2}{q_j'^2} \right) \right],
\label{mge2D}
\end{equation}
where $N$ is the number of Gaussians, the subscript $j$ denotes the $j$-th Gaussian, $L_j$ is the total luminosity, $q_j'=b_j'/a_j'$ is the axis ratio, and $\sigma_j'$ is the dispersion along the major axis $x'$. Negative Gaussians are allowed ($L_j<0$) in order to fit the BP/X-shaped structure. The fitted MGE parameters are listed in Table~\ref{table-mge-parameters}, with the corresponding surface brightness contours shown in Fig.~\ref{mge-contours}.

\begin{table}[!h]
\centering
\caption{MGE fitting parameters for NGC~1381.}
\begin{tabular}{|c|c|c|c|}
\hline
Component & $L_j\,[\rm L_{\odot}]$ & $\sigma_j'\,[\rm arcsec]$ & $q_j'$\\
\hline
\multirow{7}*{Triaxial bar} & 5.8514e+09 & 0.6156 & 0.9399 \\
~                           & 2.3452e+09 & 0.8111 & 0.2924 \\
~                           & 4.8630e+09 & 1.5880 & 0.8384 \\
~                           & 4.9462e+09 & 3.6569 & 0.5827 \\
~                           & -2.8594e+09 & 4.4236 & 0.4186 \\
~                           & 1.0836e+09 & 6.9692 & 0.5614 \\
~                           & 4.5652e+08 & 7.9081 & 0.8153 \\
\hline
\multirow{9}*{Axisymmetric disc} & 7.9384e+08 & 13.8476 & 0.1330 \\
~                               & 4.7452e+08 & 16.0688 & 0.5392 \\
~                               & -1.9755e+10 & 17.3858 & 0.1379 \\
~                               & 1.9177e+10 & 17.8108 & 0.1352 \\
~                               & 1.6929e+07 & 19.4488 & 0.9399 \\
~                               & -1.0310e+09 & 20.1765 & 0.3888 \\
~                               & 7.9930e+08 & 23.5237 & 0.2900 \\
~                               & 7.5981e+07 & 38.0095 & 0.2818 \\
~                               & 8.7859e+06 & 58.4657 & 0.6147 \\
\hline
\end{tabular}
\tablefoot{From left to right: (1) the component to which each Gaussian belongs; (2) luminosity $L_j$; (3) dispersion $\sigma_j'$; (4) axis ratio $q_j'$. These MGE parameters were calculated from Eq.~(\ref{mge2D}).}
\label{table-mge-parameters}
\end{table}

\section{Parameter space and mass profiles}
\label{sec-parameter-space}
\begin{figure*}[!h]
    \centering
    \includegraphics[width=18cm]{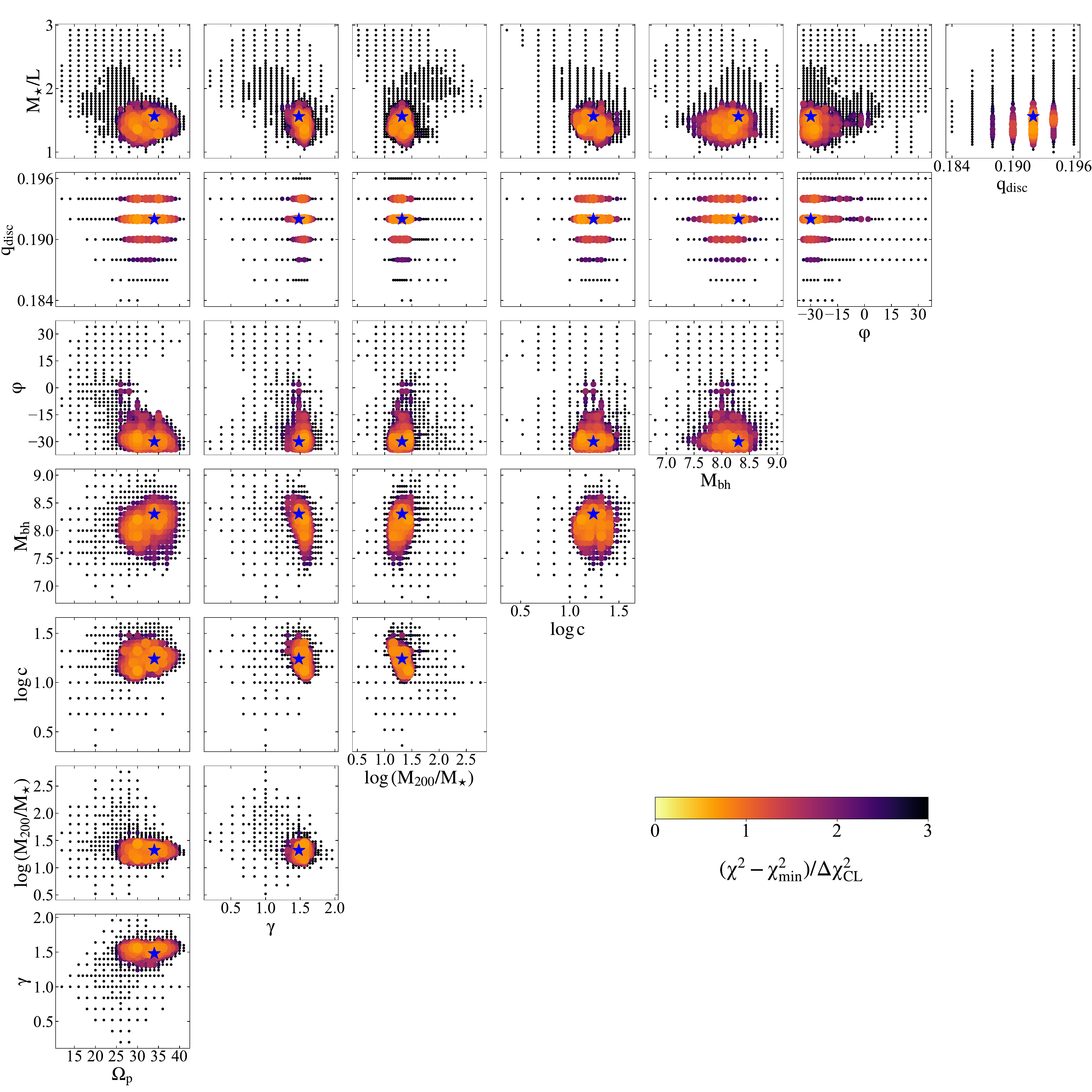}
    \caption{Parameter space we explored for NGC~1381. The eight free parameters are the disc axis ratio, $q_{\rm disc}$ (transformed from the inclination angle, $\theta$); the bar azimuthal angle, $\varphi$; the stellar mass-to-light ratio, $M_\star/L$; the bar pattern speed, $\rm\Omega_p$; the dark matter concentration, $c$; the virial mass, $M_{200}$; the inner density slope of dark matter, $\gamma$; and the black hole mass, $M_{\rm BH}$. The blue star marks the best-fitting model with minimum $\chi^2$ ($\chi^2_{\rm min}$), and the coloured dots indicate models that fall within the $3\sigma$ confidence level, with their $\chi^2$ values indicated by the colour bar. The small black dots correspond to models outside the $3\sigma$ confidence level. For formal definitions of $\chi^2$ and confidence levels, see Sects.~\ref{sec3.1} and~\ref{sec3.4}.}
    \label{parameter-grids}
\end{figure*}
\begin{figure}[!h]
    \centering
    \includegraphics[width=9cm]{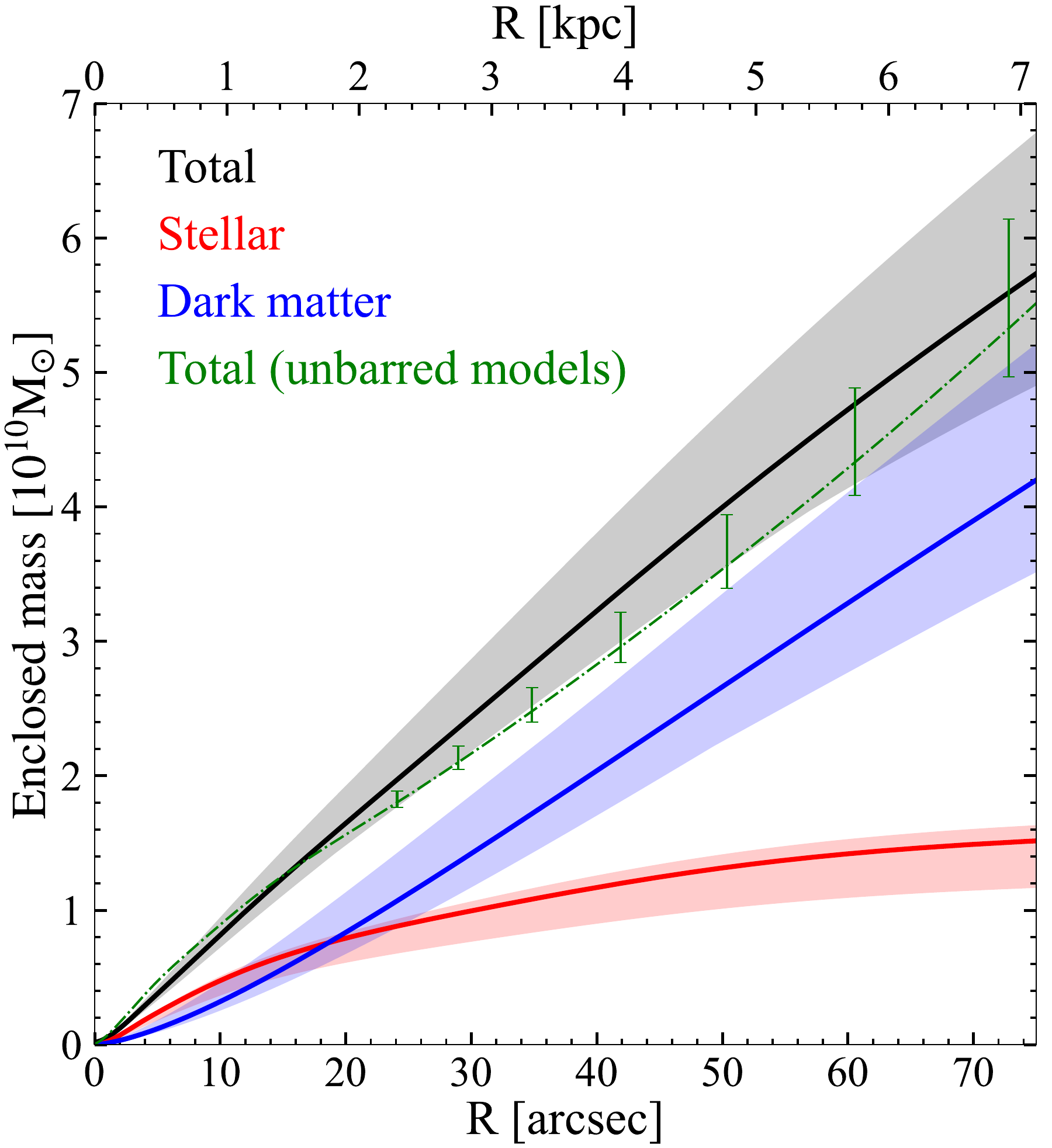}
    \caption{Enclosed mass profiles for NGC~1381. The black, red, and blue solid lines represent the profiles of total mass, stellar mass, and dark matter mass from the best-fitting model, respectively, while the shaded regions denote the model uncertainties within the $1\sigma$ confidence level. The green dashed line with error bars corresponds to the total mass profile from orbit-superposition models without considering the bar \citep{Ding2023}.}
    \label{mass-profiles}
\end{figure}

The parameter space explored in our dynamical modelling is presented in Fig.~\ref{parameter-grids}. The derived values of the free parameters are: (1) a disc axis ratio of $q_{\rm disc}=0.192_{-0.002}^{+0.002}$ (corresponding to an inclination angle of $\theta\approx87$--$88^\circ$); (2) a bar azimuthal angle of $|\varphi|={30_{-14}^{+4}}^\circ$; (3) a stellar mass-to-light ratio of $M_\star/L=1.56_{-0.36}^{+0.12}$ (corresponding to a stellar mass of $M_\star=1.65_{-0.38}^{+0.13}\times10^{10}\rm\,M_\odot$); (4) a bar pattern speed of $\rm\Omega_p=34_{-7}^{+4}\,km\,s^{-1}\,kpc^{-1}$; (5) a dark matter concentration of $\log c=1.24_{-0.16}^{+0.16}$; (6) a virial mass of $M_{200}=3.45_{-1.55}^{+0.91}\times10^{11}\rm\,M_\odot$; (7) a dark matter inner density slope of $\gamma=1.48_{-0.04}^{+0.16}$; and (8) a black hole mass of $M_{\rm BH}=2.0_{-1.5}^{+1.2}\times10^{8}\rm\,M_\odot$.

In our dynamical modelling, the gravitational potential comprises contributions from stellar mass, dark matter, and a supermassive black hole. This leads to a degeneracy between these three components, especially in the innermost regions ($\lesssim5\rm\,arcsec$). \citet{MartinNavarro2021} reported that the IMF becomes more bottom-heavy towards the innermost regions, leading to a higher $M_\star/L$ there. Therefore, assuming a constant $M_\star/L$ likely results in an underestimation of the stellar mass in the central regions. However, this bias could be masked by models that adopt a more cuspy dark matter halo and/or a more massive black hole.

The enclosed stellar, dark matter, and total mass profiles are shown in Fig.~\ref{mass-profiles}. We compared our total mass profile with the result from \citet{Ding2023}, who constructed orbit-superposition models without considering the bar. Regardless of whether the bar is included in the models, the total mass profiles, which are directly constrained by the kinematic data, are consistent.

\section{Circularity, $\lambda_z$: Definitions with and without cross terms in $V_{\rm rms}$}
\label{sec-circularity-definitions}
\begin{figure}[!h]
    \centering
    \includegraphics[width=9cm]{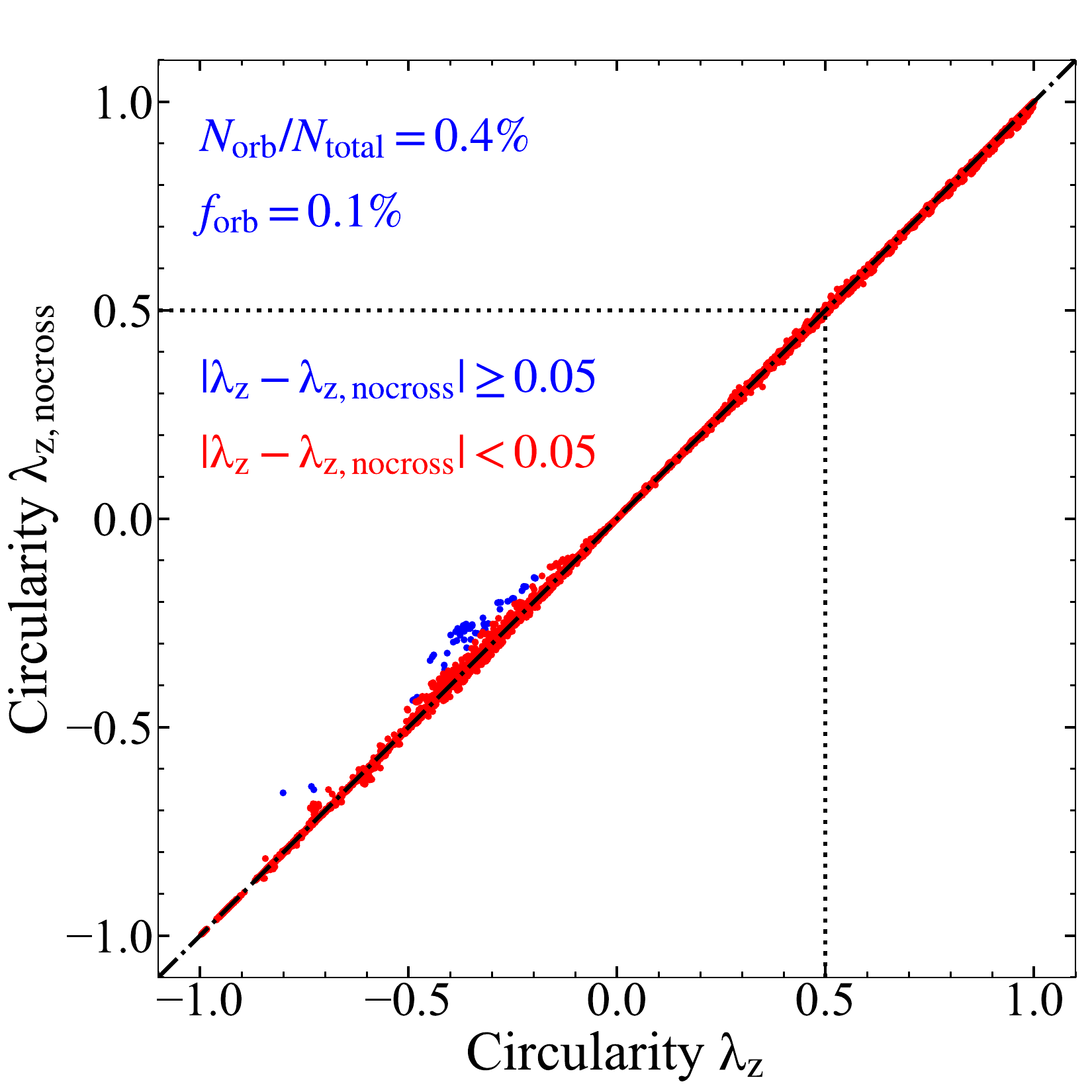}
    \caption{One-to-one comparison of the circularity for each orbit under two definitions. The default circularity $\lambda_z$ adopted in this work (Eq.~\ref{circularity-with-cross-terms}) is defined using the $V_{\rm rms}$ containing cross terms, while $\lambda_{z,\rm nocross}$ (Eq.~\ref{circularity-without-cross-terms}) is based on the $V_{\rm rms,nocross}$ where cross terms are excluded. Each orbit is represented by a dot. Red (blue) dots indicate orbits with $|\lambda_z-\lambda_{z,\rm nocross}|<0.05$ ($\ge0.05$). The orbits with $|\lambda_z-\lambda_{z,\rm nocross}|\ge0.05$ only account for $0.4\%$ of all orbits and contribute $0.1\%$ of the total luminosity. The dashed line represents equality between the two definitions, and the dotted lines correspond to $\lambda_z=0.5$ and $\lambda_{z,\rm nocross}=0.5$.}
    \label{circularity-definition-comparison}
\end{figure}

In this work, we adopted the circularity ($\lambda_z$) defined in Eq.~(\ref{circularity-with-cross-terms}) following previous works on orbit-superposition models (e.g. \citealp{Zhu2018a,Tahmasebzadeh2022,Jin2025a}). This definition of $\lambda_z$ employs a root-mean-square velocity that includes cross terms ($\overline{V_{\rm rms}}=\sqrt{\overline{v_{x0}^2+v_{y0}^2+v_z^2+2v_{x0}v_{y0}+2v_{x0}v_z+2v_{y0}v_z}}$). For comparison, we also considered an alternative definition that excludes these cross terms ($\overline{V_{\rm rms,nocross}}=\sqrt{\overline{v_{x0}^2+v_{y0}^2+v_z^2}}$). The corresponding circularity is then written as
\begin{equation}
    \lambda_{z,\rm nocross}=\overline{L_z}/(R\times\overline{V_{\rm rms,nocross}}).
\label{circularity-without-cross-terms}    
\end{equation}

Figure~\ref{circularity-definition-comparison} presents a one-to-one comparison of the circularity for each orbit in the best-fitting model under two definitions: $\lambda_z$ and $\lambda_{z,\rm nocross}$. For most orbits, the two circularities have very similar values, especially for orbits with $\lambda_z>0$. Only a small fraction of orbits, which account for $0.4\%$ of all orbits and contribute $0.1\%$ of the total luminosity, have a difference of $|\lambda_z-\lambda_{z,\rm nocross}|\ge0.05$. Since the orbit classification criteria we adopted use a threshold of $\lambda_z=0.5$ (see Fig.~\ref{orbit-classification-criteria}), the definition of circularity does not affect our dynamical decompositions.

\section{Stellar orbit distributions in the $\lambda_z$--$E_{\rm bind}$ phase space}
\label{sec-Ebind}
\begin{figure*}
    \centering
    \includegraphics[width=18cm]{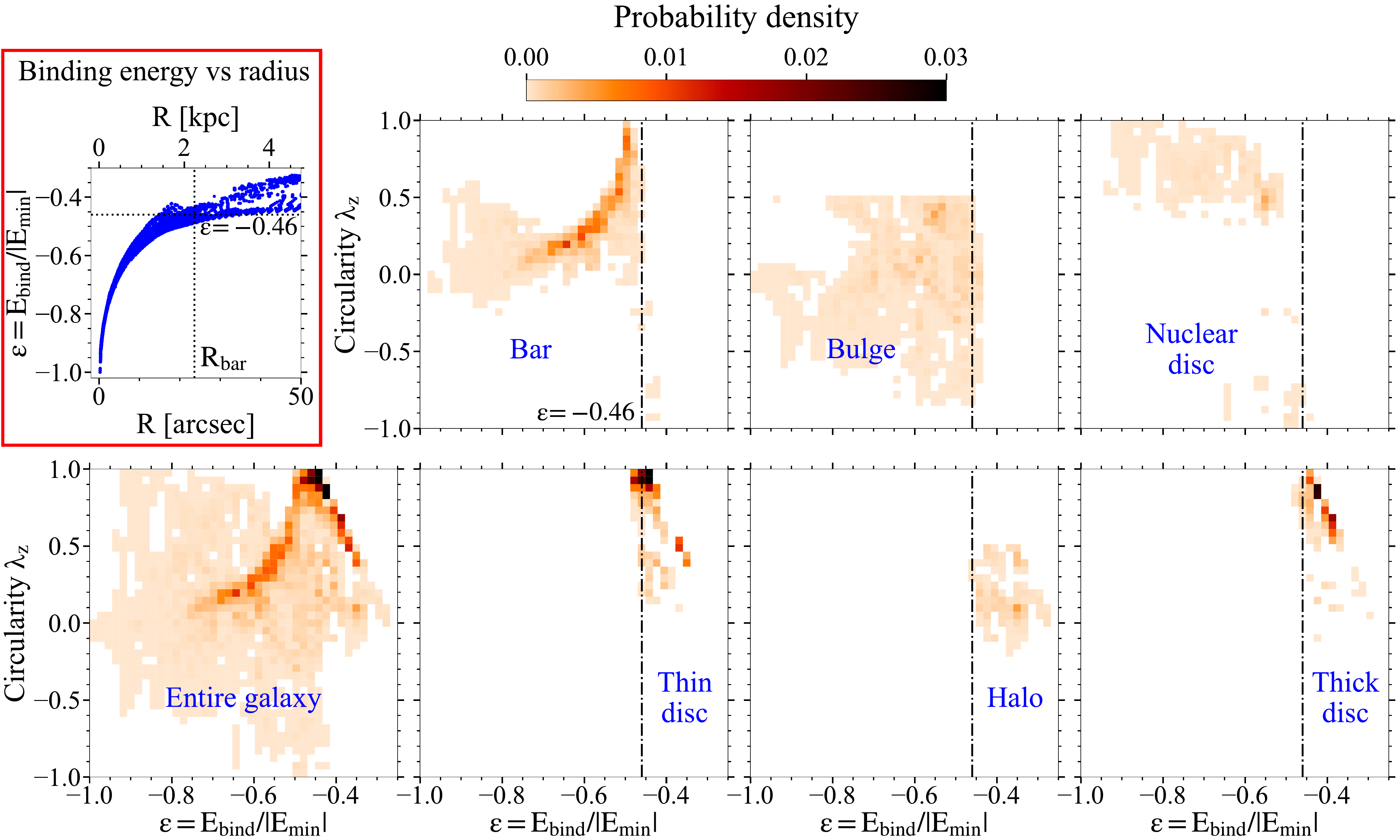}
    \caption{Probability density distributions of stellar orbits in the $\lambda_z$--$E_{\rm bind}$ phase space for the best-fitting model of NGC~1381. \textit{Top left}: Relation between the orbital binding energy ($E_{\rm bind}$, normalised by its minimum value over all orbits in the model) and the time-averaged radius ($R$), with each orbit represented by a blue dot. The vertical dotted line represents the bar length ($R_{\rm bar}$), which is adopted to separate the inner components (nuclear disc, bar, and bulge) from the outer components (thin disc, thick disc, and stellar halo) of the galaxy. The horizontal dotted line indicates the average binding energy ($\varepsilon=E_{\rm bind}/|E_{\rm min}|=-0.46$) at $R_{\rm bar}$. \textit{Other panels}: Stellar orbit distributions of the entire galaxy and of each dynamical component, corresponding to Fig.~\ref{orbit-distributions} but with the $x$-axis changed from radius ($R$) to normalised binding energy ($\varepsilon=E_{\rm bind}/|E_{\rm min}|$). The vertical dashed line represents $\varepsilon=-0.46$.}
    \label{orbit-distributions-Ebind}
\end{figure*}

As mentioned in Sect.~\ref{sec3.2}, we adopted the bar length ($R_{\rm bar}$) to separate the inner components (nuclear disc, bar, and bulge) from the outer components (thin disc, thick disc, and stellar halo) of the galaxy. The reasons for this classification are as follows. First, the bar redistributes angular momentum within $R_{\rm bar}$; for example, there is a lack of dynamically cold orbits at $\sim1.5\,\mathrm{kpc}$ (see Fig.~\ref{orbit-distributions}), which clearly separates the inner (nuclear) disc from the outer (thin and thick) discs. Second, the time-averaged radius is highly correlated with the orbital binding energy. Therefore, for dynamically hot components, using a radius criterion is similar to using a binding-energy criterion. Fig.~\ref{orbit-distributions-Ebind} illustrates this: the separation in normalised binding energy ($\varepsilon=E_{\rm bind}/|E_{\rm min}|=-0.46$) is generally consistent with that based on $R_{\rm bar}$. 

Furthermore, we did not find an obvious bimodal distribution of dynamically hot orbits in the $\lambda_z$--$E_{\rm bind}$ plane, so it is hard to derive a distribution-based classification as \citet{Du2019,Du2020} did. Using a fixed value to separate the different components is therefore a reasonable choice, and the value of $R_{\rm bar}=2.24_{-0.22}^{+0.43}\rm\,kpc$ is generally consistent with the physical scale of dynamical bulges ($\sim2$--$3.5\rm\,kpc$) found in simulations (e.g. \citealp{Du2020,Du2021}) and adopted in previous dynamical models (e.g. \citealp{Zhu2022a,Zhu2022b,Wu2025}). Since the bulge and halo have similar metallicities and [Mg/Fe] abundances, using a separating radius different from $R_{\rm bar}$ will not affect our main conclusion that both the bulge and halo are likely related to early minor mergers.

\section{Stellar orbit distributions in the $\lambda_z$--$R$ phase space for barred and unbarred models}
\label{sec-barred-vs-unbarred}
\begin{figure*}
    \sidecaption
    \includegraphics[width=12cm]{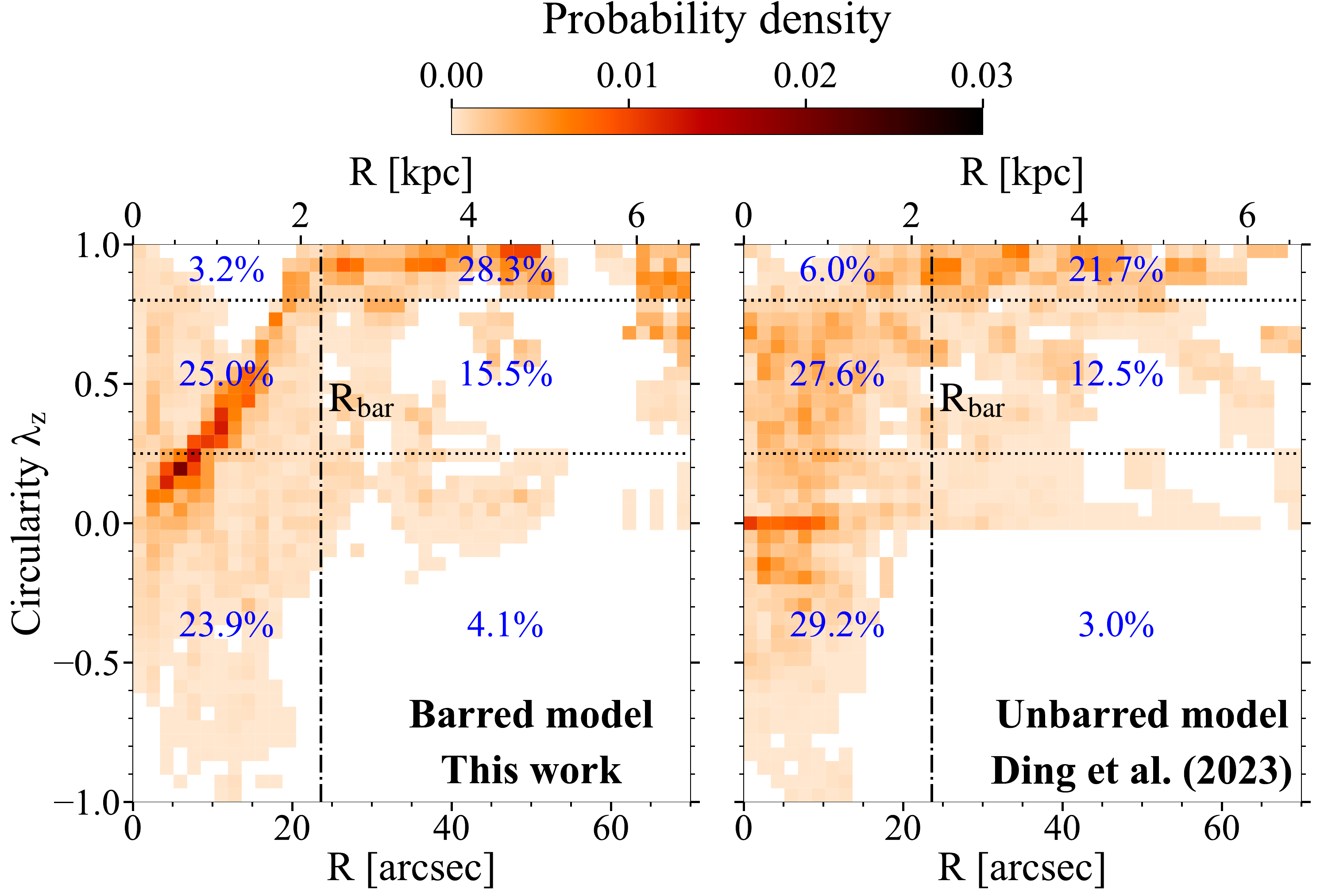}
    \caption{Probability density distributions of stellar orbits in the $\lambda_z$--$R$ phase space for the best-fitting barred (\textit{left}; this work) and unbarred (\textit{right}; \citealp{Ding2023}) models of NGC~1381. The vertical dashed lines represent the dynamical bar length ($R_{\rm bar}=2.24\rm\,kpc$), which separates the orbits in the galaxy's inner ($R\le R_{\rm bar}$) and outer ($R>R_{\rm bar}$) regions. The horizontal dotted lines denote $\lambda_z=0.8$ and $\lambda_z=0.25$, further classifying the orbits into cold ($\lambda_z\ge0.8$), warm ($0.25<\lambda_z<0.8$), and hot ($\lambda_z\le0.25$) components. This results in six distinct components in total, with the luminosity fraction of each component labelled.}
    \label{orbit-distributions-compared-to-Ding2023}
\end{figure*}

Figure~\ref{orbit-distributions-compared-to-Ding2023} shows a comparison between the stellar orbit distributions derived from the barred (this work) and unbarred \citep{Ding2023} models. Since \citet{Ding2023} did not analyse the orbital morphological parameters ($p_{\rm orb}$, $q_{\rm orb,0}$, and $z_{\rm orb}$) adopted in this work, we directly compared different components using only the criteria based on $\lambda_z$ and $R$.

In the galaxy's outer regions ($R>R_{\rm bar}$), the results from the barred and unbarred models are generally consistent: cold orbits ($\lambda_z\ge0.8$) dominate, while warm ($0.25<\lambda_z<0.8$) and hot ($\lambda_z\le0.25$) orbits account for moderate and small fractions, respectively. For each component, the fraction is slightly lower in the unbarred model than in the barred model.

In the galaxy's inner regions ($R\le R_{\rm bar}$), the unbarred model cannot reproduce the dynamical signature of a rigidly rotating bar (the strong $\lambda_z$--$R$ correlation), resulting in a mixture of bar and bulge orbits. Nevertheless, the fractions of cold, warm, and hot orbits remain generally consistent, with the unbarred model yielding slightly higher fractions than the barred model.
 
\section{Spatial distributions of mean velocity and velocity dispersion}
\begin{figure*}
    \centering
    \includegraphics[width=18cm]{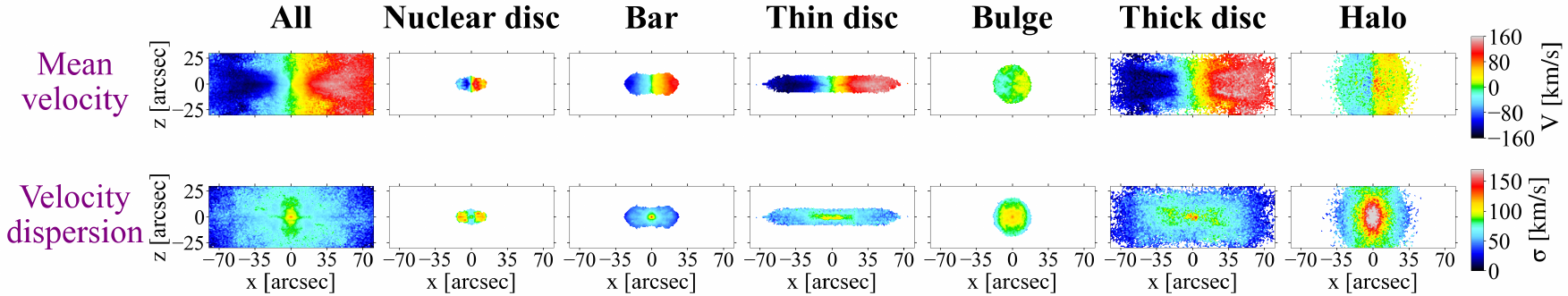}
    \caption{Spatial distributions of luminosity-weighted mean velocities (\textit{top}) and velocity dispersions (\textit{bottom}) on the $x$-$z$ planes for dynamical components in the best-fitting model of NGC~1381. \textit{From left to right}: Entire galaxy, nuclear disc, bar, thin disc, bulge, thick disc, and stellar halo. For each component, only pixels above specified brightness thresholds are plotted.}
    \label{velocity-sigma-distributions}
\end{figure*}

Using the orbital information from the best-fitting model, we derived the 3D spatial distributions of luminosity-weighted mean velocity and velocity dispersion for each decomposed component. We then projected them onto the $x$-$z$ plane and presented the resulting 2D distributions in Fig.~\ref{velocity-sigma-distributions}. The bulge and stellar halo are dominated by random motions, while the other components are dominated by rotation.

\section{Stellar metallicity and [Mg/Fe] abundance profiles}
\begin{figure*}
    \sidecaption
    \includegraphics[width=12cm]{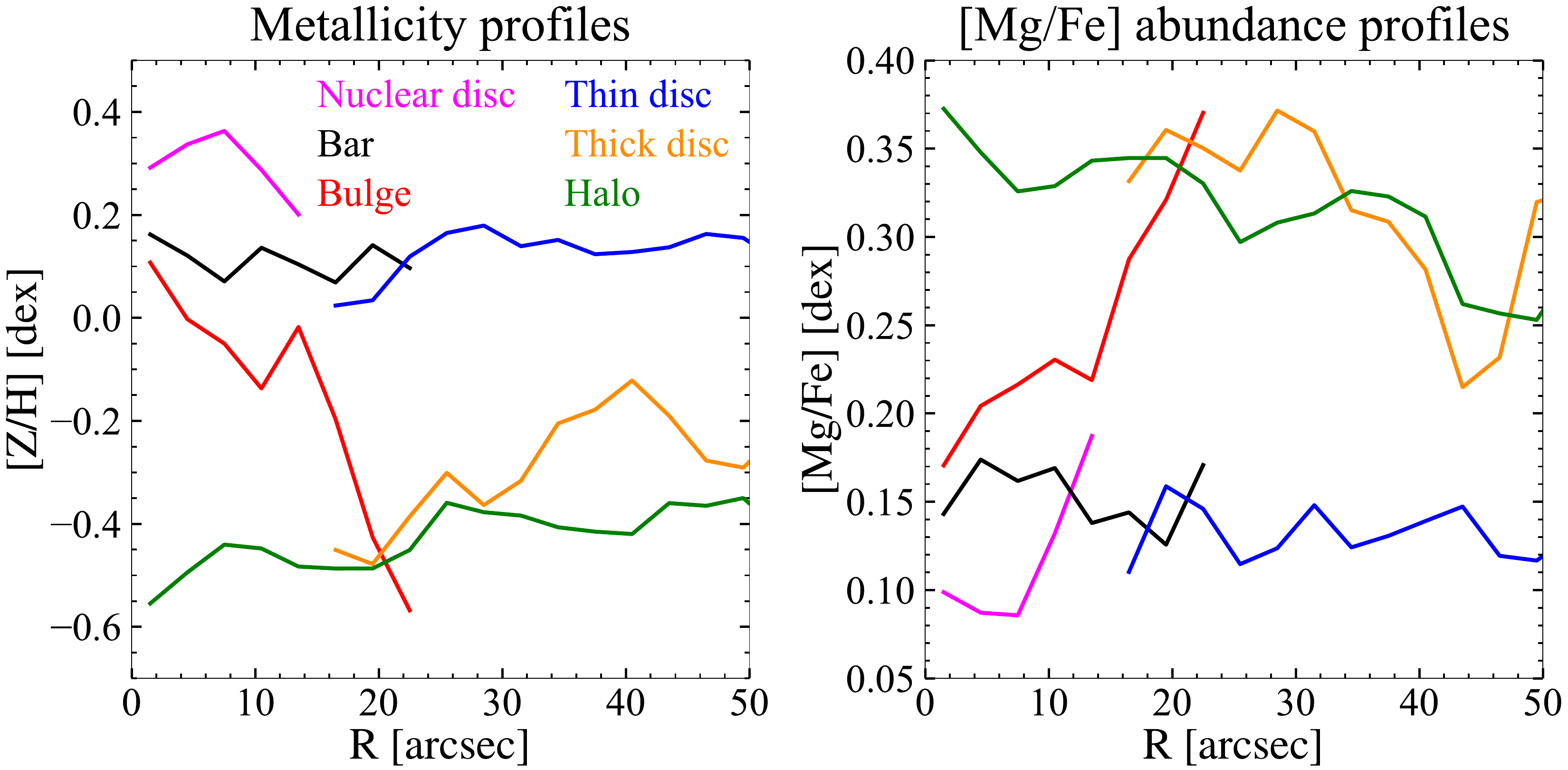}
    \caption{Stellar metallicity (\textit{left}) and [Mg/Fe] abundance (\textit{right}) profiles for the dynamical components in the best-fitting model of NGC~1381. The different colours represent different components: the nuclear disc (magenta), bar (black), bulge (red), thin disc (blue), thick disc (orange), and stellar halo (green).}
    \label{metallicity-alpha-profiles}
\end{figure*}

Using the orbital information from the best-fitting model, we calculated the luminosity-weighted stellar metallicity and [Mg/Fe] abundance profiles and presented them in Fig.~\ref{metallicity-alpha-profiles}. For the best-fitting model, the bulge exhibits a negative metallicity gradient and a positive [Mg/Fe] gradient, while the gradients of other components are not significant.

\end{appendix}

\begin{thebibliography}{142}
\expandafter\ifx\csname natexlab\endcsname\relax\def\natexlab#1{#1}\fi

\bibitem[{Aguerri {et~al.}(2015)Aguerri, M{\'e}ndez-Abreu, Falc{\'o}n-Barroso, Amorin, Barrera-Ballesteros, Cid~Fernandes, Garc{\'\i}a-Benito, Garc{\'\i}a-Lorenzo, Gonz{\'a}lez~Delgado, Husemann, Kalinova, Lyubenova, Marino, M{\'a}rquez, Mast, P{\'e}rez, S{\'a}nchez, van~de Ven, Walcher, Backsmann, Cortijo-Ferrero, Bland-Hawthorn, del Olmo, Iglesias-P{\'a}ramo, P{\'e}rez, S{\'a}nchez-Bl{\'a}zquez, Wisotzki, \& Ziegler}]{Aguerri2015}
Aguerri, J.~A.~L., M{\'e}ndez-Abreu, J., Falc{\'o}n-Barroso, J., {et~al.} 2015, \aap, 576, A102

\bibitem[{Athanassoula(1992{\natexlab{a}})}]{Athanassoula1992a}
Athanassoula, E. 1992{\natexlab{a}}, \mnras, 259, 345

\bibitem[{Athanassoula(1992{\natexlab{b}})}]{Athanassoula1992b}
Athanassoula, E. 1992{\natexlab{b}}, \mnras, 259, 328

\bibitem[{Athanassoula(2003)}]{Athanassoula2003}
Athanassoula, E. 2003, \mnras, 341, 1179

\bibitem[{Athanassoula(2005)}]{Athanassoula2005}
Athanassoula, E. 2005, \mnras, 358, 1477

\bibitem[{Athanassoula(2013)}]{Athanassoula2013b}
Athanassoula, E. 2013, in Secular Evolution of Galaxies, ed. J.~{Falc{\'o}n-Barroso} \& J.~H. {Knapen}, 305

\bibitem[{Bacon {et~al.}(2010)Bacon, Accardo, Adjali, Anwand, Bauer, Biswas, Blaizot, Boudon, Brau-Nogue, Brinchmann, Caillier, Capoani, Carollo, Contini, Couderc, Daguis{\'e}, Deiries, Delabre, Dreizler, Dubois, Dupieux, Dupuy, Emsellem, Fechner, Fleischmann, Fran{\c{c}}ois, Gallou, Gharsa, Glindemann, Gojak, Guiderdoni, Hansali, Hahn, Jarno, Kelz, Koehler, Kosmalski, Laurent, Le~Floch, Lilly, Lizon, Loupias, Manescau, Monstein, Nicklas, Olaya, Pares, Pasquini, P{\'e}contal-Rousset, Pell{\'o}, Petit, Popow, Reiss, Remillieux, Renault, Roth, Rupprecht, Serre, Schaye, Soucail, Steinmetz, Streicher, Stuik, Valentin, Vernet, Weilbacher, Wisotzki, \& Yerle}]{Bacon2010}
Bacon, R., Accardo, M., Adjali, L., {et~al.} 2010, in Society of Photo-Optical Instrumentation Engineers (SPIE) Conference Series, Vol. 7735, Ground-based and Airborne Instrumentation for Astronomy III, ed. I.~S. {McLean}, S.~K. {Ramsay}, \& H.~{Takami}, 773508

\bibitem[{Bacon {et~al.}(2017)Bacon, Conseil, Mary, Brinchmann, Shepherd, Akhlaghi, Weilbacher, Piqueras, Wisotzki, Lagattuta, Epinat, Guerou, Inami, Cantalupo, Courbot, Contini, Richard, Maseda, Bouwens, Bouch{\'e}, Kollatschny, Schaye, Marino, Pello, Herenz, Guiderdoni, \& Carollo}]{Bacon2017}
Bacon, R., Conseil, S., Mary, D., {et~al.} 2017, \aap, 608, A1

\bibitem[{Bacon {et~al.}(2001)Bacon, Copin, Monnet, Miller, Allington-Smith, Bureau, Carollo, Davies, Emsellem, Kuntschner, Peletier, Verolme, \& de~Zeeuw}]{Bacon2001}
Bacon, R., Copin, Y., Monnet, G., {et~al.} 2001, \mnras, 326, 23

\bibitem[{Barnab{\`e} {et~al.}(2012)Barnab{\`e}, Dutton, Marshall, Auger, Brewer, Treu, Bolton, Koo, \& Koopmans}]{Barnabe2012}
Barnab{\`e}, M., Dutton, A.~A., Marshall, P.~J., {et~al.} 2012, \mnras, 423, 1073

\bibitem[{Binney \& Tremaine(2008)}]{Binney2008}
Binney, J. \& Tremaine, S. 2008, Galactic Dynamics: Second Edition

\bibitem[{Blakeslee {et~al.}(2009)Blakeslee, Jord{\'a}n, Mei, C{\^o}t{\'e}, Ferrarese, Infante, Peng, Tonry, \& West}]{Blakeslee2009}
Blakeslee, J.~P., Jord{\'a}n, A., Mei, S., {et~al.} 2009, \apj, 694, 556

\bibitem[{Bournaud {et~al.}(2007)Bournaud, Elmegreen, \& Elmegreen}]{Bournaud2007}
Bournaud, F., Elmegreen, B.~G., \& Elmegreen, D.~M. 2007, \apj, 670, 237

\bibitem[{Breda {et~al.}(2020)Breda, Papaderos, \& Gomes}]{Breda2020}
Breda, I., Papaderos, P., \& Gomes, J.-M. 2020, \aap, 640, A20

\bibitem[{Bryant {et~al.}(2015)Bryant, Owers, Robotham, Croom, Driver, Drinkwater, Lorente, Cortese, Scott, Colless, Schaefer, Taylor, Konstantopoulos, Allen, Baldry, Barnes, Bauer, Bland-Hawthorn, Bloom, Brooks, Brough, Cecil, Couch, Croton, Davies, Ellis, Fogarty, Foster, Glazebrook, Goodwin, Green, Gunawardhana, Hampton, Ho, Hopkins, Kewley, Lawrence, Leon-Saval, Leslie, McElroy, Lewis, Liske, L{\'o}pez-S{\'a}nchez, Mahajan, Medling, Metcalfe, Meyer, Mould, Obreschkow, O'Toole, Pracy, Richards, Shanks, Sharp, Sweet, Thomas, Tonini, \& Walcher}]{Bryant2015}
Bryant, J.~J., Owers, M.~S., Robotham, A. S.~G., {et~al.} 2015, \mnras, 447, 2857

\bibitem[{Bundy {et~al.}(2015)Bundy, Bershady, Law, Yan, Drory, MacDonald, Wake, Cherinka, S{\'a}nchez-Gallego, Weijmans, Thomas, Tremonti, Masters, Coccato, Diamond-Stanic, Arag{\'o}n-Salamanca, Avila-Reese, Badenes, Falc{\'o}n-Barroso, Belfiore, Bizyaev, Blanc, Bland-Hawthorn, Blanton, Brownstein, Byler, Cappellari, Conroy, Dutton, Emsellem, Etherington, Frinchaboy, Fu, Gunn, Harding, Johnston, Kauffmann, Kinemuchi, Klaene, Knapen, Leauthaud, Li, Lin, Maiolino, Malanushenko, Malanushenko, Mao, Maraston, McDermid, Merrifield, Nichol, Oravetz, Pan, Parejko, Sanchez, Schlegel, Simmons, Steele, Steinmetz, Thanjavur, Thompson, Tinker, van~den Bosch, Westfall, Wilkinson, Wright, Xiao, \& Zhang}]{Bundy2015}
Bundy, K., Bershady, M.~A., Law, D.~R., {et~al.} 2015, \apj, 798, 7

\bibitem[{Bureau {et~al.}(2006)Bureau, Aronica, Athanassoula, Dettmar, Bosma, \& Freeman}]{Bureau2006}
Bureau, M., Aronica, G., Athanassoula, E., {et~al.} 2006, \mnras, 370, 753

\bibitem[{Bureau \& Athanassoula(2005)}]{Bureau2005}
Bureau, M. \& Athanassoula, E. 2005, \apj, 626, 159

\bibitem[{Buta {et~al.}(2015)Buta, Sheth, Athanassoula, Bosma, Knapen, Laurikainen, Salo, Elmegreen, Ho, Zaritsky, Courtois, Hinz, Mu{\~n}oz-Mateos, Kim, Regan, Gadotti, Gil~de Paz, Laine, Men{\'e}ndez-Delmestre, Comer{\'o}n, Erroz~Ferrer, Seibert, Mizusawa, Holwerda, \& Madore}]{Buta2015}
Buta, R.~J., Sheth, K., Athanassoula, E., {et~al.} 2015, \apjs, 217, 32

\bibitem[{Buta \& Zhang(2009)}]{Buta2009}
Buta, R.~J. \& Zhang, X. 2009, \apjs, 182, 559

\bibitem[{Buttitta {et~al.}(2022)Buttitta, Corsini, Cuomo, Aguerri, Coccato, Costantin, Dalla~Bont{\`a}, Debattista, Iodice, M{\'e}ndez-Abreu, Morelli, \& Pizzella}]{Buttitta2022}
Buttitta, C., Corsini, E.~M., Cuomo, V., {et~al.} 2022, \aap, 664, L10

\bibitem[{Cappellari(2002)}]{Cappellari2002}
Cappellari, M. 2002, \mnras, 333, 400

\bibitem[{Cappellari(2017)}]{Cappellari2017}
Cappellari, M. 2017, \mnras, 466, 798

\bibitem[{Cappellari \& Copin(2003)}]{Cappellari2003}
Cappellari, M. \& Copin, Y. 2003, \mnras, 342, 345

\bibitem[{Cappellari \& Emsellem(2004)}]{Cappellari2004}
Cappellari, M. \& Emsellem, E. 2004, \pasp, 116, 138

\bibitem[{Cappellari {et~al.}(2011)Cappellari, Emsellem, Krajnovi{\'c}, McDermid, Scott, Verdoes~Kleijn, Young, Alatalo, Bacon, Blitz, Bois, Bournaud, Bureau, Davies, Davis, de~Zeeuw, Duc, Khochfar, Kuntschner, Lablanche, Morganti, Naab, Oosterloo, Sarzi, Serra, \& Weijmans}]{Cappellari2011}
Cappellari, M., Emsellem, E., Krajnovi{\'c}, D., {et~al.} 2011, \mnras, 413, 813

\bibitem[{Cappellari {et~al.}(2013)Cappellari, Scott, Alatalo, Blitz, Bois, Bournaud, Bureau, Crocker, Davies, Davis, de~Zeeuw, Duc, Emsellem, Khochfar, Krajnovi{\'c}, Kuntschner, McDermid, Morganti, Naab, Oosterloo, Sarzi, Serra, Weijmans, \& Young}]{Cappellari2013}
Cappellari, M., Scott, N., Alatalo, K., {et~al.} 2013, \mnras, 432, 1709

\bibitem[{Chemin \& Hernandez(2009)}]{Chemin2009}
Chemin, L. \& Hernandez, O. 2009, \aap, 499, L25

\bibitem[{Chung \& Bureau(2004)}]{Chung2004}
Chung, A. \& Bureau, M. 2004, \aj, 127, 3192

\bibitem[{Contopoulos \& Mertzanides(1977)}]{Contopoulos1977}
Contopoulos, G. \& Mertzanides, C. 1977, \aap, 61, 477

\bibitem[{Cristiani {et~al.}(2024)Cristiani, Abadi, Taverna, Cabral, Benelli, \& S{\'a}nchez}]{Cristiani2024}
Cristiani, V.~A., Abadi, M.~G., Taverna, A., {et~al.} 2024, \aap, 692, A63

\bibitem[{Cuomo {et~al.}(2020)Cuomo, Aguerri, Corsini, \& Debattista}]{Cuomo2020}
Cuomo, V., Aguerri, J.~A.~L., Corsini, E.~M., \& Debattista, V.~P. 2020, \aap, 641, A111

\bibitem[{Dattathri {et~al.}(2024)Dattathri, Valluri, Vasiliev, Wheeler, \& Erwin}]{Dattathri2024}
Dattathri, S., Valluri, M., Vasiliev, E., Wheeler, V., \& Erwin, P. 2024, \mnras, 530, 1195

\bibitem[{de~S{\'a}-Freitas {et~al.}(2023)de~S{\'a}-Freitas, Fragkoudi, Gadotti, Falc{\'o}n-Barroso, Bittner, S{\'a}nchez-Bl{\'a}zquez, van~de Ven, Bieri, Coccato, Coelho, Fahrion, Gon{\c{c}}alves, Kim, de~Lorenzo-C{\'a}ceres, Martig, Mart{\'\i}n-Navarro, Mendez-Abreu, Neumann, \& Querejeta}]{deSaFreitas2023}
de~S{\'a}-Freitas, C., Fragkoudi, F., Gadotti, D.~A., {et~al.} 2023, \aap, 671, A8

\bibitem[{de~S{\'a}-Freitas {et~al.}(2025)de~S{\'a}-Freitas, Gadotti, Fragkoudi, Coelho, de~Lorenzo-C{\'a}ceres, Falc{\'o}n-Barroso, S{\'a}nchez-Bl{\'a}zquez, Kim, Mendez-Abreu, Neumann, Querejeta, \& van~de Ven}]{deSaFreitas2025}
de~S{\'a}-Freitas, C., Gadotti, D.~A., Fragkoudi, F., {et~al.} 2025, \aap, 698, A5

\bibitem[{Debattista {et~al.}(2004)Debattista, Carollo, Mayer, \& Moore}]{Debattista2004}
Debattista, V.~P., Carollo, C.~M., Mayer, L., \& Moore, B. 2004, \apjl, 604, L93

\bibitem[{Debattista \& Sellwood(2000)}]{Debattista2000}
Debattista, V.~P. \& Sellwood, J.~A. 2000, \apj, 543, 704

\bibitem[{Ding {et~al.}(2023)Ding, Zhu, van~de Ven, Coccato, Corsini, Costantin, Fahrion, Falc{\'o}n-Barroso, Gadotti, Iodice, Lyubenova, Mart{\'\i}n-Navarro, McDermid, Pinna, \& Sarzi}]{Ding2023}
Ding, Y., Zhu, L., van~de Ven, G., {et~al.} 2023, \aap, 672, A84

\bibitem[{Drinkwater {et~al.}(2001)Drinkwater, Gregg, \& Colless}]{Drinkwater2001}
Drinkwater, M.~J., Gregg, M.~D., \& Colless, M. 2001, \apjl, 548, L139

\bibitem[{Du {et~al.}(2021)Du, Ho, Debattista, Pillepich, Nelson, Hernquist, \& Weinberger}]{Du2021}
Du, M., Ho, L.~C., Debattista, V.~P., {et~al.} 2021, \apj, 919, 135

\bibitem[{Du {et~al.}(2020)Du, Ho, Debattista, Pillepich, Nelson, Zhao, \& Hernquist}]{Du2020}
Du, M., Ho, L.~C., Debattista, V.~P., {et~al.} 2020, \apj, 895, 139

\bibitem[{Du {et~al.}(2019)Du, Ho, Zhao, Shi, Debattista, Hernquist, \& Nelson}]{Du2019}
Du, M., Ho, L.~C., Zhao, D., {et~al.} 2019, \apj, 884, 129

\bibitem[{Efstathiou {et~al.}(1982)Efstathiou, Lake, \& Negroponte}]{Efstathiou1982}
Efstathiou, G., Lake, G., \& Negroponte, J. 1982, \mnras, 199, 1069

\bibitem[{Emsellem {et~al.}(1994)Emsellem, Monnet, \& Bacon}]{Emsellem1994}
Emsellem, E., Monnet, G., \& Bacon, R. 1994, \aap, 285, 723

\bibitem[{Erwin \& Debattista(2017)}]{Erwin2017}
Erwin, P. \& Debattista, V.~P. 2017, \mnras, 468, 2058

\bibitem[{Erwin {et~al.}(2021)Erwin, Seth, Debattista, Seidel, Mehrgan, Thomas, Saglia, de~Lorenzo-C{\'a}ceres, Maciejewski, Fabricius, M{\'e}ndez-Abreu, Hopp, Kluge, Beckman, Bender, Drory, \& Fisher}]{Erwin2021}
Erwin, P., Seth, A., Debattista, V.~P., {et~al.} 2021, \mnras, 502, 2446

\bibitem[{Ferguson(1989)}]{Ferguson1989}
Ferguson, H.~C. 1989, \aj, 98, 367

\bibitem[{Fisher \& Drory(2008)}]{Fisher2008}
Fisher, D.~B. \& Drory, N. 2008, \aj, 136, 773

\bibitem[{Fraser-McKelvie {et~al.}(2025)Fraser-McKelvie, van~de Sande, Gadotti, Emsellem, Brown, Fisher, Martig, Bureau, Gerhard, Battisti, Bland-Hawthorn, Boecker, Catinella, Combes, Cortese, Croom, Davis, Falc{\'o}n-Barroso, Fragkoudi, Freeman, Hayden, McDermid, Mazzilli~Ciraulo, Mendel, Pinna, Poci, Rutherford, de~S{\'a}-Freitas, Silva-Lima, Valenzuela, van~de Ven, Wang, \& Watts}]{FraserMcKelvie2025}
Fraser-McKelvie, A., van~de Sande, J., Gadotti, D.~A., {et~al.} 2025, \aap, 700, A237

\bibitem[{Gadotti(2009)}]{Gadotti2009}
Gadotti, D.~A. 2009, \mnras, 393, 1531

\bibitem[{Gadotti(2011)}]{Gadotti2011}
Gadotti, D.~A. 2011, \mnras, 415, 3308

\bibitem[{Gadotti(2026)}]{Gadotti2026}
Gadotti, D.~A. 2026, \mnras, 545, staf2072

\bibitem[{Gadotti {et~al.}(2020)Gadotti, Bittner, Falc{\'o}n-Barroso, M{\'e}ndez-Abreu, Kim, Fragkoudi, de~Lorenzo-C{\'a}ceres, Leaman, Neumann, Querejeta, S{\'a}nchez-Bl{\'a}zquez, Martig, Mart{\'\i}n-Navarro, P{\'e}rez, Seidel, \& van~de Ven}]{Gadotti2020}
Gadotti, D.~A., Bittner, A., Falc{\'o}n-Barroso, J., {et~al.} 2020, \aap, 643, A14

\bibitem[{Garma-Oehmichen {et~al.}(2022)Garma-Oehmichen, Hern{\'a}ndez-Toledo, Aquino-Ort{\'\i}z, Martinez-Medina, Puerari, Cano-D{\'\i}az, Valenzuela, V{\'a}zquez-Mata, G{\'e}ron, Mart{\'\i}nez-V{\'a}zquez, \& Lane}]{GarmaOehmichen2022}
Garma-Oehmichen, L., Hern{\'a}ndez-Toledo, H., Aquino-Ort{\'\i}z, E., {et~al.} 2022, \mnras, 517, 5660

\bibitem[{Gerhard(1993)}]{Gerhard1993}
Gerhard, O.~E. 1993, \mnras, 265, 213

\bibitem[{Grand {et~al.}(2024)Grand, Fragkoudi, G{\'o}mez, Jenkins, Marinacci, Pakmor, \& Springel}]{Grand2024}
Grand, R. J.~J., Fragkoudi, F., G{\'o}mez, F.~A., {et~al.} 2024, \mnras, 532, 1814

\bibitem[{Grand {et~al.}(2017)Grand, G{\'o}mez, Marinacci, Pakmor, Springel, Campbell, Frenk, Jenkins, \& White}]{Grand2017}
Grand, R. J.~J., G{\'o}mez, F.~A., Marinacci, F., {et~al.} 2017, \mnras, 467, 179

\bibitem[{Guo {et~al.}(2019)Guo, Mao, Athanassoula, Li, Ge, Long, Merrifield, \& Masters}]{Guo2019}
Guo, R., Mao, S., Athanassoula, E., {et~al.} 2019, \mnras, 482, 1733

\bibitem[{Guo {et~al.}(2025)Guo, Jogee, Wise, Pritchett, McGrath, Finkelstein, Iyer, Arrabal~Haro, Bagley, Dickinson, Kartaltepe, Koekemoer, Papovich, Pirzkal, Yung, Backhaus, Bell, Bhatawdekar, Cheng, Costantin, de~la Vega, Giavalisco, Hathi, Holwerda, Kurczynski, Lucas, Mobasher, P{\'e}rez-Gonz{\'a}lez, \& Pacucci}]{Guo2025}
Guo, Y., Jogee, S., Wise, E., {et~al.} 2025, \apj, 985, 181

\bibitem[{Habibi {et~al.}(2024)Habibi, Roshan, Hosseinirad, Khosroshahi, Aguerri, Cuomo, \& Abbassi}]{Habibi2024}
Habibi, A., Roshan, M., Hosseinirad, M., {et~al.} 2024, \aap, 691, A122

\bibitem[{Helmi {et~al.}(2018)Helmi, Babusiaux, Koppelman, Massari, Veljanoski, \& Brown}]{Helmi2018}
Helmi, A., Babusiaux, C., Koppelman, H.~H., {et~al.} 2018, \nat, 563, 85

\bibitem[{Hohl(1971)}]{Hohl1971}
Hohl, F. 1971, \apj, 168, 343

\bibitem[{Hopkins {et~al.}(2010)Hopkins, Bundy, Croton, Hernquist, Keres, Khochfar, Stewart, Wetzel, \& Younger}]{Hopkins2010}
Hopkins, P.~F., Bundy, K., Croton, D., {et~al.} 2010, \apj, 715, 202

\bibitem[{Iodice {et~al.}(2016)Iodice, Capaccioli, Grado, Limatola, Spavone, Napolitano, Paolillo, Peletier, Cantiello, Lisker, Wittmann, Venhola, Hilker, D'Abrusco, Pota, \& Schipani}]{Iodice2016}
Iodice, E., Capaccioli, M., Grado, A., {et~al.} 2016, \apj, 820, 42

\bibitem[{Iodice {et~al.}(2019b)Iodice, Sarzi, Bittner, Coccato, Costantin, Corsini, van~de Ven, de~Zeeuw, Falc{\'o}n-Barroso, Gadotti, Lyubenova, Mart{\'\i}n-Navarro, McDermid, Nedelchev, Pinna, Pizzella, Spavone, \& Viaene}]{Iodice2019b}
Iodice, E., Sarzi, M., Bittner, A., {et~al.} 2019b, \aap, 627, A136

\bibitem[{Iodice {et~al.}(2019a)Iodice, Spavone, Capaccioli, Peletier, van~de Ven, Napolitano, Hilker, Mieske, Smith, Pasquali, Limatola, Grado, Venhola, Cantiello, Paolillo, Falcon-Barroso, D'Abrusco, \& Schipani}]{Iodice2019a}
Iodice, E., Spavone, M., Capaccioli, M., {et~al.} 2019a, \aap, 623, A1

\bibitem[{Jethwa {et~al.}(2020)Jethwa, Thater, Maindl, \& Van~de Ven}]{Jethwa2020}
Jethwa, P., Thater, S., Maindl, T., \& Van~de Ven, G. 2020, DYNAMITE: DYnamics, Age and Metallicity Indicators Tracing Evolution, Astrophysics Source Code Library, record ascl:2011.007

\bibitem[{Jin {et~al.}(2020)Jin, Zhu, Long, Mao, Wang, \& van~de Ven}]{Jin2020}
Jin, Y., Zhu, L., Long, R.~J., {et~al.} 2020, \mnras, 491, 1690

\bibitem[{Jin {et~al.}(2019)Jin, Zhu, Long, Mao, Xu, Li, \& van~de Ven}]{Jin2019}
Jin, Y., Zhu, L., Long, R.~J., {et~al.} 2019, \mnras, 486, 4753

\bibitem[{Jin {et~al.}(2025b)Jin, Zhu, Tahmasebzadeh, Mao, van~de Ven, \& Davis}]{Jin2025b}
Jin, Y., Zhu, L., Tahmasebzadeh, B., {et~al.} 2025b, \aap, 704, A262

\bibitem[{Jin {et~al.}(2025a)Jin, Zhu, Tahmasebzadeh, Mao, van~de Ven, Guo, \& Cai}]{Jin2025a}
Jin, Y., Zhu, L., Tahmasebzadeh, B., {et~al.} 2025a, \aap, 700, A249

\bibitem[{Jin {et~al.}(2024)Jin, Zhu, Zibetti, Costantin, van~de Ven, \& Mao}]{Jin2024}
Jin, Y., Zhu, L., Zibetti, S., {et~al.} 2024, \aap, 681, A95

\bibitem[{Kormendy \& Kennicutt(2004)}]{Kormendy2004}
Kormendy, J. \& Kennicutt, Robert~C., J. 2004, \araa, 42, 603

\bibitem[{Kroupa(2001)}]{Kroupa2001}
Kroupa, P. 2001, \mnras, 322, 231

\bibitem[{Lawson \& Hanson(1974)}]{Lawson1974}
Lawson, C.~L. \& Hanson, R.~J. 1974, Solving least squares problems

\bibitem[{Le~Conte {et~al.}(2024)Le~Conte, Gadotti, Ferreira, Conselice, de~S{\'a}-Freitas, Kim, Neumann, Fragkoudi, Athanassoula, \& Adams}]{LeConte2024}
Le~Conte, Z.~A., Gadotti, D.~A., Ferreira, L., {et~al.} 2024, \mnras, 530, 1984

\bibitem[{Lee {et~al.}(2025)Lee, Hwang, Cuomo, Park, Kim, Hwang, Ann, Kim, Kim, Seok, Lee, \& Choi}]{Lee2025}
Lee, Y.~H., Hwang, H.~S., Cuomo, V., {et~al.} 2025, \apj, 989, 55

\bibitem[{Lehnert {et~al.}(2014)Lehnert, Di~Matteo, Haywood, \& Snaith}]{Lehnert2014}
Lehnert, M.~D., Di~Matteo, P., Haywood, M., \& Snaith, O.~N. 2014, \apjl, 789, L30

\bibitem[{Li {et~al.}(2017)Li, Ho, \& Barth}]{Li2017}
Li, Z.-Y., Ho, L.~C., \& Barth, A.~J. 2017, \apj, 845, 87

\bibitem[{Li {et~al.}(2018)Li, Shen, Bureau, Zhou, Du, \& Debattista}]{LiZhaoyu2018}
Li, Z.-Y., Shen, J., Bureau, M., {et~al.} 2018, \apj, 854, 65

\bibitem[{{\L}okas(2018)}]{Lokas2018}
{\L}okas, E.~L. 2018, \apj, 857, 6

\bibitem[{Long \& Mao(2018)}]{Long2018}
Long, R.~J. \& Mao, S. 2018, Research in Astronomy and Astrophysics, 18, 145

\bibitem[{L{\"u}tticke {et~al.}(2000)L{\"u}tticke, Dettmar, \& Pohlen}]{Luetticke2000}
L{\"u}tticke, R., Dettmar, R.~J., \& Pohlen, M. 2000, \aaps, 145, 405

\bibitem[{Martig {et~al.}(2021)Martig, Pinna, Falc{\'o}n-Barroso, Gadotti, Husemann, Minchev, Neumann, Ruiz-Lara, \& van~de Ven}]{Martig2021}
Martig, M., Pinna, F., Falc{\'o}n-Barroso, J., {et~al.} 2021, \mnras, 508, 2458

\bibitem[{Martig {et~al.}(2026)Martig, Pinna, Falc{\'o}n-Barroso, Mart{\'\i}n-Navarro, Minchev, \& Ding}]{Martig2026}
Martig, M., Pinna, F., Falc{\'o}n-Barroso, J., {et~al.} 2026, \mnras, 547, stag458

\bibitem[{Mart{\'\i}n-Navarro {et~al.}(2021)Mart{\'\i}n-Navarro, Pinna, Coccato, Falc{\'o}n-Barroso, van~de Ven, Lyubenova, Corsini, Fahrion, Gadotti, Iodice, McDermid, Poci, Sarzi, Spriggs, Viaene, de~Zeeuw, \& Zhu}]{MartinNavarro2021}
Mart{\'\i}n-Navarro, I., Pinna, F., Coccato, L., {et~al.} 2021, \aap, 654, A59

\bibitem[{M{\'e}ndez-Abreu {et~al.}(2017)M{\'e}ndez-Abreu, Ruiz-Lara, S{\'a}nchez-Menguiano, de~Lorenzo-C{\'a}ceres, Costantin, Catal{\'a}n-Torrecilla, Florido, Aguerri, Bland-Hawthorn, Corsini, Dettmar, Galbany, Garc{\'\i}a-Benito, Marino, M{\'a}rquez, Ortega-Minakata, Papaderos, S{\'a}nchez, S{\'a}nchez-Blazquez, Spekkens, van~de Ven, Wild, \& Ziegler}]{MendezAbreu2017}
M{\'e}ndez-Abreu, J., Ruiz-Lara, T., S{\'a}nchez-Menguiano, L., {et~al.} 2017, \aap, 598, A32

\bibitem[{Molaeinezhad {et~al.}(2016)Molaeinezhad, Falc{\'o}n-Barroso, Mart{\'\i}nez-Valpuesta, Khosroshahi, Balcells, \& Peletier}]{Molaeinezhad2016}
Molaeinezhad, A., Falc{\'o}n-Barroso, J., Mart{\'\i}nez-Valpuesta, I., {et~al.} 2016, \mnras, 456, 692

\bibitem[{Morokuma-Matsui {et~al.}(2022)Morokuma-Matsui, Bekki, Wang, Serra, Koyama, Morokuma, Egusa, For, Nakanishi, Koribalski, Okamoto, Kodama, Lee, Maccagni, Miura, Espada, Takeuchi, Yang, Lee, Ueda, \& Matsushita}]{MorokumaMatsui2022}
Morokuma-Matsui, K., Bekki, K., Wang, J., {et~al.} 2022, \apjs, 263, 40

\bibitem[{Navarro {et~al.}(1996)Navarro, Frenk, \& White}]{Navarro1996}
Navarro, J.~F., Frenk, C.~S., \& White, S. D.~M. 1996, \apj, 462, 563

\bibitem[{Neumayer {et~al.}(2020)Neumayer, Seth, \& B{\"o}ker}]{Neumayer2020}
Neumayer, N., Seth, A., \& B{\"o}ker, T. 2020, \aapr, 28, 4

\bibitem[{Neureiter {et~al.}(2021)Neureiter, Thomas, Saglia, Bender, Finozzi, Krukau, Naab, Rantala, \& Frigo}]{Neureiter2021}
Neureiter, B., Thomas, J., Saglia, R., {et~al.} 2021, \mnras, 500, 1437

\bibitem[{Noguchi(1996)}]{Noguchi1996}
Noguchi, M. 1996, \apj, 469, 605

\bibitem[{Ostriker \& Peebles(1973)}]{Ostriker1973}
Ostriker, J.~P. \& Peebles, P.~J.~E. 1973, \apj, 186, 467

\bibitem[{Pietrinferni {et~al.}(2004)Pietrinferni, Cassisi, Salaris, \& Castelli}]{Pietrinferni2004}
Pietrinferni, A., Cassisi, S., Salaris, M., \& Castelli, F. 2004, \apj, 612, 168

\bibitem[{Pietrinferni {et~al.}(2006)Pietrinferni, Cassisi, Salaris, \& Castelli}]{Pietrinferni2006}
Pietrinferni, A., Cassisi, S., Salaris, M., \& Castelli, F. 2006, \apj, 642, 797

\bibitem[{Pinna {et~al.}(2019b)Pinna, Falc{\'o}n-Barroso, Martig, Coccato, Corsini, de~Zeeuw, Gadotti, Iodice, Leaman, Lyubenova, Mart{\'\i}n-Navarro, Morelli, Sarzi, van~de Ven, Viaene, \& McDermid}]{Pinna2019b}
Pinna, F., Falc{\'o}n-Barroso, J., Martig, M., {et~al.} 2019b, \aap, 625, A95

\bibitem[{Pinna {et~al.}(2019a)Pinna, Falc{\'o}n-Barroso, Martig, Sarzi, Coccato, Iodice, Corsini, de~Zeeuw, Gadotti, Leaman, Lyubenova, McDermid, Minchev, Morelli, van~de Ven, \& Viaene}]{Pinna2019a}
Pinna, F., Falc{\'o}n-Barroso, J., Martig, M., {et~al.} 2019a, \aap, 623, A19

\bibitem[{Poci {et~al.}(2021)Poci, McDermid, Lyubenova, Zhu, van~de Ven, Iodice, Coccato, Pinna, Corsini, Falc{\'o}n-Barroso, Gadotti, Grand, Fahrion, Mart{\'\i}n-Navarro, Sarzi, Viaene, \& de~Zeeuw}]{Poci2021}
Poci, A., McDermid, R.~M., Lyubenova, M., {et~al.} 2021, \aap, 647, A145

\bibitem[{Poci {et~al.}(2019)Poci, McDermid, Zhu, \& van~de Ven}]{Poci2019}
Poci, A., McDermid, R.~M., Zhu, L., \& van~de Ven, G. 2019, \mnras, 487, 3776

\bibitem[{Pulsoni {et~al.}(2020)Pulsoni, Gerhard, Arnaboldi, Pillepich, Nelson, Hernquist, \& Springel}]{Pulsoni2020}
Pulsoni, C., Gerhard, O., Arnaboldi, M., {et~al.} 2020, \aap, 641, A60

\bibitem[{Quenneville {et~al.}(2022)Quenneville, Liepold, \& Ma}]{Quenneville2022}
Quenneville, M.~E., Liepold, C.~M., \& Ma, C.-P. 2022, \apj, 926, 30

\bibitem[{Rautiainen {et~al.}(2008)Rautiainen, Salo, \& Laurikainen}]{Rautiainen2008}
Rautiainen, P., Salo, H., \& Laurikainen, E. 2008, \mnras, 388, 1803

\bibitem[{Rix {et~al.}(1997)Rix, de~Zeeuw, Cretton, van~der Marel, \& Carollo}]{Rix1997}
Rix, H.-W., de~Zeeuw, P.~T., Cretton, N., van~der Marel, R.~P., \& Carollo, C.~M. 1997, \apj, 488, 702

\bibitem[{Ruiz-Garc{\'\i}a {et~al.}(2024)Ruiz-Garc{\'\i}a, Querejeta, Garc{\'\i}a-Burillo, Emsellem, Meidt, Sormani, Schinnerer, Williams, Bazzi, Colombo, Gleis, Gnedin, Klessen, Leroy, S{\'a}nchez-Bl{\'a}zquez, \& Stuber}]{RuizGarcia2024}
Ruiz-Garc{\'\i}a, M., Querejeta, M., Garc{\'\i}a-Burillo, S., {et~al.} 2024, \aap, 691, A351

\bibitem[{S{\'a}nchez {et~al.}(2012)S{\'a}nchez, Kennicutt, Gil~de Paz, van~de Ven, V{\'\i}lchez, Wisotzki, Walcher, Mast, Aguerri, Albiol-P{\'e}rez, Alonso-Herrero, Alves, Bakos, Bart{\'a}kov{\'a}, Bland-Hawthorn, Boselli, Bomans, Castillo-Morales, Cortijo-Ferrero, de~Lorenzo-C{\'a}ceres, Del~Olmo, Dettmar, D{\'\i}az, Ellis, Falc{\'o}n-Barroso, Flores, Gallazzi, Garc{\'\i}a-Lorenzo, Gonz{\'a}lez~Delgado, Gruel, Haines, Hao, Husemann, Igl{\'e}sias-P{\'a}ramo, Jahnke, Johnson, Jungwiert, Kalinova, Kehrig, Kupko, L{\'o}pez-S{\'a}nchez, Lyubenova, Marino, M{\'a}rmol-Queralt{\'o}, M{\'a}rquez, Masegosa, Meidt, Mendez-Abreu, Monreal-Ibero, Montijo, Mour{\~a}o, Palacios-Navarro, Papaderos, Pasquali, Peletier, P{\'e}rez, P{\'e}rez, Quirrenbach, Rela{\~n}o, Rosales-Ortega, Roth, Ruiz-Lara, S{\'a}nchez-Bl{\'a}zquez, Sengupta, Singh, Stanishev, Trager, Vazdekis, Viironen, Wild, Zibetti, \& Ziegler}]{Sanchez2012}
S{\'a}nchez, S.~F., Kennicutt, R.~C., Gil~de Paz, A., {et~al.} 2012, \aap, 538, A8

\bibitem[{S{\'a}nchez-Bl{\'a}zquez {et~al.}(2006)S{\'a}nchez-Bl{\'a}zquez, Peletier, Jim{\'e}nez-Vicente, Cardiel, Cenarro, Falc{\'o}n-Barroso, Gorgas, Selam, \& Vazdekis}]{SanchezBlazquez2006}
S{\'a}nchez-Bl{\'a}zquez, P., Peletier, R.~F., Jim{\'e}nez-Vicente, J., {et~al.} 2006, \mnras, 371, 703

\bibitem[{Santucci {et~al.}(2023)Santucci, Brough, van~de Sande, McDermid, Barsanti, Bland-Hawthorn, Bryant, Croom, Lagos, Lawrence, Owers, van~de Ven, Vaughan, \& Yi}]{Santucci2023}
Santucci, G., Brough, S., van~de Sande, J., {et~al.} 2023, \mnras, 521, 2671

\bibitem[{Santucci {et~al.}(2022)Santucci, Brough, van~de Sande, McDermid, van~de Ven, Zhu, D'Eugenio, Bland-Hawthorn, Barsanti, Bryant, Croom, Davies, Green, Lawrence, Lorente, Owers, Poci, Richards, Thater, \& Yi}]{Santucci2022}
Santucci, G., Brough, S., van~de Sande, J., {et~al.} 2022, \apj, 930, 153

\bibitem[{Sarzi {et~al.}(2018)Sarzi, Iodice, Coccato, Corsini, de~Zeeuw, Falc{\'o}n-Barroso, Gadotti, Lyubenova, McDermid, van~de Ven, Fahrion, Pizzella, \& Zhu}]{Sarzi2018}
Sarzi, M., Iodice, E., Coccato, L., {et~al.} 2018, \aap, 616, A121

\bibitem[{Sattler {et~al.}(2025)Sattler, Pinna, Comer{\'o}n, Martig, Falc{\'o}n-Barroso, Mart{\'\i}n-Navarro, \& Neumayer}]{Sattler2025}
Sattler, N., Pinna, F., Comer{\'o}n, S., {et~al.} 2025, \aap, 698, A235

\bibitem[{Sattler {et~al.}(2023)Sattler, Pinna, Neumayer, Falc{\'o}n-Barroso, Martig, Gadotti, van~de Ven, \& Minchev}]{Sattler2023}
Sattler, N., Pinna, F., Neumayer, N., {et~al.} 2023, \mnras, 520, 3066

\bibitem[{Schwarzschild(1979)}]{Schwarzschild1979}
Schwarzschild, M. 1979, \apj, 232, 236

\bibitem[{Schwarzschild(1982)}]{Schwarzschild1982}
Schwarzschild, M. 1982, \apj, 263, 599

\bibitem[{Schwarzschild(1993)}]{Schwarzschild1993}
Schwarzschild, M. 1993, \apj, 409, 563

\bibitem[{Semczuk {et~al.}(2024)Semczuk, Dehnen, Sch{\"o}nrich, \& Athanassoula}]{Semczuk2024}
Semczuk, M., Dehnen, W., Sch{\"o}nrich, R., \& Athanassoula, E. 2024, \aap, 692, A159

\bibitem[{Sersic(1968)}]{Sersic1968}
Sersic, J.~L. 1968, Atlas de Galaxias Australes

\bibitem[{Shaw(1987)}]{Shaw1987}
Shaw, M.~A. 1987, \mnras, 229, 691

\bibitem[{Spriggs {et~al.}(2021)Spriggs, Sarzi, Gal{\'a}n-de Anta, Napiwotzki, Viaene, Nedelchev, Coccato, Corsini, Fahrion, Falc{\'o}n-Barroso, Gadotti, Iodice, Lyubenova, Mart{\'\i}n-Navarro, McDermid, Morelli, Pinna, van~de Ven, de~Zeeuw, \& Zhu}]{Spriggs2021}
Spriggs, T.~W., Sarzi, M., Gal{\'a}n-de Anta, P.~M., {et~al.} 2021, \aap, 653, A167

\bibitem[{Tahmasebzadeh {et~al.}(2024)Tahmasebzadeh, Zhu, Shen, Gadotti, Valluri, Thater, van~de Ven, Jin, Gerhard, Erwin, Jethwa, Zocchi, Lilley, Fragkoudi, de~Lorenzo-C{\'a}ceres, M{\'e}ndez-Abreu, Neumann, \& Guo}]{Tahmasebzadeh2024}
Tahmasebzadeh, B., Zhu, L., Shen, J., {et~al.} 2024, \mnras, 534, 861

\bibitem[{Tahmasebzadeh {et~al.}(2021)Tahmasebzadeh, Zhu, Shen, Gerhard, \& Qin}]{Tahmasebzadeh2021}
Tahmasebzadeh, B., Zhu, L., Shen, J., Gerhard, O., \& Qin, Y. 2021, \mnras, 508, 6209

\bibitem[{Tahmasebzadeh {et~al.}(2022)Tahmasebzadeh, Zhu, Shen, Gerhard, \& van~de Ven}]{Tahmasebzadeh2022}
Tahmasebzadeh, B., Zhu, L., Shen, J., Gerhard, O., \& van~de Ven, G. 2022, \apj, 941, 109

\bibitem[{Thater {et~al.}(2025)Thater, Jethwa, Lilley, Zocchi, Santucci, \& van~de Ven}]{Thater2025}
Thater, S., Jethwa, P., Lilley, E.~J., {et~al.} 2025, in IAU Symposium, Vol. 379, Dynamical Masses of Local Group Galaxies, ed. P.~{Bonifacio}, M.-R. {Cioni}, F.~{Hammer}, M.~S. {Pawloswki}, \& S.~{Taibi}, 314--318

\bibitem[{Tremaine \& Weinberg(1984b)}]{Tremaine1984b}
Tremaine, S. \& Weinberg, M.~D. 1984b, \mnras, 209, 729

\bibitem[{van~de Ven {et~al.}(2025)van~de Ven, Falc{\'o}n-Barroso, \& Lyubenova}]{vdV2025}
van~de Ven, G., Falc{\'o}n-Barroso, J., \& Lyubenova, M. 2025, \araa, 63, 259

\bibitem[{van~den Bosch {et~al.}(2008)van~den Bosch, van~de Ven, Verolme, Cappellari, \& de~Zeeuw}]{vdB2008}
van~den Bosch, R. C.~E., van~de Ven, G., Verolme, E.~K., Cappellari, M., \& de~Zeeuw, P.~T. 2008, \mnras, 385, 647

\bibitem[{van~der Marel \& Franx(1993)}]{vdMarel1993}
van~der Marel, R.~P. \& Franx, M. 1993, \apj, 407, 525

\bibitem[{Vasiliev \& Valluri(2020)}]{Vasiliev2020}
Vasiliev, E. \& Valluri, M. 2020, \apj, 889, 39

\bibitem[{Vazdekis {et~al.}(2015)Vazdekis, Coelho, Cassisi, Ricciardelli, Falc{\'o}n-Barroso, S{\'a}nchez-Bl{\'a}zquez, La~Barbera, Beasley, \& Pietrinferni}]{Vazdekis2015}
Vazdekis, A., Coelho, P., Cassisi, S., {et~al.} 2015, \mnras, 449, 1177

\bibitem[{Vazdekis {et~al.}(2010)Vazdekis, S{\'a}nchez-Bl{\'a}zquez, Falc{\'o}n-Barroso, Cenarro, Beasley, Cardiel, Gorgas, \& Peletier}]{Vazdekis2010}
Vazdekis, A., S{\'a}nchez-Bl{\'a}zquez, P., Falc{\'o}n-Barroso, J., {et~al.} 2010, \mnras, 404, 1639

\bibitem[{Venhola {et~al.}(2018)Venhola, Peletier, Laurikainen, Salo, Iodice, Mieske, Hilker, Wittmann, Lisker, Paolillo, Cantiello, Janz, Spavone, D'Abrusco, van~de Ven, Napolitano, Verdoes~Kleijn, Maddox, Capaccioli, Grado, Valentijn, Falc{\'o}n-Barroso, \& Limatola}]{Venhola2018}
Venhola, A., Peletier, R., Laurikainen, E., {et~al.} 2018, \aap, 620, A165

\bibitem[{Weinberg(1985)}]{Weinberg1985}
Weinberg, M.~D. 1985, \mnras, 213, 451

\bibitem[{Wu {et~al.}(2025)Wu, Zhu, Chang, Zeng, \& Lei}]{Wu2025}
Wu, X., Zhu, L., Chang, J., Zeng, G., \& Lei, Y. 2025, \aap, 699, A374

\bibitem[{Yoon {et~al.}(2019)Yoon, Im, Lee, Lee, \& Lim}]{Yoon2019}
Yoon, Y., Im, M., Lee, G.-H., Lee, S.-K., \& Lim, G. 2019, Nature Astronomy, 3, 844

\bibitem[{Yu {et~al.}(2021)Yu, Bullock, Klein, Stern, Wetzel, Ma, Moreno, Hafen, Gurvich, Hopkins, Kere{\v{s}}, Faucher-Gigu{\`e}re, Feldmann, \& Quataert}]{Yu2021}
Yu, S., Bullock, J.~S., Klein, C., {et~al.} 2021, \mnras, 505, 889

\bibitem[{Zhao(1996)}]{Zhao1996}
Zhao, H. 1996, \mnras, 278, 488

\bibitem[{Zhu {et~al.}(2022a)Zhu, Pillepich, van~de Ven, Leaman, Hernquist, Nelson, Pakmor, Vogelsberger, \& Zhang}]{Zhu2022a}
Zhu, L., Pillepich, A., van~de Ven, G., {et~al.} 2022a, \aap, 660, A20

\bibitem[{Zhu {et~al.}(2020)Zhu, van~de Ven, Leaman, Grand, Falc{\'o}n-Barroso, Jethwa, Watkins, Mao, Poci, McDermid, \& Nelson}]{Zhu2020}
Zhu, L., van~de Ven, G., Leaman, R., {et~al.} 2020, \mnras, 496, 1579

\bibitem[{Zhu {et~al.}(2022b)Zhu, van~de Ven, Leaman, Pillepich, Coccato, Ding, Falc{\'o}n-Barroso, Iodice, Navarro, Pinna, Corsini, Gadotti, Fahrion, Lyubenova, Mao, McDermid, Poci, Sarzi, \& de~Zeeuw}]{Zhu2022b}
Zhu, L., van~de Ven, G., Leaman, R., {et~al.} 2022b, \aap, 664, A115

\bibitem[{Zhu {et~al.}(2018c)Zhu, van~de Ven, M{\'e}ndez-Abreu, \& Obreja}]{Zhu2018c}
Zhu, L., van~de Ven, G., M{\'e}ndez-Abreu, J., \& Obreja, A. 2018c, \mnras, 479, 945

\bibitem[{Zhu {et~al.}(2018b)Zhu, van~de Ven, van~den Bosch, Rix, Lyubenova, Falc{\'o}n-Barroso, Martig, Mao, Xu, Jin, Obreja, Grand, Dutton, Macci{\`o}, G{\'o}mez, Walcher, Garc{\'\i}a-Benito, Zibetti, \& S{\'a}nchez}]{Zhu2018b}
Zhu, L., van~de Ven, G., van~den Bosch, R., {et~al.} 2018b, Nature Astronomy, 2, 233

\bibitem[{Zhu {et~al.}(2018a)Zhu, van~den Bosch, van~de Ven, Lyubenova, Falc{\'o}n-Barroso, Meidt, Martig, Shen, Li, Yildirim, Walcher, \& Sanchez}]{Zhu2018a}
Zhu, L., van~den Bosch, R., van~de Ven, G., {et~al.} 2018a, \mnras, 473, 3000

\end{thebibliography}
\end{document}